\documentclass[traditabstract]{aa}

\usepackage{graphicx}
\usepackage{epstopdf}
\usepackage{txfonts}
\usepackage{psfrag}

\begin{document}
%
%
\newcommand{\nc}{\newcommand}
\nc{\bea}{\begin{eqnarray}} \nc{\eea}{\end{eqnarray}}
\nc{\beq}{\begin{equation}} \nc{\eeq}{\end{equation}}
\nc{\ve}[1]{{\mathbf{#1}}}
 \nc{\beps}{\boldsymbol{\varepsilon}}
 \nc{\bP}{\mathbf{P}}
 \nc{\bQ}{\mathbf{Q}}
 \nc{\bR}{\mathbf{R}}
 \nc{\bF}{\mathbf{F}}
 \nc{\bB}{\mathbf{B}}
 \nc{\bZ}{\mathbf{Z}}
 \nc{\bX}{\mathbf{X}}
 \nc{\bI}{\mathbf{I}}
 \nc{\bIt}{\mathbf{I}_t}
 \nc{\bIm}{\mathbf{I}_m}
 \nc{\bA}{\mathbf{A}}
 \nc{\bC}{\mathbf{C}}
 \nc{\bD}{\mathbf{D}}
 \nc{\bN}{\mathbf{N}}
 \nc{\bH}{\mathbf{H}}
 \nc{\binvN}{\mathbf{N}^{-1}}
 \nc{\fk}{f_\mathrm{k}}
 \nc{\etal}{{et al.}}

\def\,{\thinspace}
\def\sup#1{$^{\rm #1}$}
\newcount\tableno  \tableno=0
\newbox\tablebox    \newdimen\tablewidth
\def\tabl #1\par{\centerline{TABLE \number\tableno}
                 \vskip 7pt
                 \centerline{#1}\nointerlineskip\noindent}
\def\enddtable{\tablewidth=\wd\tablebox
    $$\hss\copy\tablebox\hss$$
    \vskip-\lastskip\vskip -2pt}
\def\tablenote#1 #2\par{\begingroup \parindent=0.8em
    \abovedisplayshortskip=0pt\belowdisplayshortskip=0pt
    \noindent
    $$\hss\vbox{\hsize\tablewidth \hangindent=\parindent \hangafter=1 \noindent
    \hbox to \parindent{\sup{\rm #1}\hss}\strut#2\strut\par}\hss$$
    \endgroup}
\def\doubleline{\vskip 3pt\hrule \vskip 1.5pt \hrule \vskip 5pt}
\def\leaderfil{\leaders\hbox to 5pt{\hss.\hss}\hfil}
\font\csc=cmcsc10 at 11pt
\def\pdeg{\ifmmode $\setbox0=\hbox{$^{\circ}$}\rlap{\hskip.11\wd0 .}$^{\circ}
          \else \setbox0=\hbox{$^{\circ}$}\rlap{\hskip.11\wd0 .}$^{\circ}$\fi}
\def\arcs{\ifmmode {^{\scriptscriptstyle\prime\prime}}
          \else $^{\scriptscriptstyle\prime\prime}$\fi}
\def\arcm{\ifmmode {^{\scriptscriptstyle\prime}}
          \else $^{\scriptscriptstyle\prime}$\fi}
\newdimen\sa  \newdimen\sb
\def\parcs{\sa=.07em \sb=.03em
     \ifmmode \hbox{\rlap{.}}^{\scriptscriptstyle\prime\kern -\sb\prime}\hbox{\kern -\sa}
     \else \rlap{.}$^{\scriptscriptstyle\prime\kern -\sb\prime}$\kern -\sa\fi}
\def\parcm{\sa=.08em \sb=.03em
     \ifmmode \hbox{\rlap{.}\kern\sa}^{\scriptscriptstyle\prime}\hbox{\kern-\sb}
     \else \rlap{.}\kern\sa$^{\scriptscriptstyle\prime}$\kern-\sb\fi}


\title{Making Maps from Planck LFI 30GHz Data with Asymmetric Beams and Cooler Noise}

\author{%
  M.\ A.\ J.\ Ashdown\inst{1,2}
  \and
  C.\ Baccigalupi\inst{3}
  \and
  J.\ G.\ Bartlett\inst{4}
  \and
  J.\ Borrill\inst{5,6}
  \and
  C.\ Cantalupo\inst{5}
  \and
  G.\ de Gasperis\inst{7}
  \and
  G. de Troia\inst{7}
  \and
  K.\ M.\ G\'{o}rski\inst{8,9,10}
  \and
  E.\ Hivon\inst{9,11}
  \and
  K. Huffenberger\inst{8}
  \and
  E.\ Keih\"{a}nen\inst{12}
  \and
  R.\ Keskitalo\inst{12,13}
  \and
  T.\ Kisner\inst{5}
  \and
  H.\ Kurki-Suonio\inst{12,13}
  \and
  C.\ R.\ Lawrence\inst{8}
  \and
  P.\ Natoli\inst{7,14}
  \and
  T.\ Poutanen\inst{12,13,15}
  \and
  G. Pr\'ezeau\inst{8}
  \and
  M.\ Reinecke\inst{16}
  \and
  G. Rocha\inst{17}
  \and
  M.\ Sandri\inst{18}
  \and
  R.\ Stompor\inst{4}
  \and
  F.\ Villa\inst{18}
  \and
  B.\ Wandelt\inst{19,20}\\
  (The {\sc Planck} CTP Working Group\thanks{The main part of the work reported in this paper was done in May 2006 when the CTP Working Group of
  the {\sc Planck} Consortia met in Trieste. The author list reflects the CTP membership at the time. Since the Trieste meeting, the CTP group has
  received new members who were not involved in this study and whose names do not therefore appear in the author list.})
}

\institute{%
  Astrophysics Group, Cavendish Laboratory, J J Thomson Avenue,
  Cambridge CB3 0HE, United Kingdom.
  \and
  Institute of Astronomy, Madingley Road, Cambridge CB3 0HA,
  United Kingdom.
  \and
  SISSA/ISAS, Via Beirut 4, I-34014 Trieste, and INFN,
  Sezione di Trieste, Via Valerio 2, I-34127, Italy
  \and
  Laboratoire Astroparticule \& Cosmologie, 10 rue Alice Domon \&
  L\'eonie Duquet, 75205 Paris Cedex 13, France (UMR 7164 - Universit\'e
  Paris Diderot, CEA, CNRS, Observatoire de Paris)
  \and
  Computational Cosmology Center, Lawrence Berkeley National
  Laboratory, Berkeley CA 94720, U.\ S.\ A.
  \and
  Space Sciences Laboratory,
  University of California Berkeley, Berkeley CA 94720, U.\ S.\ A.
  \and
  Dipartimento di Fisica, Universit\`{a} di Roma ``Tor Vergata'',
  via della Ricerca Scientifica 1, I-00133 Roma, Italy.
  \and
  Jet Propulsion Laboratory, California Institute of Technology, 4800 Oak
  Grove Drive, Pasadena CA 91109, U.\ S.\ A.
  \and
  California Institute of Technology, Pasadena CA 91125, U.\ S.\ A.
  \and
  Warsaw University Observatory, Aleje Ujazdowskie 4, 00478 Warszawa, Poland.
  \and
  Institut d'Astrophysique de Paris, 98 bis Boulevard Arago,
  F-75014 Paris, France.
  \and
  University of Helsinki, Department of Physics,
  P. O. Box 64, FIN-00014 Helsinki, Finland.
  \and
  Helsinki Institute of Physics, P.\ O.\ Box 64, FIN-00014 Helsinki,
  Finland.
  \and
  INFN, Sezione di Tor Vergata, via della Ricerca Scientifica 1, I-00133 Roma, Italy
  \and
  Mets\"ahovi Radio Observatory, Helsinki University of Technology,
  Mets\"ahovintie 114, 02540 Kylm\"al\"a, Finland
  \and
  Max-Planck-Institut f\"{u}r Astrophysik, Karl-Schwarzschild-Str.~1,
  D-85741 Garching, Germany.
  \and
  Infrared Processing and Analysis Center, California Institute of Technology,
  Pasadena CA 91125, U. S. A.
  \and
  INAF-IASF Bologna, Via P.~Gobetti, 101, I-40129 Bologna, Italy
  \and
  Department of Physics, University of Illinois at
  Urbana-Champaign, 1110 West Green Street, Urbana IL 61801, U.\ S.\ A.
  \and
  Department of Astronomy, University of Illinois at
  Urbana-Champaign, 1002 West Green Street, Urbana IL 61801, U.\ S.\ A.
}

\date{Received 13 June 2008 / Accepted 3 November 2008}

\abstract {The \textsc{Planck} satellite will observe the full sky
at nine frequencies from 30 to 857~GHz. Temperature and polarization
frequency maps made from these observations are prime deliverables
of the \textsc{Planck} mission. The goal of this paper is to examine
the effects of four realistic instrument systematics in the 30~GHz
frequency maps: non-axially-symmetric beams, sample integration,
sorption cooler noise, and pointing errors. We simulated one year
long observations of four 30~GHz detectors. The simulated
timestreams contained cosmic microwave background (CMB) signal,
foreground components (both galactic and extra-galactic), instrument
noise (correlated and white), and the four instrument systematic
effects. We made maps from the timelines and examined the magnitudes
of the systematics effects in the maps and their angular power
spectra. We also compared the maps of different mapmaking codes to
see how they performed. We used five mapmaking codes (two destripers
and three optimal codes). None of our mapmaking codes makes an
attempt to deconvolve the beam from its output map. Therefore all
our maps had similar smoothing due to beams and sample integration.
This is a complicated smoothing, because every map pixel has its own
effective beam. Temperature to polarization cross-coupling due to
beam mismatch causes a detectable bias in the TE spectrum of the CMB
map. The effects of cooler noise and pointing errors did not appear
to be major concerns for the 30~GHz channel. The only essential
difference found so far between mapmaking codes that affects
accuracy (in terms of residual root-mean-square) is baseline length.
All optimal codes give essentially indistinguishable results. A
destriper gives the same result as the optimal codes when the
baseline is set short enough (Madam). For longer baselines
destripers (Springtide and Madam) require less computing resources
but deliver a noisier map.}


\maketitle

\section{Introduction} \label{sec:intro}

Starting in 2003, {\sc Planck} Working Group 3 (the ``CTP'' group)
undertook a comparison of mapmaking codes in increasingly realistic
situations. The approach to realism proceeded in four steps, named
after the locations of working meetings of the group (Cambridge,
Helsinki, Paris, and Trieste). Results from the Cambridge, Helsinki,
and Paris steps have been presented in previous papers (Poutanen et
al.~\cite{Pou06}, Ashdown et al.~\cite{Ash07a}, \cite{Ash07b}). Here
we present results from the Trieste simulations, designed to
determine how mapmaking codes handled four aspects of real {\sc
Planck} data not included in previous simulations. The first was
non-axially-symmetric beams. In previous simulations, we assumed
that the beams on the sky were axially symmetric Gaussians. The
second was the effect of detector sample integration, which
introduces an effective smearing of the sky signal along the
scanning direction. The third was ``cooler noise'', representing the
effect of temperature fluctuations induced in the focal plane by the
20~K sorption cooler. The fourth was the pointing errors. In
previous simulations, we assumed that the detector pointings were
known without error in the mapmaking. In this paper we present the
results of this latest round of simulations, which are realistic
enough to allow us to draw some preliminary conclusions about
mapmaking. We also outline additional work that must be done before
final conclusions can be drawn.

The organization of the paper is as follows. In Section 2 we
describe the simulations that produced the time-ordered data (TOD)
streams that were inputs to our mapmaking. In Section 3 we give the
inputs that were used in these simulations. In Section 4 we describe
the mapmaking codes we used in this study. Section 4 details the
changes that we made in those codes since our earlier Paris
simulation round. Section 5 gives the results of our Trieste
simulation round and the computational resource requirements of our
mapmaking codes are listed in Section 6. Finally we give our
conclusions and proposal for future mapmaking tests in Section 7. In
Appendix A we describe an analytic model that we used in explaining
the effects of beam mismatch in the CMB maps. Appendix A also shows
how we can use this model to correct these effects from the observed
spectrum.

\section{Simulations}\label{sec:simulations}

We used the Level-S simulations pipeline (Reinecke at
al.~\cite{Rei06}) to generate 1-year intervals of simulated detector
observations (time-ordered data streams, or TODs). As in the Paris
round (Ashdown et al.~\cite{Ash07b}) all simulations were done at
30\,GHz, the lowest {\sc Planck} frequency. This was chosen, because
the TOD and maps are the smallest in data size for this frequency,
minimizing the computer resources required for the simulations, and
because the beams are furthest from circular, emphasizing one of the
effects we are trying to study. We simulated the relevant sky
emissions (CMB, dipole, diffuse galactic foreground emissions, and
the strongest extragalactic point sources) in both temperature and
polarization, plus a number of instrumental effects: uncorrelated
(white) noise, correlated ($1/f$) noise, noise from sorption cooler
temperature fluctuations, both circular and elliptical detector
beams, sample integration, semi-realistic nutation of the satellite
spin axis, and fluctuations of the satellite spin rate.

TODs 366\,days long were generated for the four 30\,GHz LFI
detectors (Low Frequency Instrument), with $1.028\times 10^9$
samples per detector corresponding to a sampling frequency of
$f_\mathrm{s}=32.5$\,Hz.

For every sky component we made four different simulated TODs. A TOD
included the effects of either axially symmetric or asymmetric
Gaussian beams and the sample integration was either on or off. We
call these four TODs as
\begin{itemize}
\item Symmetric beams \& no sampling
\item Symmetric beams \& sampling
\item Asymmetric beams \& no sampling
\item Asymmetric beams \& sampling.
\end{itemize}
For the instrument noise (uncorrelated + correlated) we used the
noise TODs of the Paris round (Ashdown et al.~\cite{Ash07b}).
Finally, we had a TOD of the cooler noise. Maps were later made from
different combinations of these TODs.

\section{Inputs}
\label{sec:inputs}

\subsection{Scanning strategy} \label{subsec:scanning}
The correspondence between the sample sequence of the TOD and
locations on the sky is determined by the scan strategy. The {\sc
Planck} satellite will orbit the second Lagrangian point ($L_2$) of
the Earth-Sun system (Dupac \& Tauber~\cite{Dup05}), where it will
stay near the ecliptic plane and the Sun-Earth line.

{\sc Planck} will spin at $\sim1$\,rpm on an axis pointed near the
Sun-Earth line. The angle between the spin axis and the optical axis
of the telescope (telescope line-of-sight, see
Fig.~\ref{fig:focalplane}) is $85\degr$; the detectors will scan
nearly great circles on the sky.  The spin axis follows a circular
path around the anti-Sun direction with a period of six months; the
angle between the spin axis and the anti-Sun direction is 7\pdeg5.
The spin axis thus follows a cycloidal path across the sky, (like
the one we used in our Paris round, Ashdown et al.~(\cite{Ash07b})).
In this simulation the spin axis is repointed hourly. During each
repointing the projection of the spin axis onto the ecliptic moves
by a fixed 2\parcm5. Our simulation had 8784 repointings in total.
We assumed non-ideal satellite motion, with spin axis nutation and
variations in the satellite spin rate.

The scan strategy planned for flight differs from the one used here
only in that instead of repointing once per hour with a fixed offset
in ecliptic longitude, we will repoint in 2\arcm\ intervals along
the cycloid.  To maintain the 2\parcm5\,hr$^{-1}$ average rate of
motion along the ecliptic, the time spent at a given spin axis
pointing will vary somewhat.

 \begin{figure*} [!tbp]
    \begin{center}
    \includegraphics[scale=0.3,angle=90]{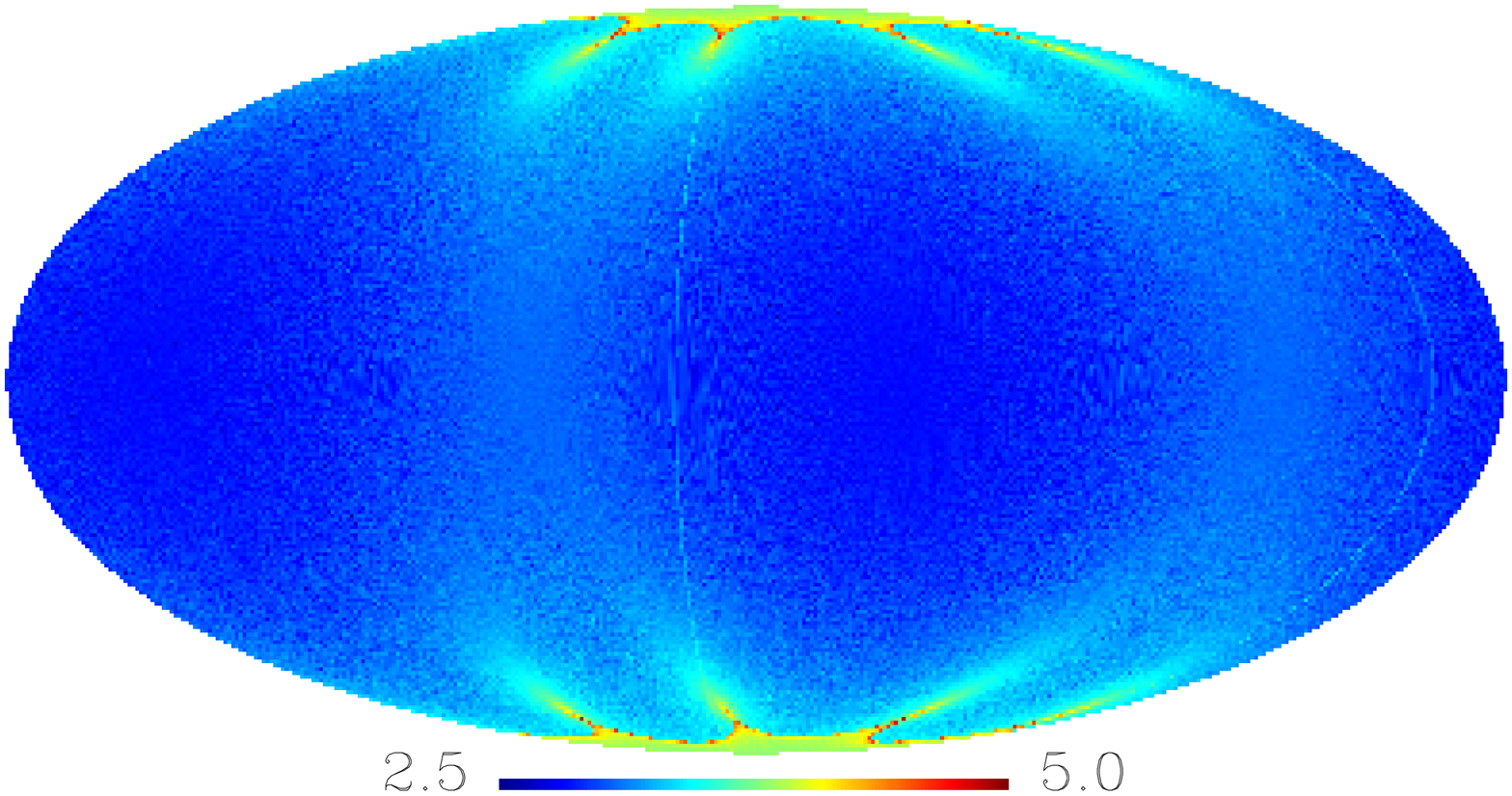}
    \includegraphics[scale=0.3,angle=90]{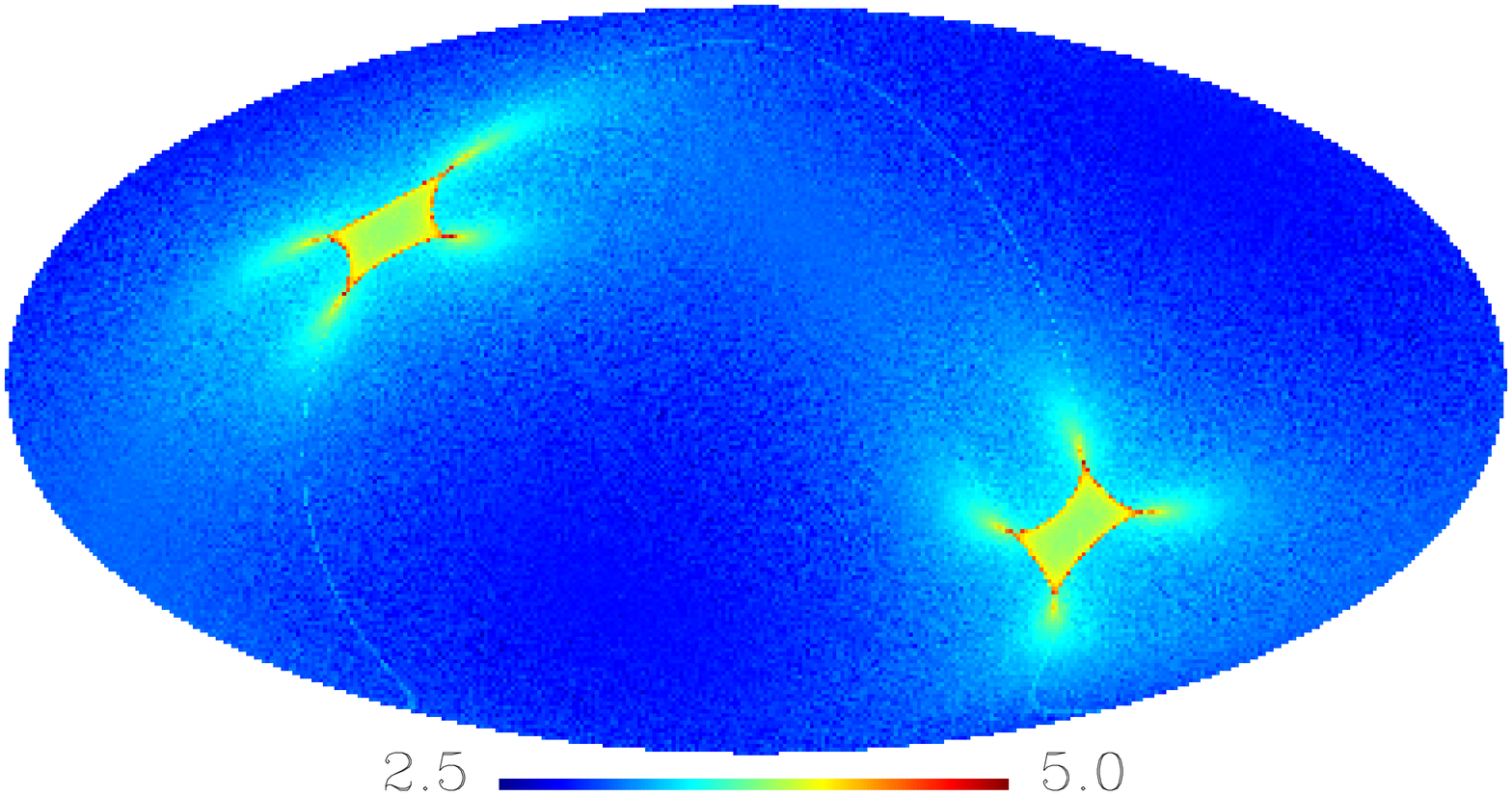}
    \end{center}
\caption{Number of hits per pixel ($n_\mathrm{hit}$) for the scan
strategy applied in this study. The hit map is shown in the ecliptic
(left) and galactic (right) coordinates. The latter map shows the
areas of the ecliptic poles more clearly. Both maps include the hits
of all four LFI 30\,GHz detectors. The scale is
$\log_{10}(n_{\rm{hit}})$.} \label{fig:hits}
 \end{figure*}

Spin rate variations were chosen randomly at every repointing from a
truncated Gaussian probability distribution with parameters
(0\pdeg1\,s$^{-1}$ RMS, 0\pdeg3\,s$^{-1}$ max). The abbreviation
``RMS'' refers to the root-mean-square.

The satellite spin axis nutated continuously according to the
satellite dynamics.  The nutation amplitude was chosen randomly at
every repointing to mimic the disturbance that the repointing
maneuver causes in the spin axis motion. In this simulation the
Gaussian distribution of nutation amplitudes ranged from
$0\parcm006$ to $16\parcm4$, with mean and standard deviation values
of $1\parcm2$ and $0\parcm8$. The nutation amplitudes of the
repointings were all $\le 3\parcm2$ except for two large excursions
(out of 8784) of $4\parcm3$ and $16\parcm4$. This level of nutation
is many times larger than is (now) expected in flight.

The discussion of the performance of the {\sc Planck} pointing
system is outside the scope of this paper. Our simulation of
pointing is based on the results of a detailed simulation of the
spacecraft pointing dynamics. In the simulations of this paper the
satellite attitude (``satellite pointing'') was sampled at 1 Hz. In
flight the {\sc Planck} attitude will be sampled at a higher rate.

We used the HEALPix\footnote{http://healpix.jpl.nasa.gov}
pixelisation scheme (G\'orski et al.~\cite{Gor05a}) with $N_{\rm
side}=512$.  A map of the full sky contains $12N_{\rm side}^2$
pixels.  The Stokes parameters Q and U at a point on the sky are
defined in a reference coordinate system
($\mathbf{e}_{\theta},\mathbf{e}_{\varphi},\mathbf{n}$), where the
unit vector $\mathbf{e}_{\theta}$ is along the increasing $\theta$
direction, $\mathbf{e}_{\varphi}$ is along the increasing $\varphi$
direction, and $\mathbf{n}$ points to the sky (G\'orski et
al.~\cite{Gor05b}). The angles $\theta$ and $\varphi$ are the polar
and azimuth angles of the spherical polar coordinate system used for
the celestial sphere.

The number of hits per pixel from all detectors is shown in
Fig.~\ref{fig:hits}.  At this resolution every pixel was hit.

\subsection{Telescope beams}
\label{subsec:beams}

\begin{figure}[!tbp]
   \begin{center}
     \resizebox{\hsize}{!}{\includegraphics*{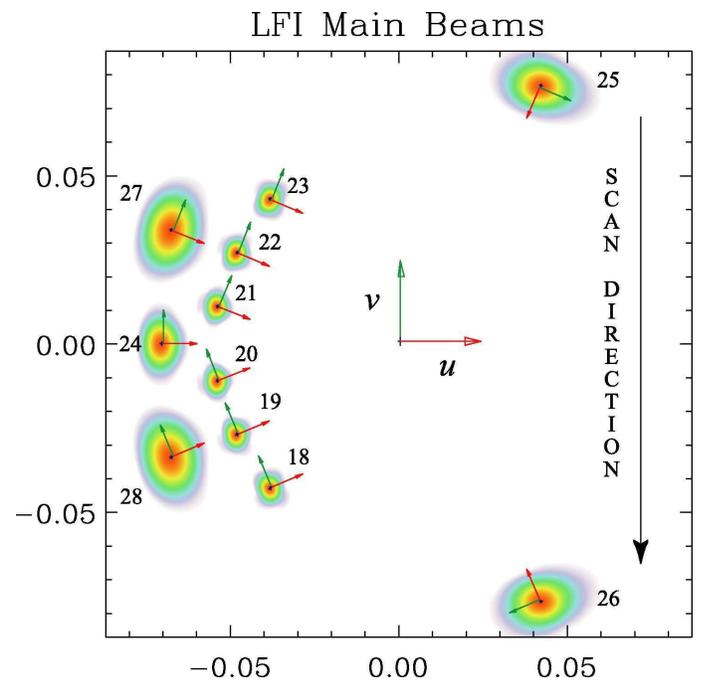}}
   \end{center}
\caption{Footprint of the {\sc Planck} LFI focalplane on the sky as
seen by an observer looking towards the satellite along its optical
axis. The origin of a right-handed $uv$-coordinate system is at the
center of the focalplane (telescope line-of-sight). The $z$-axis is
along the line-of-sight and points towards the observer. Labels
``18--23'' refer to 70\,GHz horns, ``24--26'' refer to 44\,GHz
horns, and ``27--28'' refer to 30\,GHz horns. Each beam has its own
coordinate system as shown in the figure.  The coordinate axes show
the polarization orientations for the co--polar beams of the ``a''
and ``b'' radiometers. The focalplane scans the sky as the satellite
spins. The scanning direction is indicated by an arrow. The $+u$
axis points to the spin axis of the satellite. The centers of the
30\,GHz beams sweep $\backsim$$1\degr$ from the ecliptic poles when
the spin axis is in the ecliptic plane. In Appendix~\ref{sec:model}
we need the angle from the scan direction to the $u$-axis of horn
``27''. This angle is 67\pdeg5.} \label{fig:focalplane}
\end{figure}

The horns of the LFI detectors sit in the {\sc Planck} telescope
focalplane (Fig.~\ref{fig:focalplane}). The center of the field of
view, which is empty in the figure, is populated with the beams of
the HFI bolometers (High Frequency Instrument).  There are two
30\,GHz horns in the focalplane. The corresponding beams are
labelled with ``27'' and ``28'' in Fig.~\ref{fig:focalplane}. Behind
each horn we have two detectors tuned to orthogonal linear
polarizations, called LFI-27a, LFI-27b, LFI-28a, and LFI-28b. For
simplicity, we refer hereafter the field of view as focalplane.

The time-ordered data were simulated using two sets of beams. The
first set were circular Gaussian beams of the same beamwidth for all
of the detectors. The second set were the best-fit elliptical beams
for the LFI 30\,GHz detectors. The beam parameters that we used in
our simulations, are given in Table~1. These were obtained by
fitting a bivariate Gaussian to the co--polar component of each beam
over the whole angular area in which each beam was calculated. For
the 30 GHz beams this was $-0.026 <$ $u$,$v$ $< 0.026$\footnote{$u$
and $v$ are equal to $\sin(\theta)\cos(\varphi)$ and
$\sin(\theta)\sin(\varphi)$, where $\theta$ and $\varphi$ are the
polar and azimuth angles of the spherical polar coordinates of the
beam coordinate system (see Fig.~\ref{fig:focalplane}). The
$uv$-coordinate system is applied in the antenna beam pattern
representations since it permits to map the beam from the spherical
surface to a plane.}.

\begin{figure*}[!tbp]
\centering
\begin{tabular}{cccc}
\includegraphics[scale=0.26]{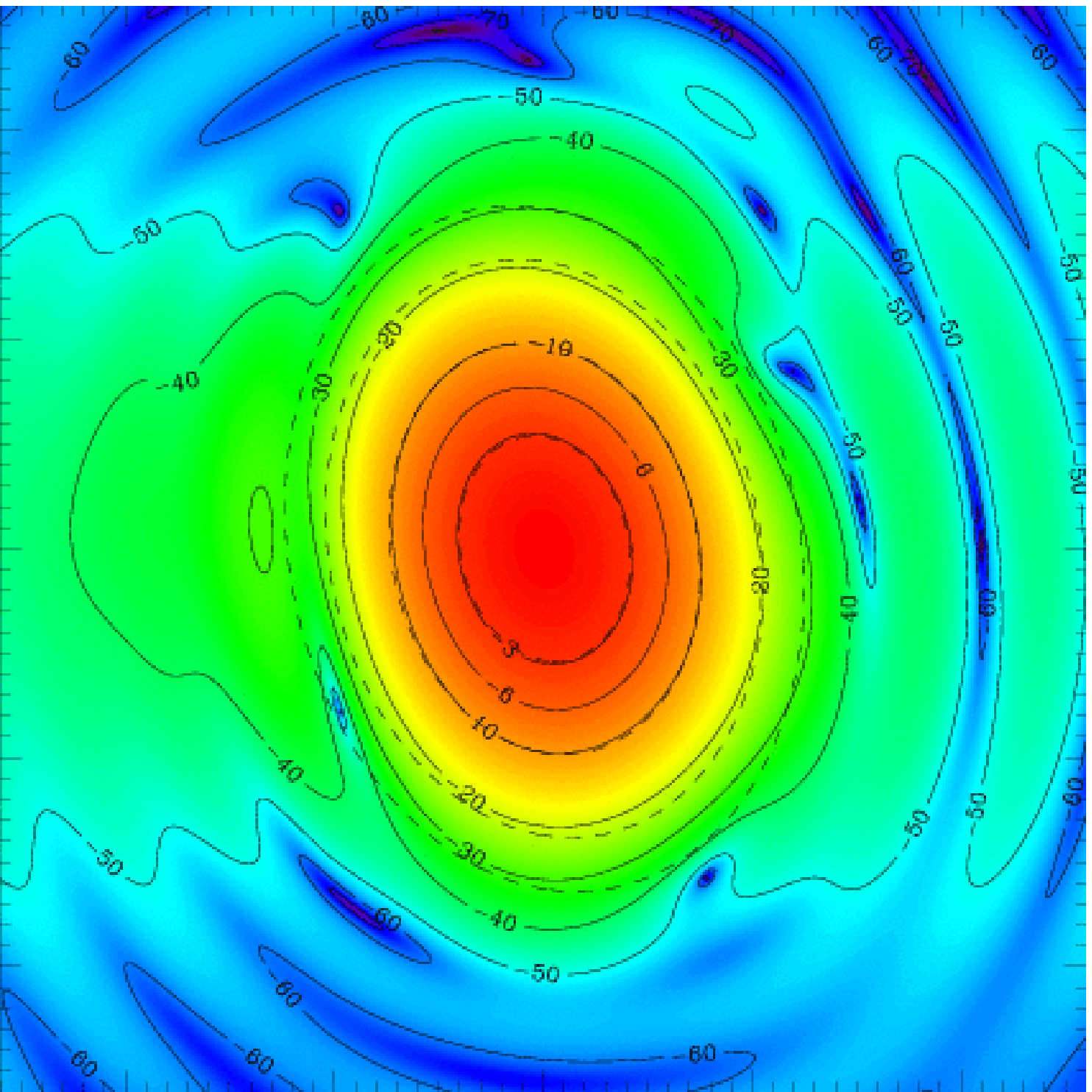} &
\includegraphics[scale=0.26]{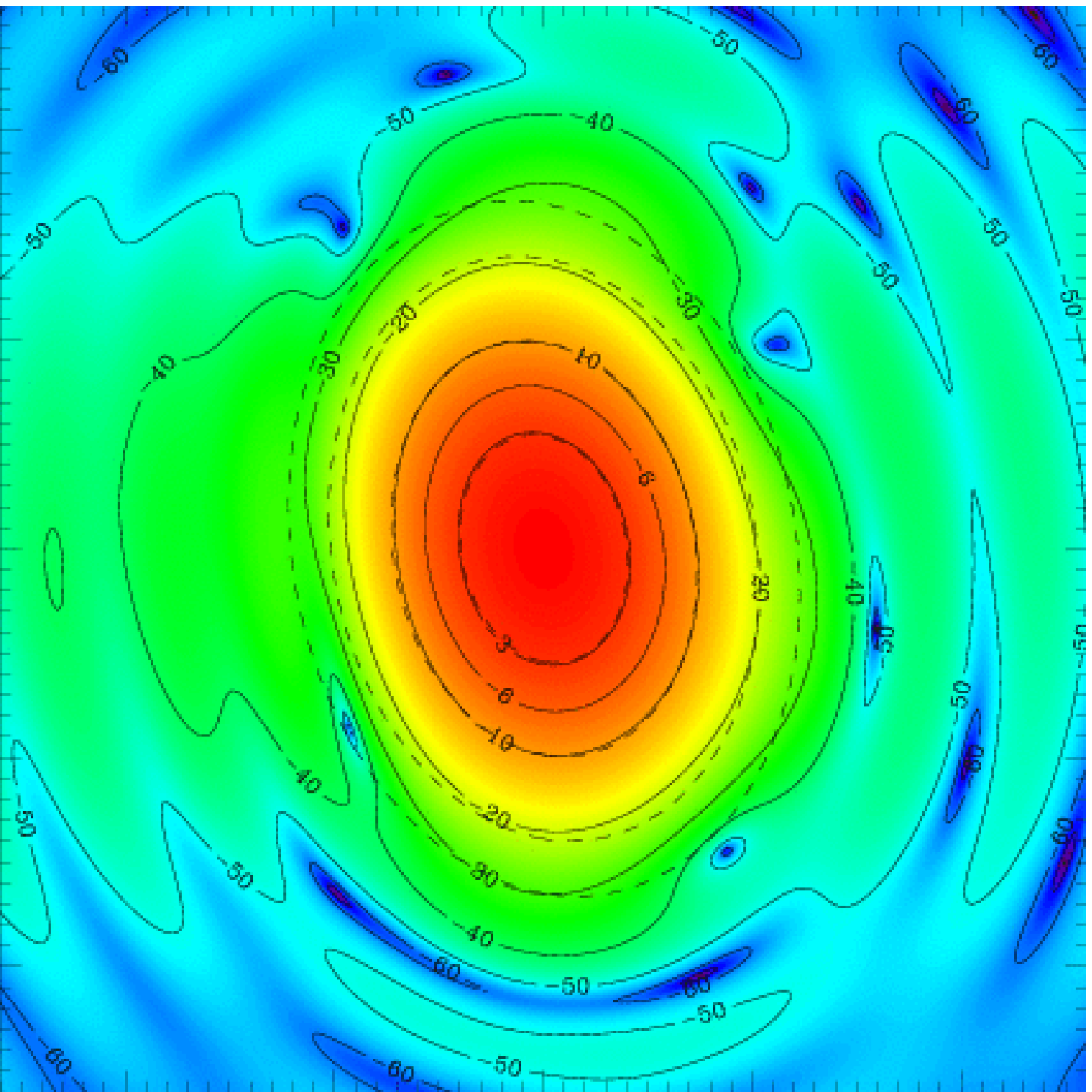} &
\includegraphics[scale=0.26]{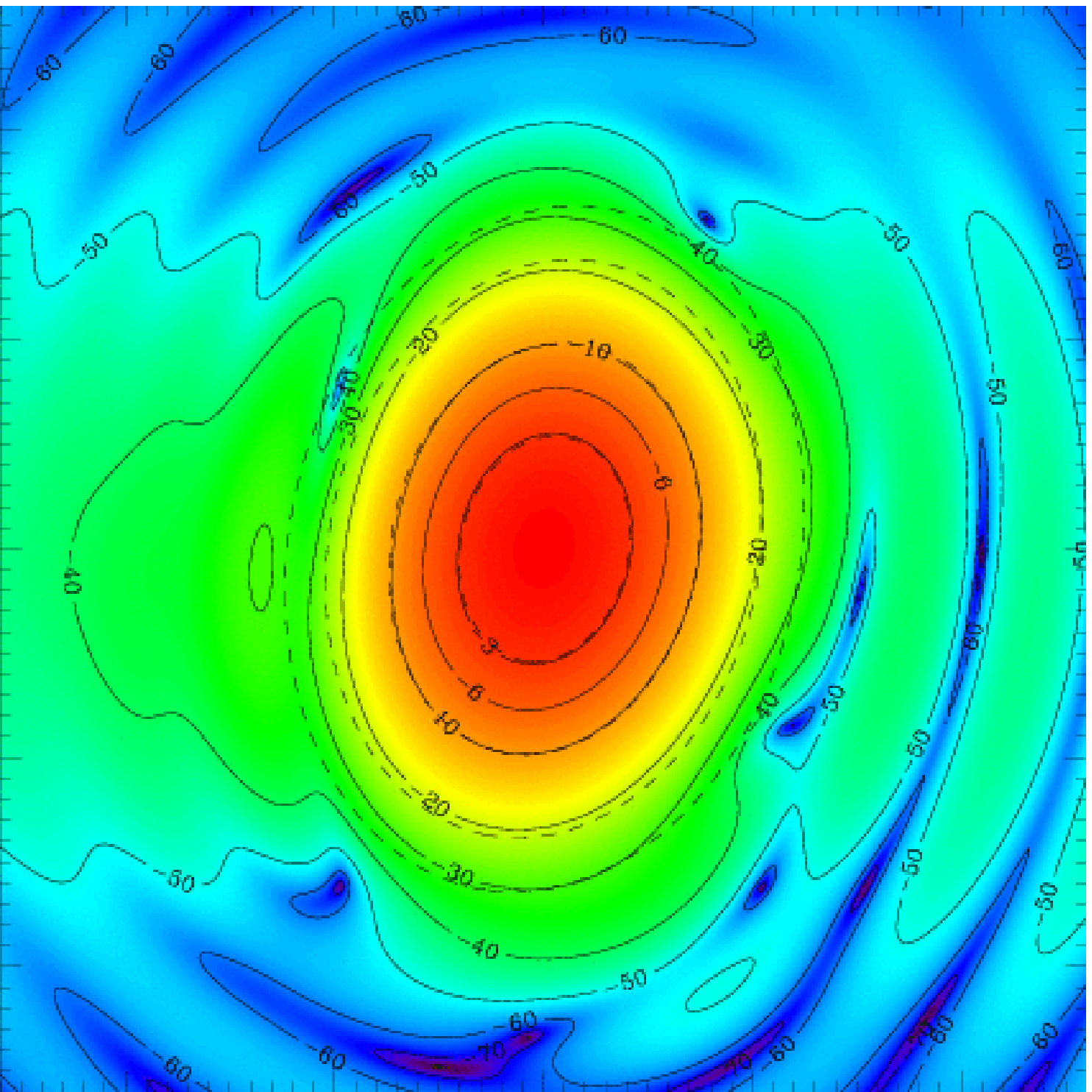} &
\includegraphics[scale=0.26]{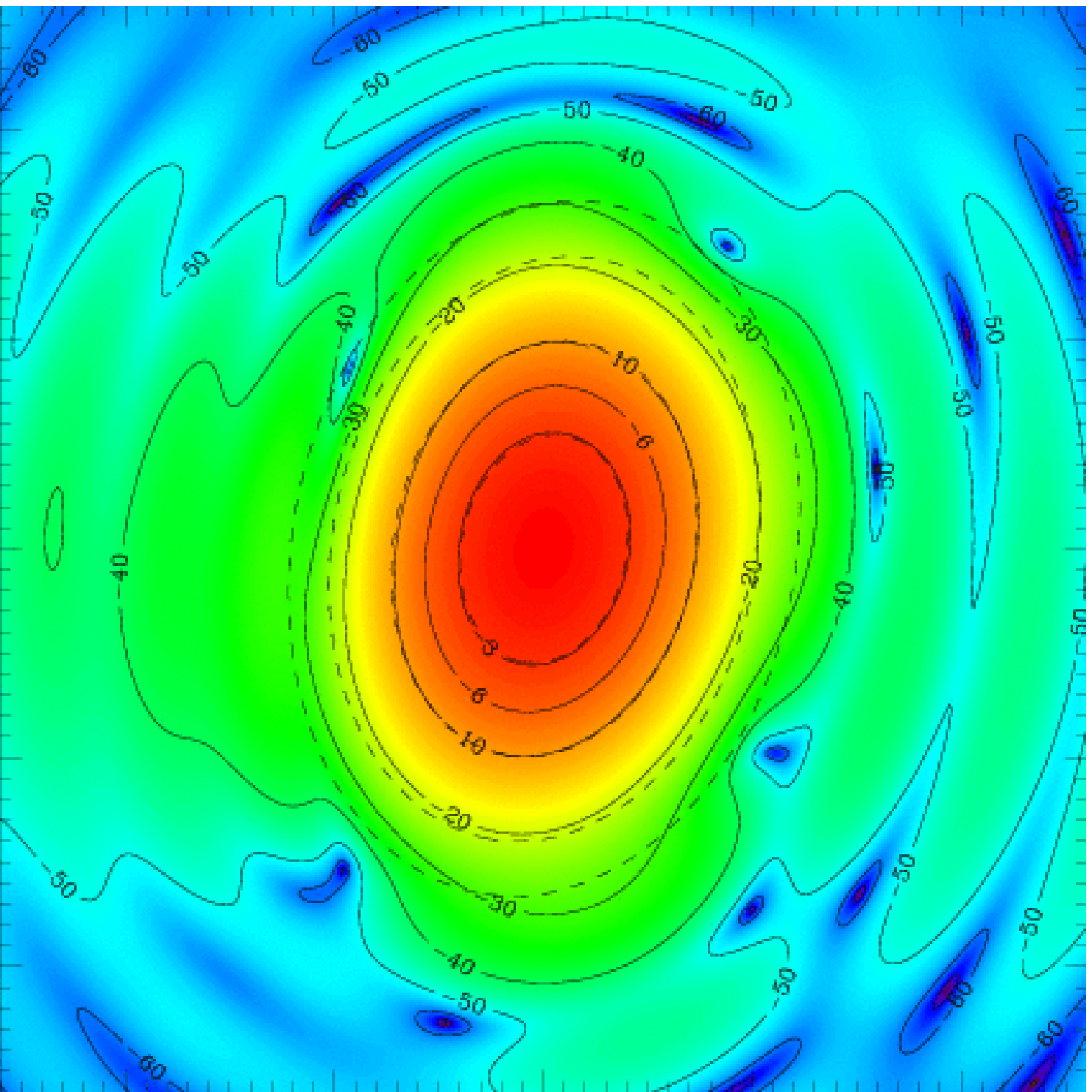}
\end{tabular}
\caption{Contour plot in the $uv$-plane ($-0.026 <$ $u$,$v$ $<
0.026$) of the main beam co-polar component computed for the 30 GHz
feed horns, assuming an ideal telescope. The color scale goes from
-90 to 0 dB. The fit bivariate Gaussian contours are superimposed
with dotted lines. From left to right the beams are for LFI-27a/b
and LFI-28a/b. They are perfectly symmetric beams with respect to
the $u$-axis because of the symmetry of the {\sc Planck} LFI
optics.} \label{fig:main_beam}
\end{figure*}

\begin{table*}[!tbp]
\global\advance\tableno by 1 \tabl {\csc Beams} \par
\setbox\tablebox=\vbox{
   \newdimen\digitwidth
   \setbox0=\hbox{\rm 0}
   \digitwidth=\wd0
   \catcode`*=\active
   \def*{\kern\digitwidth}
   \newdimen\signwidth
   \setbox0=\hbox{+}
   \signwidth=\wd0
   \catcode`!=\active
   \def!{\kern\signwidth}
\halign{\hbox to 0.7in{#\leaderfil}\tabskip=2em&
     \hfil#\hfil\tabskip=1em&
     \hfil#\hfil&
     \hfil#&
     \hfil#\hfil&
     \hfil#&
     \hfil#\hfil&
     \hfil#\tabskip=0pt\cr
\noalign{\doubleline} \omit \hfil Detector\hfil&FWHM$^a$&
   Ellipticity$^b$&$\psi_{\rm pol}^{c}$&
   $\psi_{\rm ell}^{d}$&\omit\hfil$\phi_{\rm uv}^{e}$\hfil&
   $\theta_{\rm uv}^{e}$&\omit\hfil$\psi_{\rm uv}^{e}$\hfil\cr
\noalign{\vskip 3pt\hrule\vskip 5pt} \multispan8\hfil\csc
Symmetric\hfil\cr \noalign{\vskip 4pt}
27a&$32\parcm1865$&1.0&0\pdeg2&$\ldots$&153\pdeg6074&4\pdeg3466&$-$22\pdeg5\cr
27b&$32\parcm1865$&1.0&89\pdeg9&$\ldots$&153\pdeg6074&4\pdeg3466&$-$22\pdeg5\cr
28a&$32\parcm1865$&1.0&$-$0\pdeg2&$\ldots$&$-$153\pdeg6074&4\pdeg3466&22\pdeg5\cr
28b&$32\parcm1865$&1.0&$-$89\pdeg9&$\ldots$&$-$153\pdeg6074&4\pdeg3466&22\pdeg5\cr
\noalign{\vskip 3pt\hrule \vskip 5pt} \multispan8\hfil\csc
Asymmetric\hfil\cr \noalign{\vskip 4pt}
27a&$32\parcm2352$&1.3562&0\pdeg2&101\pdeg68&153\pdeg6074&4\pdeg3466&$-$22\pdeg5\cr
27b&$32\parcm1377$&1.3929&89\pdeg9&100\pdeg89&153\pdeg6074&4\pdeg3466&$-$22\pdeg5\cr
28a&$32\parcm2352$&1.3562&$-$0\pdeg2&*78\pdeg32&$-$153\pdeg6074&4\pdeg3466&22\pdeg5\cr
28b&$32\parcm1377$&1.3929&$-$89\pdeg9&*79\pdeg11&$-$153\pdeg6074&4\pdeg3466&22\pdeg5\cr
 \noalign{\vskip 5pt\hrule\vskip 3pt}}}
\enddtable
\tablenote a Geometric mean of full width at half maximum (FWHM) of
the major and minor axes of the beam ellipse. Symmetric beam FWHM
was chosen to be the arithmetic mean of the two FWHMs of the
asymmetric beams. In practice the beamwidths will not be known to
this level of precision, but we give additional significant figures
here to show the level of variation of the widths, and to reflect
what was actually used in the simulations.
\par

\tablenote b Ratio of the FWHMs of major and minor axes.\par

\tablenote c Angle between $u$-axis and polarization sensitive
direction (see Fig.~\ref{fig:focalplane}).\par

\tablenote d Angle between $u$-axis and beam major axis (see
Fig.~\ref{fig:focalplane}). This angle is irrelevant for axially
symmetric beams.\par

\tablenote e Angles giving the position of the detectors in the
focalplane. They give the rotation of the detector $uvz$-coordinate
system from its initial pointing and orientation (aligned with the
telescope line-of-sight $uvz$-axes) to its actual pointing and
orientation in the focalplane (see Fig.~\ref{fig:focalplane}).
\par
\end{table*}

Realistic main beams have been simulated in the co- and x-polar
basis according to the Ludwig's third definition (Ludwig
\cite{Ludwig}) in $uv$-spherical grids with 301~$\times$~301 points
($\Delta u = \Delta v \simeq 10^{-4}$). Each main beam
(Fig.~\ref{fig:main_beam}) has been computed in its own coordinate
system in which the power peak falls in the center of the $uv$-grid
and the major axis of the polarization ellipse is along the
$u$-axis. In this condition, a well defined minimum appears in the
x-polar component in correspondence to the maximum of the co-polar
component.

Main beam simulations have been performed using the physical optics
considering the design telescope geometry and nominal horn location
and orientation on the focalplane as described in Sandri et
al.~(\cite{San04}). The computation was carried out with GRASP8, a
software developed by TICRA\footnote{http://www.ticra.com}
(Copenhagen, Denmark) for analysing general reflector antennas. The
field of the source (feed horn) has been propagated on the
subreflector to compute the current distribution on the surface.
These currents have been used for evaluating the radiated field from
the sub reflector. The calculation of the currents close to the edge
of the scatterer has been modeled by the physical theory of
diffraction. The radiated field from the sub reflector has been
propagated on the main reflector and the current distribution on its
surface is used to compute the final radiated field in the far
field.

In this paper we considered the effects of the co-polar main beams
only and did not include the effects of the x-polar beams in our
simulations. In making maps from data with both circular and
elliptical beams, we can quantify the effect of the elliptical beams
in the maps and in the power spectra derived from the maps.

\subsection{Signal sampling and integration}
\label{subsec:sampling}

The readout electronics of the LFI 30\,GHz channel sample the signal
measured by the detectors at 32.5\,Hz.  The value recorded in each
sample is the average of the measured signal over the period since
the last sample.  This non-zero integration time has the effect of
widening of the beam along the scan direction.  If the spin speed
remains constant throughout the mission, this effect cannot be
separated from the shape of the beam.

To quantify this effect, the TOD have been simulated using two
options for the sampling. The first is to use instantaneous
sampling, where the signal is not integrated over the past sample
period, rather the sample value is given by the sky signal at the
instant the sample is recorded. This option gives an idealized
result to compare to the realistic sampling behaviour.

In the second and realistic option, there is an additional effect
which must be taken into account. In the Level-S simulation
pipeline, the pointing of the detector is sampled at the same rate
as the signal from the detector and given at the instants the
samples are taken. However, the effect of the integration time is to
smear the sample over the past sample period; in effect, the
reported pointing lags the signal by half a sample period.  In order
to minimise the residuals in the mapmaking, some of the codes used
to produce the results in this paper can perform an interpolation to
shift the pointing back by half of a sample period to the middle of
the sample.

\subsection{Noise}
\label{subsec:noise}

\subsubsection{Detector noise}
\label{subsubsec:detnoise}

We used the instrument noise from our Paris round of simulations
(Ashdown et al.~\cite{Ash07b}). Its uncorrelated (white) noise was
simulated at the level specified in the detector database. Its
nominal standard deviation per sample time was $\sigma$ =
1350~$\mu$K (thermodynamic (CMB) scale). Correlated $1/f$ noise was
simulated from a power spectrum with a knee frequency of 50\,mHz and
slope $-1.7$. For the details of the noise generation see Ashdown et
al.~\cite{Ash07b}. Subsequent tests of the 30\,GHz flight detectors
show a lower knee frequency than 50\,mHz, so these simulations can
be taken as providing a conservative upper limit on $1/f$ noise. No
correlation was assumed between the noise TODs of different
detectors.  For the optimal and Madam mapmaking codes, perfect
knowledge of the noise parameter values was assumed in the mapmaking
phase.

\subsubsection{Sorption cooler temperature fluctuations}
\label{subsubsec:coolnoise}

The {\sc Planck} sorption cooler has two interfaces with the
instruments, LVHX1 with HFI and LVHX2 with \hbox{LFI} (LVHX =
Liquid-Vapor Heat eXchanger). The nominal temperature of LVHX1 is
18\,\hbox{K}, providing precooling for the HFI 4-K cooler. The HFI
4-K cooler in turn cools the HFI housing and the LFI reference
loads.  The temperature of the HFI housing is stabilized by a
Proportional-Integral-Differential (PID) control.   LVHX2 determines
the ambient temperature of the LFI front end.  Its nominal
temperature is 20\,\hbox{K}.

Temperature fluctuations from the coolers affect the LFI data in
three ways. First, fluctuations in LVHX2 propagate through the LFI
structure to the LFI horns, resulting in fluctuations in additive
thermal noise from the throats of the horns (where the emissivity is
highest).  Second, fluctuations in the LFI structure driven by LVHX2
propagate to HFI both by radiation and by conduction through struts
supporting the HFI, and thence the LFI reference loads.  Any
temperature fluctuations of the reference loads will appear as
spurious signals in the LFI detectors.   A significant part of these
fluctuations are suppressed by the HFI 4-K PID control, but the
30\,GHz loads are between the struts and the control stage, so
fluctuations are incompletely suppressed.  Third, temperature
fluctuations of the LFI reference loads are also driven by LVHX1,
propagated indirectly through \hbox{HFI}.  This last effect is quite
small.

A coupled LFI/HFI thermal model was not available when the work
reported in this paper was performed, so we were unable to include
all of these effects realistically.  Instead, in this paper we
consider only the direct effect of LVHX2 instabilities on the feeds.

The propagation of LVHX2 temperature fluctuations to LFI output
signals involves two transfer functions (TF):
\begin{itemize}
\item TF1 describes how the temperature fluctuations of the cold end propagate to the temperature fluctuations of the LFI front end.

\smallskip

\item TF2 describes how the fluctuations of the ambient temperature of the LFI front end translate into a variation of the output signal of a detector.

\end{itemize}

The {\sc Planck} LFI instrument team developed TF1 from the LFI
thermal model. TF2 is described in Seiffert et al.~(\cite{Sei02}).
The thermal mass of the LFI front end suppresses fast temperature
variations, so TF1 rolls off steeply at high frequencies.  TF2 is a
constant multiplier that is different for detectors of different
frequency channels.   The impact of the sorption cooler temperature
fluctuations on the output signals of the LFI detectors has been
discussed by Mennella et al.~(\cite{Men02}).

\begin{figure*}[!tbp]
    \begin{center}
    \includegraphics[scale=0.45]{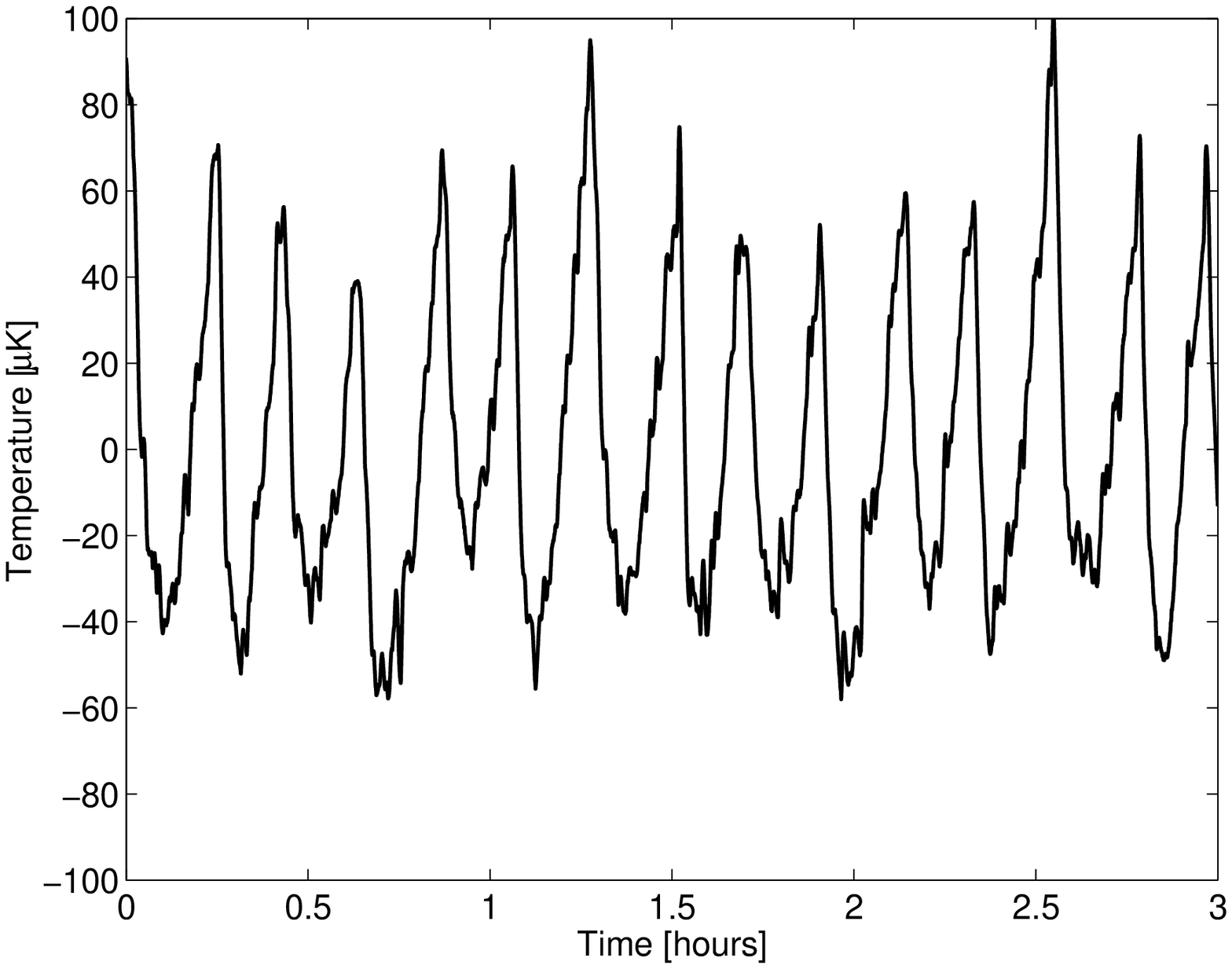}
    \includegraphics[scale=0.45]{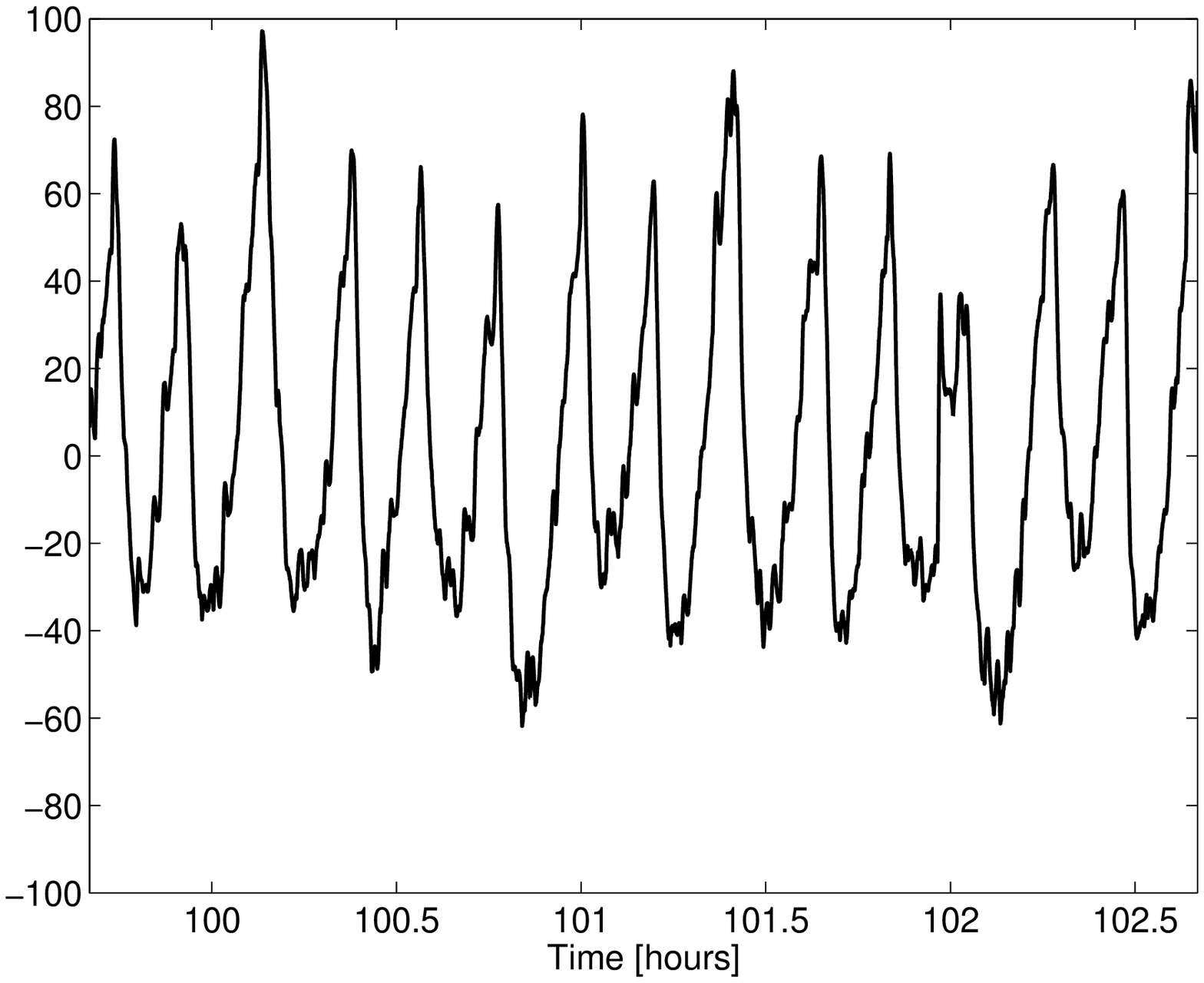}
    \end{center}
\caption{Variations of the output signal (TOD) of an LFI 30\,GHz
detector caused by the temperature fluctuations of the sorption
cooler cold end.  We show here the beginning ({\it left\/}) and the
end ({\it right\/}) of a $\backsim$100 hour segment of data. The
signal is a result of applying LFI thermal transfer functions to the
cold end temperature data measured from the sorption cooler flight
hardware. The thermal transfer functions were derived from the LFI
thermal model developed by the {\sc Planck} LFI instrument team. The
vertical axis is antenna temperature in microkelvins at 30\,GHz.}
\label{fig:cooler_tod}
\end{figure*}

For this study, simulated cooler TODs for the four LFI 30\,GHz
detectors were generated as follows.  The LFI instrument team
applied TF1 and TF2 to a $\backsim$100\,hour sequence of LVHX2 cold
end temperatures taken during cooler operation, producing a
$\backsim$100\,hour segment of data that approximates the
fluctuations as they would appear at the output of an LFI 30\,GHz
detector. Three hours of these data from the beginning and end of
the chunk are shown in Fig.~\ref{fig:cooler_tod}. The cooler signal
has a distinct periodicity, whose cycle time is $\backsim$760\,s.
Every sixth peak is stronger than the other peaks.
Fig.~\ref{fig:cooler_tod} shows that the cycle time and the
amplitude of the sorption cooler fluctuations remain stable over the
$\backsim$100\,hour period.

We used linear interpolation to increase the sampling rate of the
cooler signal (Fig.~\ref{fig:cooler_tod}) from its original 1\,Hz to
the detector sampling rate 32.5\,Hz.  After that we glued a number
of these $\backsim$100\,hour segments one after another to obtain a
one year long cooler \hbox{TOD}. We used a $\backsim$10\,hour
overlap in the boundaries of the successive segments.  The segments
were manually adjusted in the time axis to give a good alignment of
the fluctuation peaks and valleys in the overlap region.   Finally
we multiplied the end of a previous segment with linear weights
descending from 1 to 0 and the beginning of the next segment with
linearly ascending weights (ascending from 0 to 1) before summing
the segments in the overlap region.

The resolution of the LFI thermal transfer function model could not
distinguish between different detectors at 30\,GHz at the time of
this work.  Therefore, in this paper all four LFI 30\,GHz detectors
were represented by the same one-year-long cooler TOD.

\subsection{Dipole}\label{subsec:dipole}

The temperature Doppler shift arises from the (constant) motion of
the solar system relative to the last scattering surface and from
the satellite motion relative to the Sun. The latter signal is
usually used for the calibration of the CMB observations. In this
paper we assumed a perfect calibration and did not study the effects
of calibration errors in our maps. We therefore chose to include the
temperature Doppler shift of the solar system motion in our
simulations, but we did not include the part that arises from the
satellite motion relative to the Sun.

\subsection{CMB}
\label{subsec:cmb}

As in the Paris round (Ashdown et al.~\cite{Ash07b}), the CMB
template used here is WMAP (Wilkinson Microwave Anisotropy Probe)
constrained as described in the following, and included in the
version 1.1 of the {\sc Planck} reference sky\footnote{The CMB and
extragalactic components of the {\sc Planck} reference sky v1.1 used
here are available at
http://people.sissa.it/$\backsim$planck/reference\_sky.  The diffuse
Galactic components are available at
http://www.cesr.fr/$\backsim$bernard/PSM/. The most recent version,
named {\sc Planck} Sky Model, including the CMB template used here,
is available at
http://www.apc.univ-paris7.fr/APC\_CS/Recherche/Adamis/PSM/psky-en.html.}.
It is modelled in terms of the spherical harmonic coefficients, $a_{
\ell m}^{T,E,B}$, where $T$ refers to temperature, and $E$ and $B$
refer to the polarization modes. The $a_{\ell m}^{T,E,B}$ were
determined for multipoles up to $\ell = 3000$.

For $\ell < 70$, the $a_{ \ell m}^{T}$  were obtained by running the
{\it anafast\/} code of the HEALPix package on the first-year WMAP
CMB template obtained by Gibbs sampling the data (Eriksen et
al.~\cite{Eri04}). The $a_{\ell m}^{E}$ were then given by \beq
a_{\ell m}^{E} = a_{ \ell m}^{T} \frac{C_{\ell}^{TE}}{C_{\ell}^{TT}}
+ \left( \frac{x_{\ell m} + i y_{\ell m}}{\sqrt{2}} \right)
\sqrt{C_{\ell}^{EE} - \frac{C_{\ell}^{TE}}{C_{\ell}^{TT}}
C_{\ell}^{TE}},\eeq where $C_{\ell}^{XY}$ $(X,Y = T,E)$ is the best
fit angular power spectrum to the WMAP data, and $x_{\ell m}$ and
$y_{\ell m}$ are Gaussian distributed random variables with zero
mean and unit variance.  For $m=0$, the imaginary part of $y_{\ell
m}$ and the $\sqrt{2}$ were not applied.

For $\ell > 70$, we used the {\it synfast\/} code  to generate the
$a_{\ell m}^{T,E}$ as a random realization of the $C_{\ell}$
coefficients of the theoretical WMAP best fit cosmological model.

\subsection{Foreground emission} \label{subsec:fe}

With the exception of the Sunyaev-Zel'dovich (SZ) signal from
clusters of galaxies, foregrounds have been modelled according to
v1.1 of the {\sc Planck} reference sky, as for the CMB case. In this
section we describe how the various components have been modeled. SZ
sources and extra-Galactic radio sources have been added since
Ashdown et al.~(\cite{Ash07b}).

\subsubsection{Diffuse emission}
\label{subsubsec:de}

We include synchrotron emission from free electrons spiraling around
the Galactic magnetic field and bremsstrahlung emitted by electrons
scattering onto hydrogen ions. We also include the emission from
thermal dust grains; although subdominant with respect to the other
components at intermediate and high Galactic latitudes, the brighest
dusty emission regions across the Galactic plane are still relevant
at 30\,GHz. The total intensity information on these components is
obtained from non-{\sc Planck} frequencies, 408\,MHz and 3000\,GHz
for synchrotron and dust, respectively, as well as $H\alpha$ regions
tracing the bremsstrahlung.  No comparable all-sky information
exists for the linear polarization component. The latter has been
simulated by exploiting data at low and intermediate latitudes in
the radio and microwave bands (see Ashdown et al.~\cite{Ash07b} for
details).

\subsubsection{Extra-galactic radio sources}
\label{subsubsec:egrs}

Emission from unresolved extra-Galactic radio sources has been
obtained from existing catalogues as well as models,
extrapolating to 30\,GHz. The input catalogues were the NRAO
VLA Sky Survey (NVSS, Condon et al.~\cite{Con98}) and the Sydney
University Molonglo Sky Survey (SUMSS, Mauch et al.~\cite{Mau03}) at
1.4\,GHz and 0.843\,GHz, respectively, which cover only part of the
sky, as well as the Parkes-MIT-NRAO (PMN, Wright et
al.~\cite{Wri96a}) survey at 4.85\,GHz, which covers the entire sky
except for tiny regions around the poles. The catalogues were
combined by degrading and smoothing the higher resolution
observations to match those of the lower resolution surveys. To
avoid double counting of background sources, the average flux of the
NVSS and SUMSS surveys was evaluated after the removal of the one at
4.85\,GHz. That average flux was then subtracted from the summed
4.85\,GHz sources in these higher resolution surveys. In order to
obtain a uniform map and account for the fact that the NVSS and
SUMSS have only partial sky coverage, sources were copied randomly
into the survey gaps from other regions until the mean surface
density as a function of the $\sim$1\,GHz flux was equal to the
overall mean down to 5\,mJy. Note that the percentage of simulated
sources is small($\sim$4\%) and mostly located in the Galactic plane.

The frequency extrapolation proceeds as follows. Sources were
divided into two classes according to their spectral index $\alpha$,
where the flux $S$ scales as $\nu^{-\alpha}$: {\it flat spectrum\/}
with $\alpha<0.5$, and {\it steep spectrum\/} with $\alpha\ge 0.5$.
Sources measured at a single frequency were assigned randomly to a
class, with $\alpha$ drawn from two gaussian probability
distributions, one for each class, with mean and variance estimated
from the sample of sources with flux measurements at two
frequencies. In the extrapolation at 30\,GHz, corrections to the
power law approximation were accounted for by including the
multifrequency data from the WMAP (Bennett et al.~\cite{Ben03}) in
order to derive distributions of differences, $\delta\alpha$,
between spectral indices above and below 20\,GHz. For polarization,
the polarization angle was drawn randomly from a flat prior over the
$[0,\pi]$ interval, while the polarization percentage was drawn from
a probability distribution derived from observations concerning the
flat and steep spectrum sources at 20\,GHz.

In the generation of TODs, sources with a flux above 200\,mJy were
treated through the point source convolver code within the Level-S
package; the remaining sources were used to generate a sky map
that was added to the diffuse emissions.

\subsubsection{Sunyaev-Zel'dovich effect from Galaxy clusters}
\label{subsubsec:szefgc}

We used the Monte Carlo simulation package developed by Melin et
al.~(\cite{Mel06}) to generate the SZ cluster catalog in a
$\Lambda$CDM cosmology ($\Lambda$CDM = cold dark matter with dark
energy). Cluster mass $M$ and redshift $z$ were sampled according to
the mass function by Jenkins et al.~(\cite{Jen01}), and we placed
the clusters uniformly on the sky, ignoring any spatial
correlations. The primordial normalization, parametrized by
$\sigma_{8}$, the average mass variance within spheres of
$8h^{-1}$\,Mpc,  was chosen to be $0.99$. We normalized the
temperature-mass relation following Pierpaoli et al.~(\cite{Pie03})
to match the local X-ray temperature function,
$T_*=1.3$\,ke\hbox{V}, and cut the input catalog at $10^{14}$ solar
masses.

The simulation assigns velocities to the cluster halos from the
velocity distribution with variance calculated according to
linear theory. These velocities could be used to calculate the
polarized SZ signal, although this feature was not implemented
in this work. The SZ simulations in this paper are therefore
unpolarized. Future simulations will include SZ polarization.

We attributed to each cluster halo an isothermal $\beta$-model gas
profile at the temperature given by our adopted $T$-$M$ relation,
(see Melin et al.~\cite{Mel06} for details). We fixed $\beta=2/3$
and the core radius of each cluster to $r_{\rm c}=0.1r_{\rm vir}$,
i.e., one tenth of the virial radius $r_{\rm vir}$; the latter was
calculated using the spherical collapse model. The remaining
quantity is the total gas mass (or central density), which we
determined by setting the gas mass fraction $f_{\rm gas}=0.9\cdot
\Omega_{\rm B}/\Omega_{\rm M}$ (baryons and total matter). Thus the
catalogue is characterized by mass, redshift, position on the sky,
gas temperature, and density profile. From this information we
calculate the total integrated SZ flux density, $S_\nu$, at the
observation frequency, and then divide the catalog at 10\,mJy into a
set of bright and faint sources. The bright catalog contained
$\sim$20,000 sources that were used by the point source convolver
code in the Level-S package to generate beam smoothed SZ
point-sources in the TODs. We combined the catalog of fainter
clusters into a sky map that was added to the other diffuse
emissions.

\section{Mapmaking codes}

Two characteristics of mapmaking codes are important. One is {\it
accuracy\/}, that is, how close a given code comes to recovering the
input sky signal in the presence of noise and other mission and
instrumental effects.   The other is {\it resources required\/},
that is, how much processor time, input/output time, memory, and
disk space are required to produce the map.

Ideally, accuracy could be maximized and resource requirements
minimized in one and the same code. Not surprisingly, this is not
the case.  However, one can imagine different regimes of mapmaking,
with different requirements.  On the one hand,  {\it
high-accuracy\/} will be of paramount importance for the {\sc
Planck} legacy maps.  Because such maps need be produced
infrequently, the code can be quite demanding of resources if
necessary.  On the other hand, {\it resources required\/} will be
critical in the intermediate steps of the {\sc Planck\/} data
analysis (e.g., in systematics detection, understanding, and
removal), where a great many maps must be made, and where Monte
Carlo methods will be needed to characterise noise, errors, and
uncertainties.

We used mapmaking codes of two basic types, ``destripers'', and
``optimal'' codes (sometimes called generalized least squares or GLS
codes, notwithstanding the fact that destriping codes also solve GLS
equations). Key features of the mapmaking codes are summarized in
Table~2.

\begin{table*}[!tbp]
\global\advance\tableno by 1 \tabl {\csc Features} \par
\setbox\tablebox=\vbox{
   \newdimen\digitwidth
   \setbox0=\hbox{\rm 0}
   \digitwidth=\wd0
   \catcode`*=\active
   \def*{\kern\digitwidth}
   \newdimen\signwidth
   \setbox0=\hbox{+}
   \signwidth=\wd0
   \catcode`!=\active
   \def!{\kern\signwidth}
\halign{\hbox to 1.8in{#\leaderfil}\tabskip=1em&
     \hfil#\hfil\tabskip=1em&
     \hfil#\hfil&
     \hfil#\hfil&
     \hfil#\hfil&
     \hfil#\hfil\tabskip=0pt\cr
\noalign{\doubleline} \omit\hfil
Code\hfil&Madam&MADmap&MapCUMBA&ROMA&Springtide\cr \noalign{\vskip
3pt\hrule\vskip 5pt}
Algorithm&Destriping&Optimal&Optimal&Optimal&Destriping\cr Noise
estimate needed&Optional$^{\rm a}$&Yes&Yes&Yes&No\cr
Baselines&$>1$\,s&$\ldots^{\rm b}$&$\ldots^{\rm b}$&$\ldots^{\rm
b}$&$>1$\,min\cr Compressed pointing&Yes&Yes&No&No&Yes\cr Shifted
pointing$^{\rm c}$&Yes&Yes&Yes&No&Yes\cr Small memory
mode&Yes&Yes&No&No&Yes\cr Used in DPC$^{\rm d}$
pipeline&LFI&$\ldots$&HFI&LFI&HFI\cr \noalign{\vskip 5pt\hrule\vskip
3pt}}}
\enddtable
\tablenote a Noise estimate is needed for short ($<1$\,min) baselines. \par

\tablenote b Optimal codes may be considered as destripers with a baseline given by the detector sampling rate.\par

\tablenote c To correct the pointing shift caused by the sample
integration.\par

\tablenote d Data Processing Center.\par
\end{table*}

MADmap, MapCUMBA, and ROMA employ optimal algorithms, in the sense
that they compute the minimum-variance map for Gaussian-distributed,
stationary detector noise (see Wright~\cite{Wri96b},
Borrill~\cite{Bor99}, Dor\'e et al.~\cite{Dor01}, Natoli et
al.~\cite{Nat01}, Yvon \& Mayet~\cite{Yvo05}, and de Gasperis et
al.~\cite{deG05} for earlier work on optimal mapmaking). The three
codes operate from similar principles and solve the GLS mapmaking
equation efficiently using iterative conjugate gradient descent and
fast Fourier transform (FFT) techniques.  To be accurate, these
codes require a good estimate of the power spectrum of noise
fluctuations.

Springtide and Madam employ destriping algorithms. They remove
low-frequency correlated noise from the TOD by fitting a sequence of
constant offsets or ``baselines'' to the data, subtracting the
fitted offsets from the TOD, and binning the map from the cleaned
\hbox{TOD} (see Burigana et al.~\cite{Bur97},
Delabrouille~\cite{Del98}, Maino et al.~\cite{Mai99}, \cite{Mai02},
Revenu et al.~\cite{Rev00}, Sbarra et al.~\cite{Sba03}, Keih\"anen
et al.~\cite{Kei04}, \cite{Kei05} for earlier work on destriping and
Efstathiou~\cite{Efs05}, \cite{Efs07} for destriping errors). As we
will see, baseline length is a key parameter for mapmaking codes.
Baseline length is adjustable in destriping codes. In the short
baseline limit, the destriping algorithm (with priors on the
low-frequency noise) is equivalent to the optimal algorithm.
Similarly, optimal codes may be considered as destripers with a
baseline given by the detector sampling rate.

The destriper Springtide operates on scanning rings. First, it
compresses the data by binning them in 1-hour ring maps.  It then
solves for  and subtracts an offset for each ring map, and
constructs the final output map.  Due to compression of data to
rings, Springtide can run in small memory, but its long baselines
(1\,hour) leave larger residuals in the map at small angular scales.
A recent feature allows Springtide to compute the hour-long ring
maps using a one minute baseline destriper, then the final output is
constructed as before.  This double-destriping improves the maps at
the cost of longer runtime.

MADmap, Springtide, and Madam can use compressed pointing
information, meaning that they can interpolate the detector
orientation from the sparsely sampled (1\,Hz) satellite attitude
measurements. The other mapmaking codes require the full set of
detector pointings sampled at the detector sampling rate. The main
benefits of the compressed pointing are significant savings in disk
space and I/O.

Full descriptions of the codes have been given in our previous
papers (Poutanen et al.~\cite{Pou06}, Ashdown et al.~\cite{Ash07a},
\cite{Ash07b}). Changes from previous versions are detailed below.

\subsection{Madam} \label{subsec_madam}

Madam is a destriping code with a noise filter. The user has the
option of turning the noise filter off, in which case no prior
information on noise properties is used.  Mapmaking with Madam for
the case of noise filter turned off is discussed in Keih\"anen et
al.~\cite{Kei08}.

The baseline length is a key input parameter in Madam. The shorter
the baseline, the more accurate are the output maps. It can be shown
theoretically that when the baseline length approaches the inverse
of the sampling frequency, the output map approaches the optimal result.

A number of improvements have been made to Madam since the Paris
round of simulation (Ashdown et al.~\cite{Ash07b}). The code
constructs the detector pointing from satellite pointing, saving
disk space and I/O time. In case two detectors have identical
pointing, as is the case for a pair of LFI detectors sharing a horn
antenna, pointing is stored only once, dropping the memory
requirement to half.

Further, the code uses a lossless compression algorithm which
greatly decreases the memory consumption at long baselines. Madam
also allows a ``split-mode'', where the data are first destriped in
small chunks (e.g., 1 month) using short baselines. These chunks are
then combined and re-destriped, using longer baselines. The
split-mode decreases memory consumption substantially. The cost is
that run time increases and map quality decreases somewhat as
compared to the standard mode. The split-mode can be used in many
ways. One may for instance destripe data from 12 detectors in 3
parts, each consisting of data from four detectors.

With these improvements, Madam offers wide flexibility in terms of
computational resources used. The most accurate maps are obtained
with a short baseline length and fitting all the data
simultaneously. This alternative requires the maximum memory. The
memory requirement can be reduced either by choosing a longer
baseline, or by using the split-mode.

The Madam maps of this study were destriped using a single set of
short baselines (i.e., the split-mode was not used).  Unless otherwise
noted the baseline length was 1.2\,s.

The earlier Polar code (Ashdown et al.~\cite{Ash07a},
\cite{Ash07b}), corresponding to Madam with noise filter turned
off, is now merged into Madam.

\subsection{Springtide}
\label{sec:springtide}

A number of improvements have been made to Springtide since the work
reported in Ashdown et al.~(\cite{Ash07a}) and (\cite{Ash07b}).

Springtide now uses the M3 data abstraction library to read TOD and
pointing (see section~\ref{sec:madmap}).  Instead of reading the
detector pointing information from disk, M3 can use the Generalised
and Compressed Pointing (GCP) library to perform an on-the-fly
calculation of the positions of the detectors from the satellite
attitude data.

Springtide is now capable of making maps at a number of resolutions
in the same run.  This requires that a hierarchical pixelation such
as HEALPix be used for the maps. The destriping must be performed at
a sufficiently high resolution so that the sky signal is
approximately constant across a pixel. Once the offsets describing
the low-frequency noise are subtracted from the rings, they can be
binned to make the output map at any resolution equal to or lower
than that used for the destriping.

\subsection{MapCUMBA}

The current version (2.2) has been modified in several ways relevant
to this study.

While the pointing information is generally provided in spherical
coordinates, the mapmaking algorithm only requires the pixel
indexing.  To reduce the memory expense of storing both forms of
pointing information while mapping one to the other, the pointing is
read from disk into a small buffer (whose size can be adjusted by
the user) and is immediately mapped into the final pixel index
stream. The drawback of this scheme is that it is generally much
faster to read the same amount of data from disk in one piece than
in several small pieces.  In the configuration tested, the I/O
buffer has to be kept larger than $10^7$ samples for the I/O not to
dominate the total run time.

Since the preconditioned conjugate gradient (PCG) algorithm used in
MapCUMBA involves repeated overlap-add Fourier transforms of fixed
length, it is beneficial to use highly optimized FFT algorithm such
as fftw-3\footnote{http://www.fftw.org}. We found the 1D Fourier
transform offered by fftw-3.0.0, with a `measured' plan selection
algorithm and a length of 262144, to be twice as fast as the one
implemented in fftw-2.1.5, offsetting both the more cumbersome
interface of fftw-3, and the overhead associated with the `measured'
plan selection over the `estimated' one.

\subsection{MADmap}
\label{sec:madmap}

MADmap is the optimal mapmaking component of the MADCAP suite of
tools, specifically designed to analyse large CMB data sets on the
most massively parallel high performance computers. Recent
refinements to MADmap include options to reduce the memory
requirement (at the cost of some additional computations), and to
improve the computational efficiency by choosing the distribution of
the time ordered data over the processors to match the requirements
of a particular analysis.  Like Springtide, MADmap uses the M3 data
abstraction and the GCP libraries.

M3 allows an applications programmer to make a request to read a
data subset that is independent of the file format of the data and
the way the data are  distributed across files. In addition it
supports ``virtual files'', which do not exist on disk and whose
data are constructed on the fly---specifically used here by MADmap
to construct the inverse time-time noise correlation functions from
a spectral parametrization of the noise. This abstraction is
mediated through the use of an XML description of the data called a
run configuration file (or runConfig), which also provides a
convenient way of ensuring that {\em exactly} the same analysis is
executed by different applications---MADmap and Springtide here.

GCP provides a way to reduce the disk space and IO requirements of
mapmaking codes.  Instead of storing the explicit pointing solution
for every sample of every detector, we store only the pointing of
the satellite (generalized) every second (compressed) and
reconstruct via M3 the full pointing for a particular set of samples
for a particular detector through on-the-fly interpolation and
translation only when it is requested by the application. In this
analysis---mapping only the four slowest-sampled of 72 {\sc Planck}
detectors, comprising a little over 1\% of the data---the use of the
GCP library reduced the disk space and IO requirements for MADmap
and Springtide from 92 to 1.5\,GB.

MADmap uses the PCG algorithm to solve the GLS problem. This
requires applying the pointing matrix and its transpose once on each
iteration of the PCG routine.  When running in high memory mode the
sparse satellite pointing is expanded once using GCP, and the entire
portion of the pointing matrix for the time samples assigned to a
processor is stored in its memory (in packed sparse form) to be
reused in each PCG iteration. When running in low memory mode only
small portions of the pointing matrix that fit into a buffer are
expanded in sequence and the buffer sized pointing matrix is used
and then overwritten.  This exchanges memory used for computation
time spent within the GCP library, but this allows for the analysis
of very large data sets on systems where the memory requirements
would otherwise be prohibitive.  The GCP library has been optimized
to be very computationally efficient, including the use of
vectorized math libraries, and typically consumes about 1/3 of the
run time in a MADmap job in low memory mode. In low memory mode the
memory consumption scales with the number of pixels, essentially
independent of the (very much larger) number of time samples.

The other feature recently added to MADmap is an alternative
distribution of time ordered data over the processors.  The only
distribution available in previous versions of MADmap was to
concatenate all the detectors' time streams into a single vector and
distribute this over the processors so that each processor analyses
the same number of contiguous detector samples.  This distribution
is still an option in MADmap, but there is now an alternative which
saves memory and cycles in certain circumstances by reducing the
amount of compressed pointing data that each processor must
calculate and store.  In the alternative distribution each processor
analyses data for a distinct interval of time for all of the
detector samples that occur in that time interval.  These time
intervals are chosen so that each processor has the same total
number of samples (regardless of gaps in detector data).  This
implies that each processor analyses data from every detector.  Note
that each processor stores a distinct portion of the compressed
pointing (modulo small overlaps to account for noise correlations).
This distribution will also allow future versions of MADmap to
include the analysis of inter-channel noise correlations.  The runs
described in this paper were all done with the original concatenated
distribution for time ordered data.

\subsection{ROMA}

ROMA is now stable at version 5.1 (the same as was employed for
Ashdown et al.~(\cite{Ash07a}), (\cite{Ash07b})), which makes use of
fftw-3.1.2. Nonetheless, a few minor improvements have been
performed to optimize speed (by tuning some fftw parameters) and
memory usage. In addition, a new I/O module has been developed for
the sake of the present simulations that allows timelines containing
different sky components to be read in quickly and then mixed
together.

\section{Results of mapmaking} \label{sec:results}

In this section we quantify the results of our mapmaking exercise.
Our main goal was to examine the effects of detector main beams. To
compare the maps and the effects of systematics upon them, we define
three auxilliary maps:
\begin{itemize}
\item \textit{input map} represents the true sky. In some
cases it contains the CMB alone; other cases it includes the dipole
and foregrounds as well. The CMB part of the input map contains no
$B$-mode power.
\item \textit{smoothed input map} is the input
map smoothed with an axially symmetric Gaussian beam and a pixel
window function\footnote{For the axially symmetric beam, we used
FWHM = $32\farcm1865$, which is the width of the symmetric beams of
this study (see Sect.~\ref{subsec:beams}). For the pixel window we
used the HEALPix pixel window function of $N_{\rm side} = 512$ pixel
size (G\'orski et al.~\cite{Gor05b}).}.
\item \textit{binned noiseless map} refers to the ($I,Q,U$) map obtained by
summing up the noiseless time-ordered data, accounting for the
detector orientation. This is the best map of an ideal noiseless
situation.
\end{itemize}

The binned noiseless map ($\ve{m}^B$) is produced from the noiseless
TOD ($\ve{s}$) as
 \beq
 \ve{m}^B =
 \left(\mathbf{P}^T\mathbf{P}\right)^{-1}\mathbf{P}^T\ve{s},
 \label{eq:binned_map}
 \eeq
where $\mathbf{P}$ is the pointing matrix, which describes the linear
combination coefficients for the $(I,Q,U)$ pixel triplet to produce
a sample of the observed \hbox{TOD}. Each row of the pointing matrix has
three non-zero elements.

The temperature of a pixel of the smoothed input map is an integral
of the (beam smoothed) sky temperature over the pixel area. In the
binned noiseless map the corresponding pixel temperature is not a
perfect integral, but a mean of the observations falling in that
pixel. This pixel sampling is not as uniform as the integration in
the smoothed input map. Therefore the difference of the binned
noiseless map and the smoothed input map has some pixel scale power
due to this difference in pixel sampling. We call this difference
\textit{pixelization error} in this paper. For asymmetric beams
there is the additional effect that different observations centered
on the same pixel may fall on it with a different orientation of the
beam, resulting in a different measured signal.

For CMB, dipole, and foreground emissions we made four different
simulated TODs, depending on whether the beams were axially
symmetric or asymmetric Gaussians, and whether the sample
integration was on or off (see Sect.~\ref{sec:simulations}).

The widths of the symmetric beams were identical in all four LFI
30\,GHz detectors, whereas the widths and orientations of the
asymmetric beams were different. The difference of the detector beam
responses is called the \textit{beam mismatch}. For the case of
sample integration, the time stamps of the detector pointings were
assigned to the middle of the sample integration interval. When the
sample integration was turned off, the observations of the sky
signal were considered instantaneous and the timings of the detector
pointings coincided with them.

The mapmaking methods discussed here utilize the detector pointing
information only to the accuracy given by the output map pixel size.
The methods use the pointing matrix $\mathbf{P}$ (see Eq.
(\ref{eq:binned_map})) to encode the pointings of the detector beam
centers and the directions of their polarization sensitive axes.
None of our mapmaking codes makes an attempt to remove the beam
convolution from its output map. Therefore an output map pixel is
convolved with its own specific response (\textit{effective beam}),
i.e., the mean of the beams (accounting for their orientations)
falling in that pixel. Recently, deconvolution mapmaking algorithms
have been developed that can produce maps in which the smoothing of
the beam response has been deconvolved. These methods lead to maps
that approximate the true sky (Burigana \& S\'aez~\cite{Bur03},
Armitage \& Wandelt~\cite{Arm04}, Harrison et al.~\cite{Har08}).
These methods are sufficiently different from the ones considered
here that different methods of comparison must be used, and we
therefore do not include detailed description of them in this paper.

An output map of a mapmaking code can be considered as a sum of
three components: the \textit{binned noiseless map}, the
\textit{residual noise map}; and an error map that arises from the
small-scale (subpixel) signal structure that couples to the output
map through the mapmaking (Poutanen et al.~\cite{Pou06}, Ashdown et
al.~\cite{Ash07b}). We call the last error map the \textit{signal
error map}. We use the term \textit{residual map}, when we refer to
the sum of the signal error and residual noise maps. Because the
signal error arises from the signal gradients inside the output map
pixels, smaller pixel size (i.e., higher map resolution) leads to a
smaller signal error. Our earlier studies have shown that for a
typical {\sc Planck} map (e.g., $N_{\rm side}$ = 512 or smaller
pixels), the signal error is a tiny effect compared to the CMB
signal itself or to the residual noise (Poutanen et
al.~\cite{Pou06}, Ashdown et al.~\cite{Ash07b}). Therefore a {\sc
Planck} signal map is nearly the same as the corresponding binned
noiseless map. It is a common characteristic of all our maps.

Generally, the binned noiseless map contribution in the output map
is (cf. Eq. (\ref{eq:binned_map}))
 \beq
 \ve{m}^B =
 \left(\mathbf{P}^T\mathbf{C}_n^{-1}\mathbf{P}\right)^{-1}\mathbf{P}^T\mathbf{C}_n^{-1}\ve{s},
 \label{eq:binned_map_in_map}
 \eeq
where $\mathbf{C}_n$ is the diagonal time-domain covariance matrix
of the detector white noise floor.  Its diagonal elements will not
equal in general; for example, the white noise RMS of the detectors
can be different. However, in this study all detectors are assumed
to have the same white noise and therefore the matrix $\mathbf{C}_n$
can be ignored and Eq. (\ref{eq:binned_map}) gives the correct
binned noiseless map contribution.

In this study we assume that the residual noise map contains
residues of the uncorrelated (white) and correlated ($1/f$)
instrument noise and the residues of the sorption cooler
fluctuations.

The ($I,Q,U$) maps we made in this study were pixelized at $N_{\rm
side}$ = 512. At this resolution every pixel was observed (full sky
maps) and their polarization directions were well sampled. The
\textit{rcond}'s of the 3$\times$3 $\mathbf{N}_{\rm obs}$ matrices
were larger than 0.3\footnote{The quantity \textit{rcond}, the
reciprocal of the condition number, is the ratio of the absolute
values of the smallest and largest eigenvalue of the 3$\times$3
$\mathbf{N}_{\rm obs}$ matrix of a pixel. The matrix
$\mathbf{P}^T\mathbf{P}$ is block-diagonal, made up of these
$\mathbf{N}_{\rm obs}$ matrices. For a set of polarized detectors
with identical noise spectra (like the LFI 30\,GHz detectors of this
study) \textit{rcond} is $\le 0.5$}.

Unless otherwise noted the maps are presented in ecliptic
coordinates and thermodynamic (CMB) microkelvins.  The units of
angular power spectra are thermodynamic microkelvins squared.

\begin{figure*} [!tbp]
\begin{center}
\includegraphics[scale=0.3,angle=90]{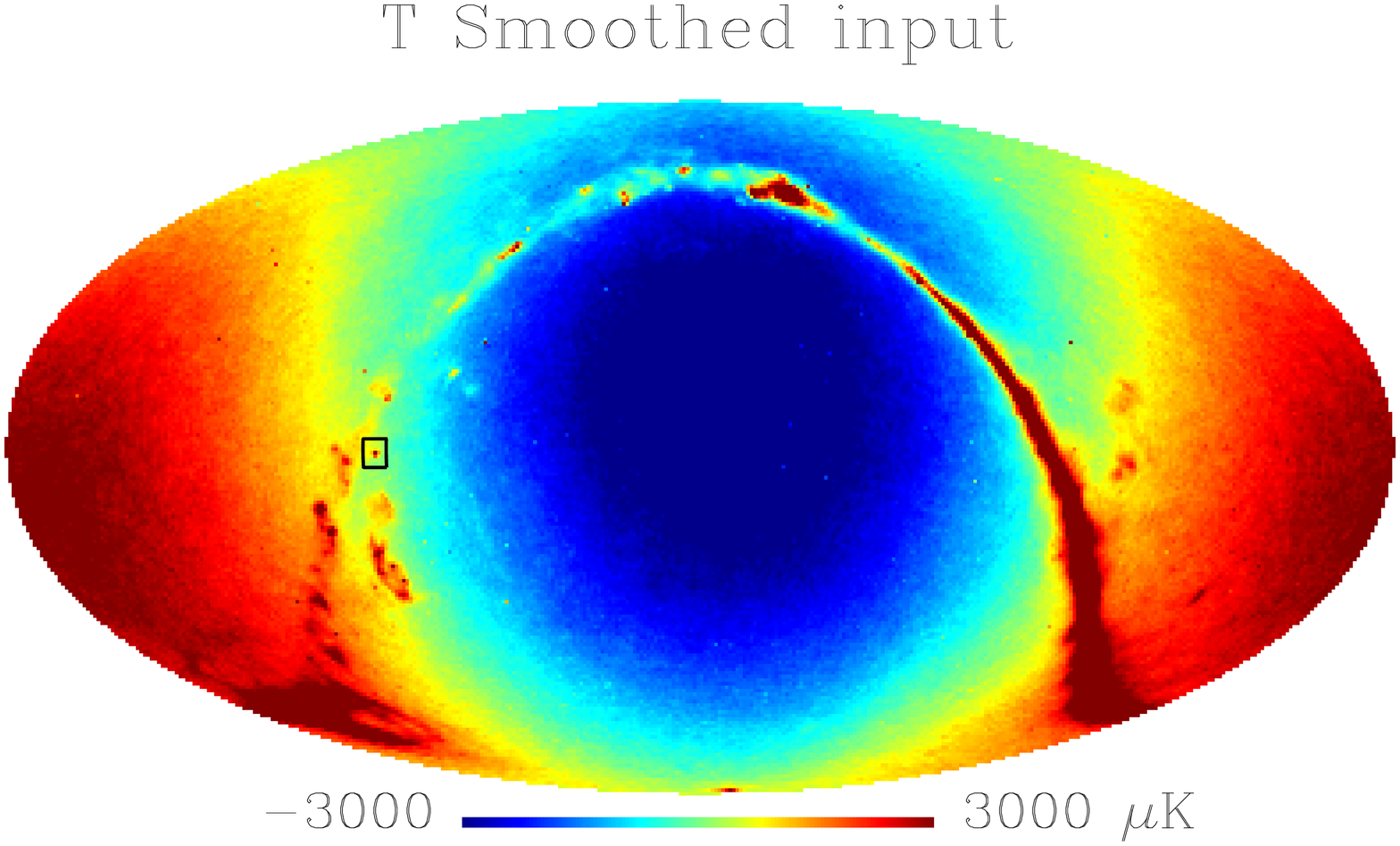}
\includegraphics[scale=0.3,angle=90]{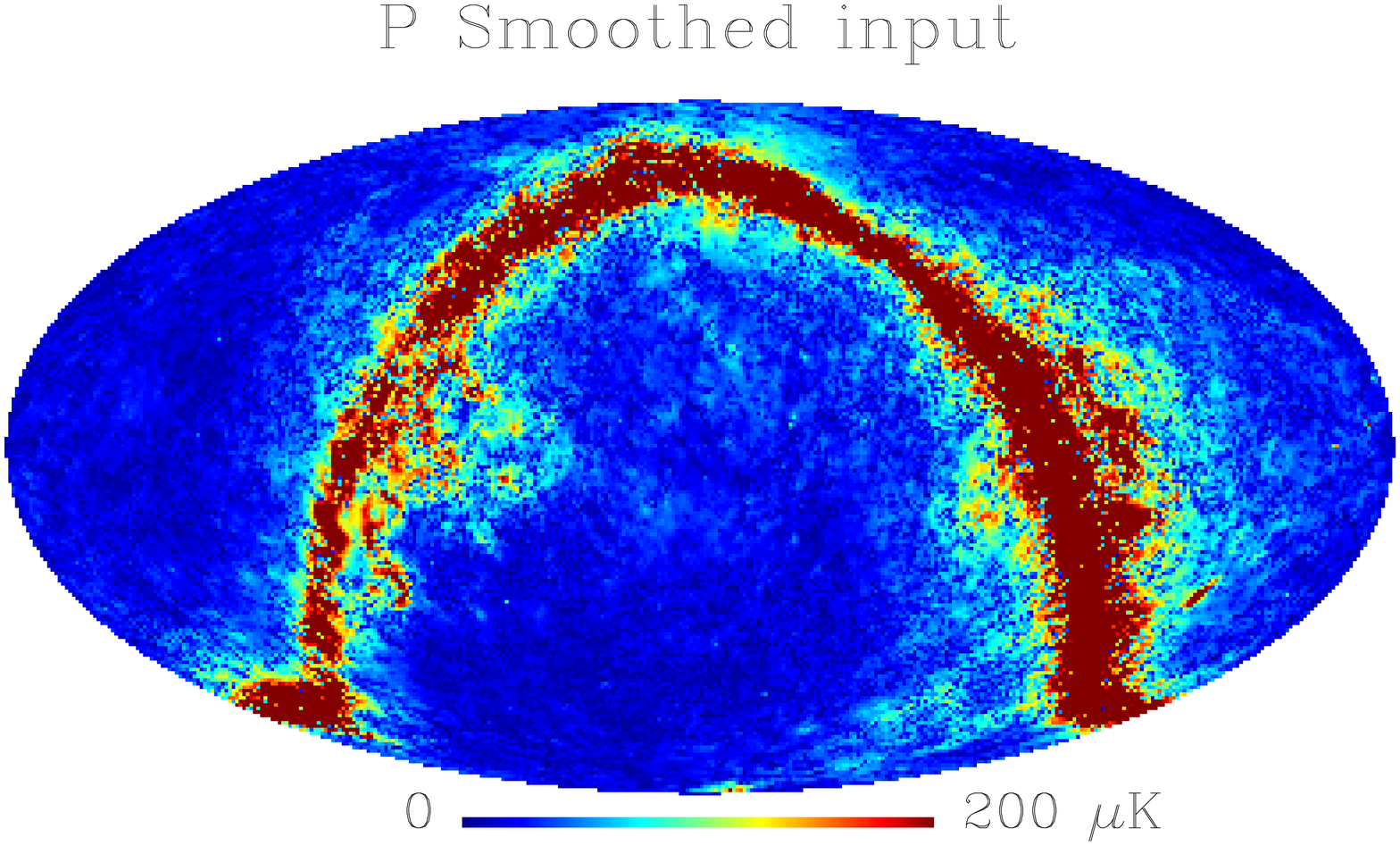}
\includegraphics[scale=0.3,angle=90]{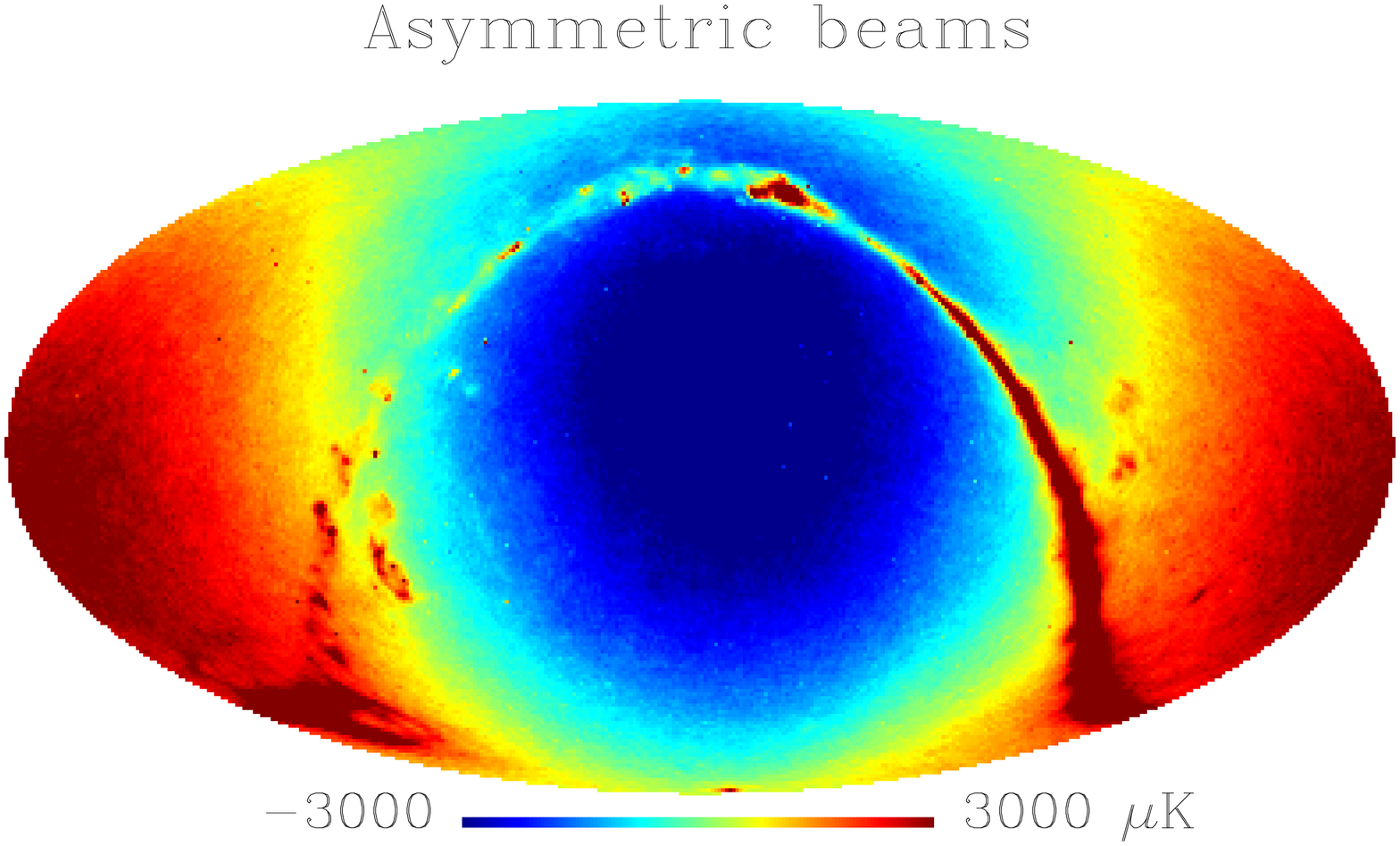}
\includegraphics[scale=0.3,angle=90]{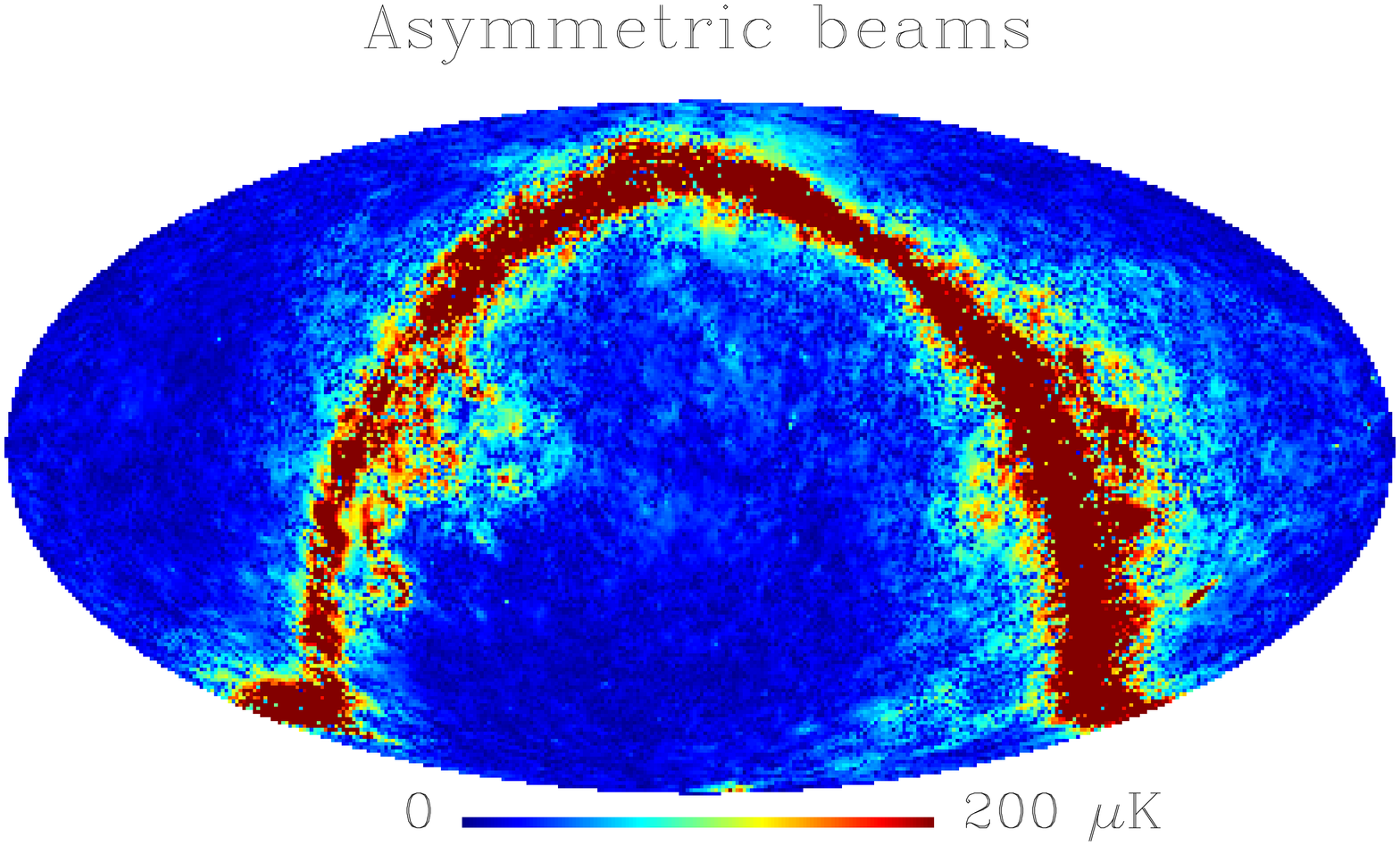}
\includegraphics[scale=0.3,angle=90]{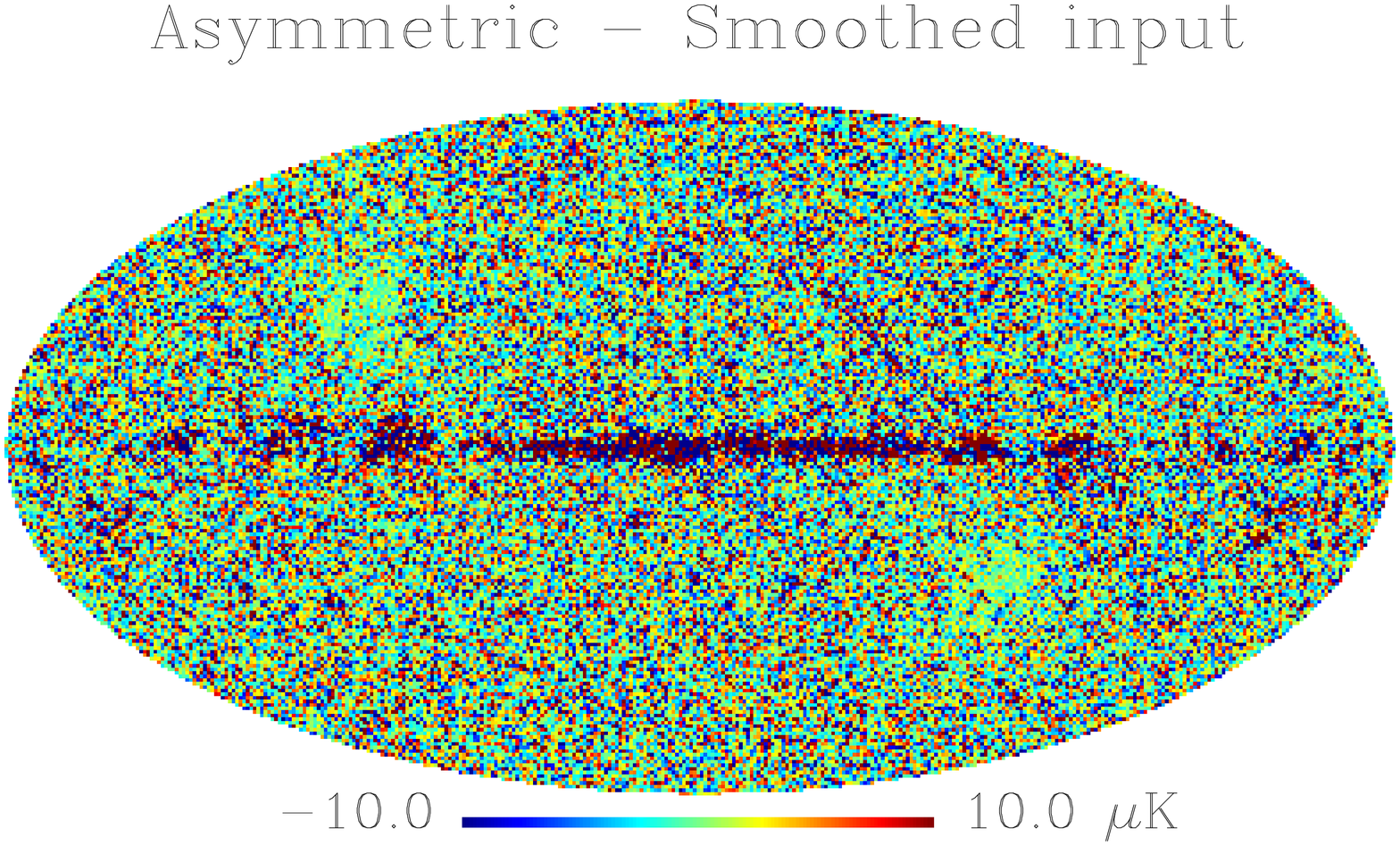}
\includegraphics[scale=0.3,angle=90]{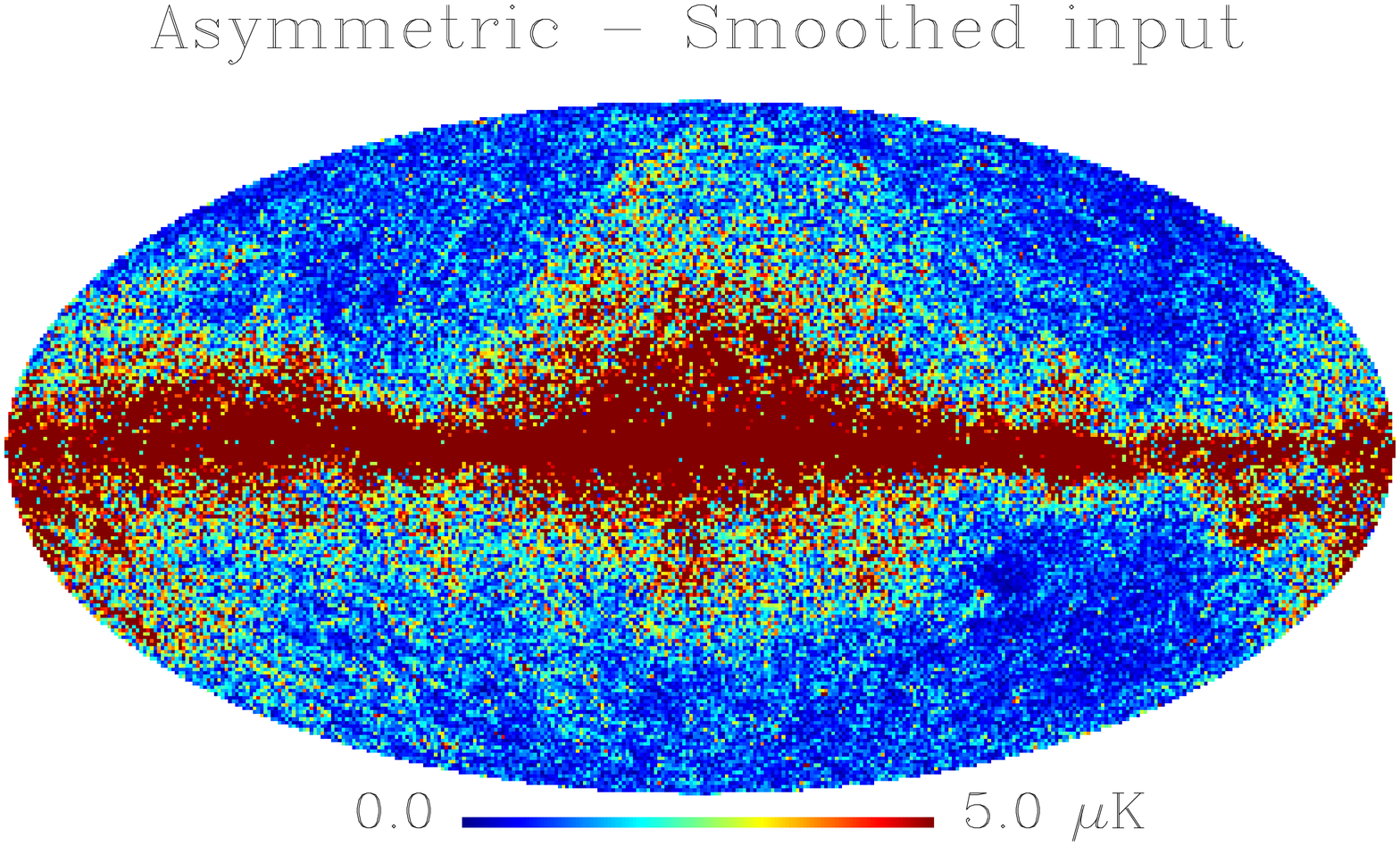}
\includegraphics[scale=0.3,angle=90]{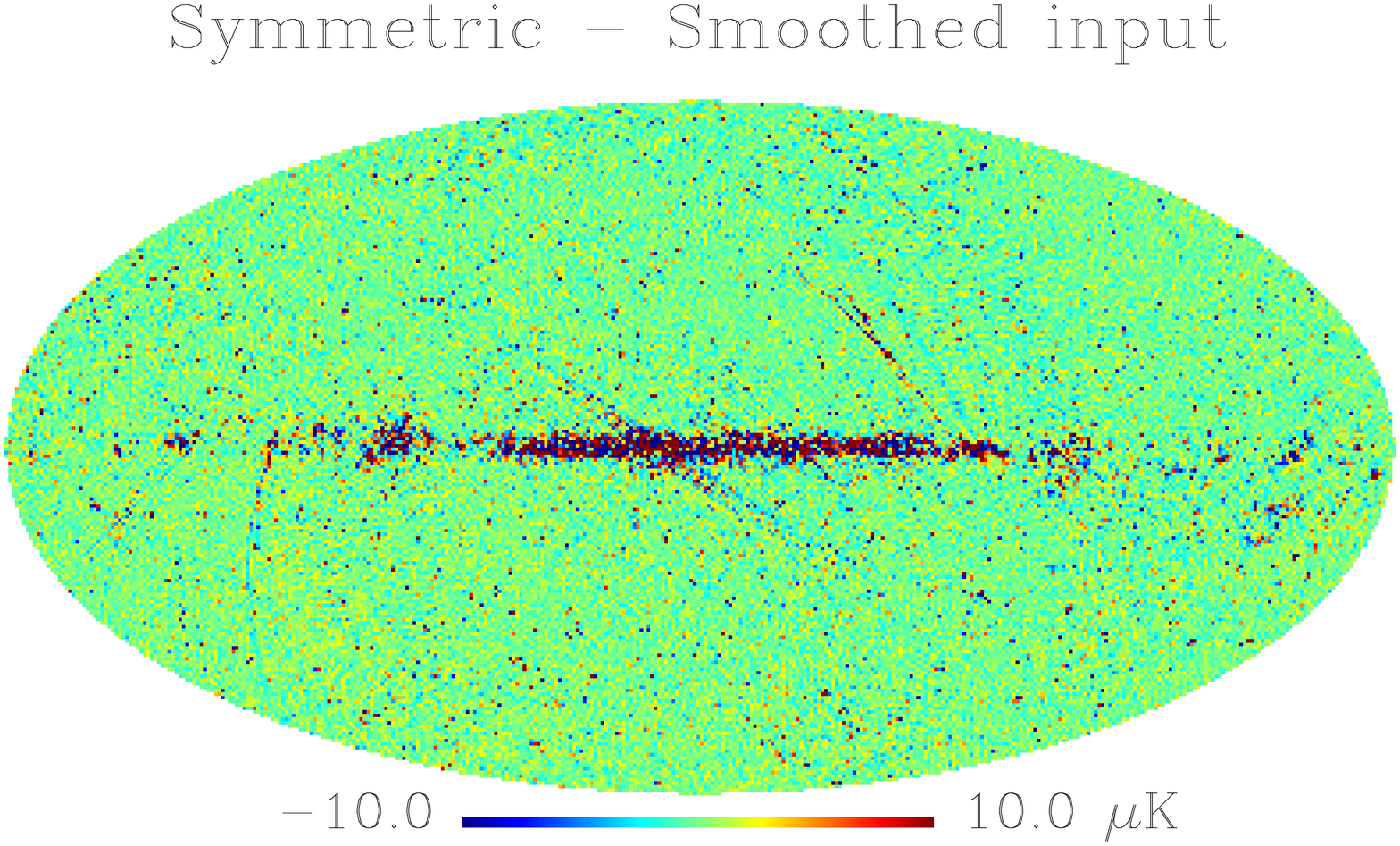}
\includegraphics[scale=0.3,angle=90]{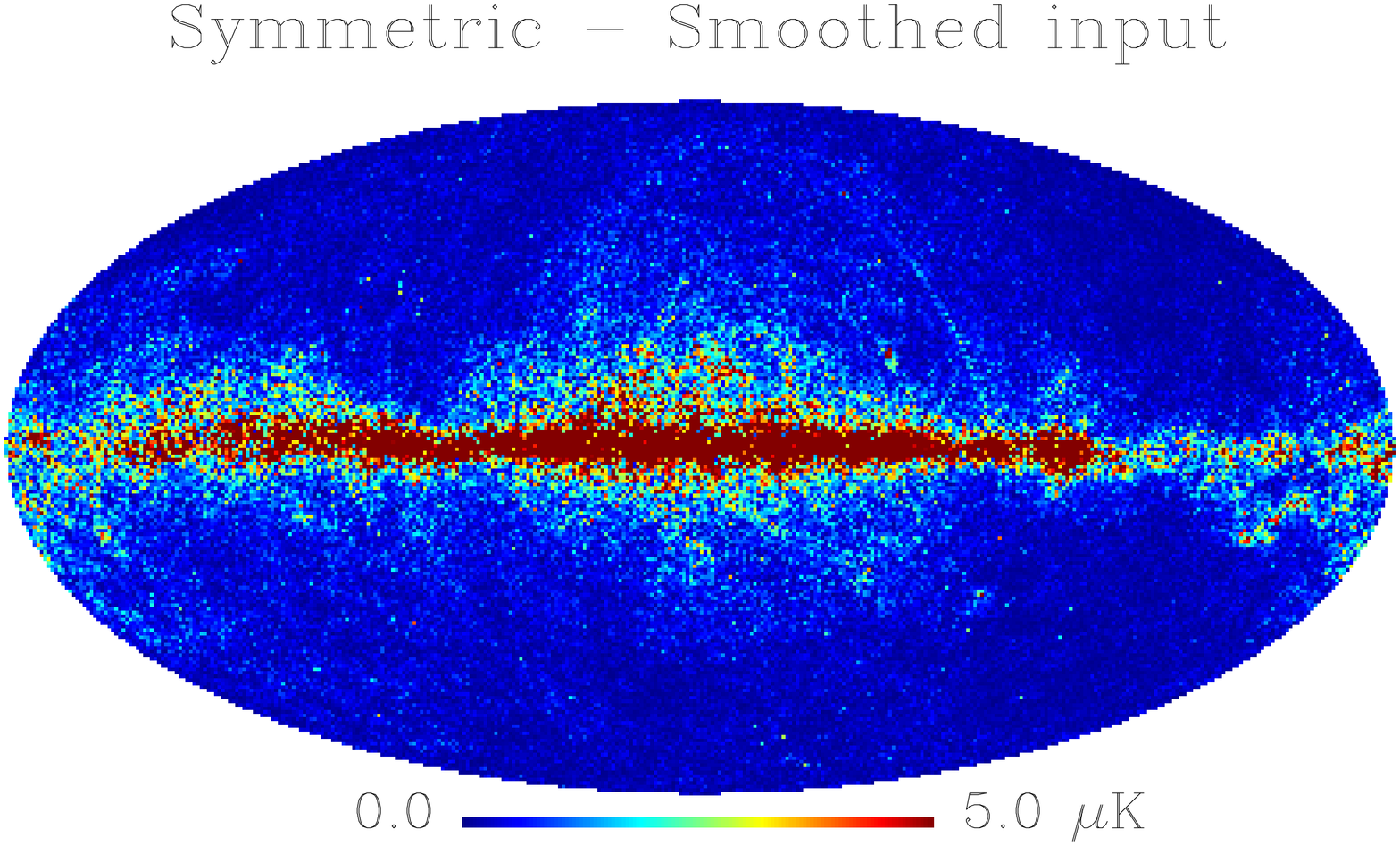}
\end{center}
\caption{Madam maps made from simulated noiseless TODs. Maps on the
left are temperature maps. Those on the right give the magnitude of
the polarization vector ($P = \sqrt{Q^2+U^2}$).   All maps contain
CMB, dipole and foreground emissions. \textit{Top row:} Smoothed
input maps. \textit{Second row:} Madam maps. These maps include the
effects of asymmetric beams and sample integration. \textit{Third
row:} Difference of the above Madam map and the smoothed input map.
\textit{Bottom row}: Same as above, but for symmetric beams. The
maps of the top and second rows are in ecliptic coordinates, whereas
the difference maps are in galactic coordinates. The latter
coordinates were chosen to give a clearer view to the ecliptic pole
regions (see the light green areas of the asymmetric difference
map). Small patches of noiseless Madam temperature maps are shown in
Figs.~\ref{fig:maps1} and \ref{fig:maps2}.} \label{fig:maps}
\end{figure*}

To demonstrate the effects of beams in our maps, we made maps from
the noiseless TODs containing CMB, dipole, and foreground emissions.
We show the Madam maps as an example in Fig.~\ref{fig:maps}.
Corresponding maps of the other mapmaking codes would look similar.
We can hardly see any differences between the noiseless output map
and the smoothed input map (see the two upper rows of
Fig.~\ref{fig:maps}). To reveal the differences, we subtract the
smoothed input map from the output map (bottom two rows of
Fig.~\ref{fig:maps}). The beam window functions of the symmetric and
asymmetric beams differ mainly at high-$\ell$ (see
Sect.~\ref{subsubsec:window}), which makes mainly small angular
(pixel) scale differences in the maps. Therefore the difference map
of the asymmetric beams contains more small-scale residuals than the
difference map of the symmetric beams. The ecliptic pole regions of
the sky are scanned in several directions, which makes the effective
beam of the asymmetric case more symmetric and therefore closer to
the effective beam of the symmetric case. Therefore the difference
of the noiseless output map of the asymmetric beams and the smoothed
input map becomes small in the vicinity of the ecliptic poles (see
the light green areas of the difference map of the third row of
Fig.~\ref{fig:maps}). The angular diameter of these areas is
$\backsim$$30^{\circ}$.

Some point source residues are visible in the temperature difference
maps of Fig.~\ref{fig:maps} (see especially the lower left corner
map). A likely reason for these residues is the difference in how
the pixel areas are sampled in the output maps and in the smoothed
input map.  In the latter map the pixel temperature is an integral
of the sky temperature over the pixel area, whereas in the output
map the pixel temperature is an average of the observations falling
in the pixel. The observations do not necessarily sample the pixel
area uniformly, which leads to a different pixel temperature than in
the uniform integration.

The difference maps of Fig.~\ref{fig:maps} have some stripes that
align with the scan paths between the ecliptic poles. These stripes
are most noticeable close to the galactic regions in the symmetric
case difference map.  In the asymmetric case the stripes do not
stand out from the larger pixel scale residuals. They arise from the
fact that signal differences (gradients) inside a map pixel create
non-zero baselines in Madam that show up as stripes in the Madam
map.  Because the signal gradients are largest in the galactic
regions, the stripes are strongest there.  All our mapmaking codes
produce such signal errors, which are stronger in the optimal codes
than in the destripers (Poutanen et al.~\cite{Pou06}, Ashdown et
al.~\cite{Ash07b}).

\begin{figure} [!tbp]
\begin{center}
\resizebox{\hsize}{!}{
\begin{tabular}{cc}
\includegraphics{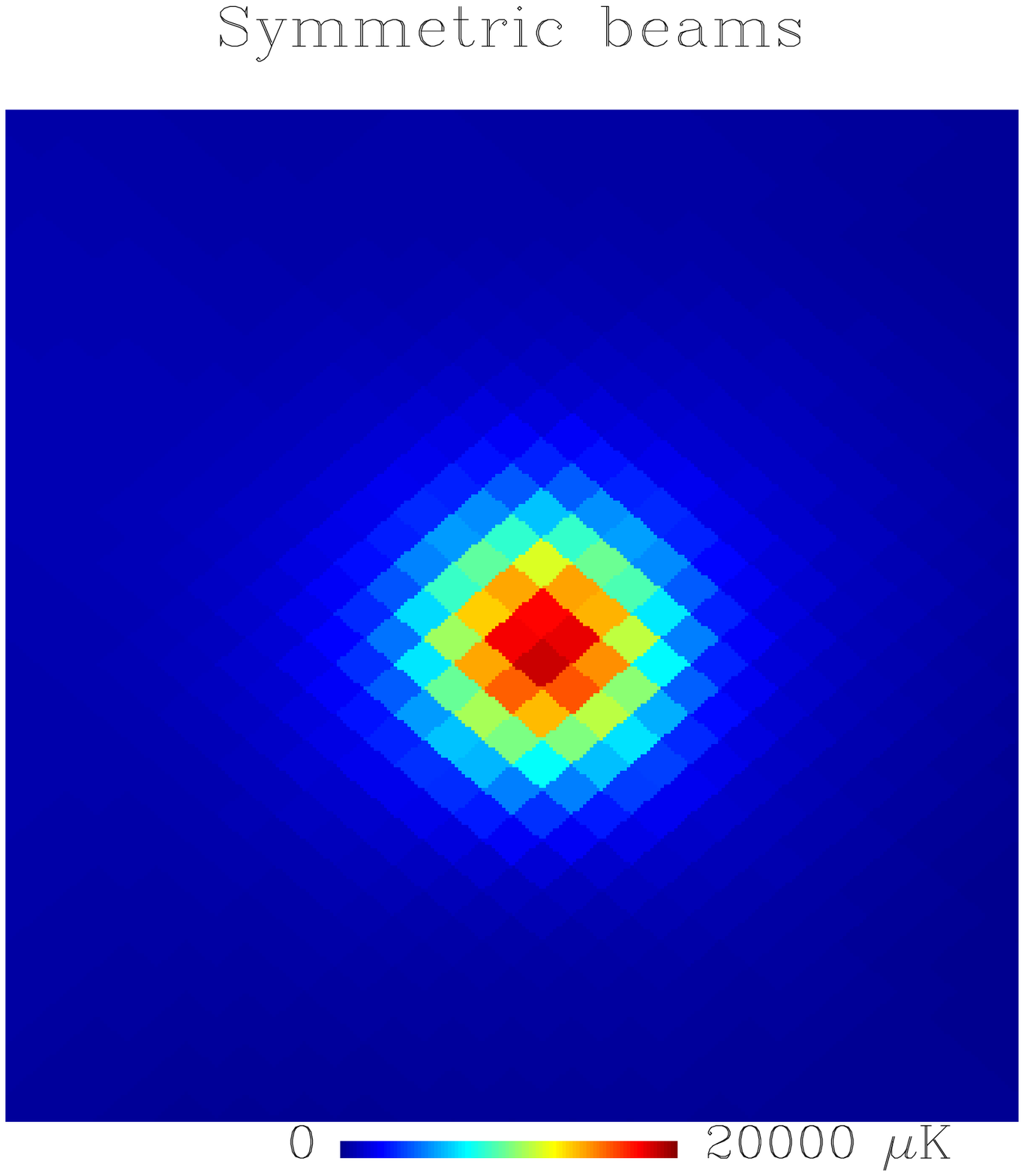}&
\includegraphics{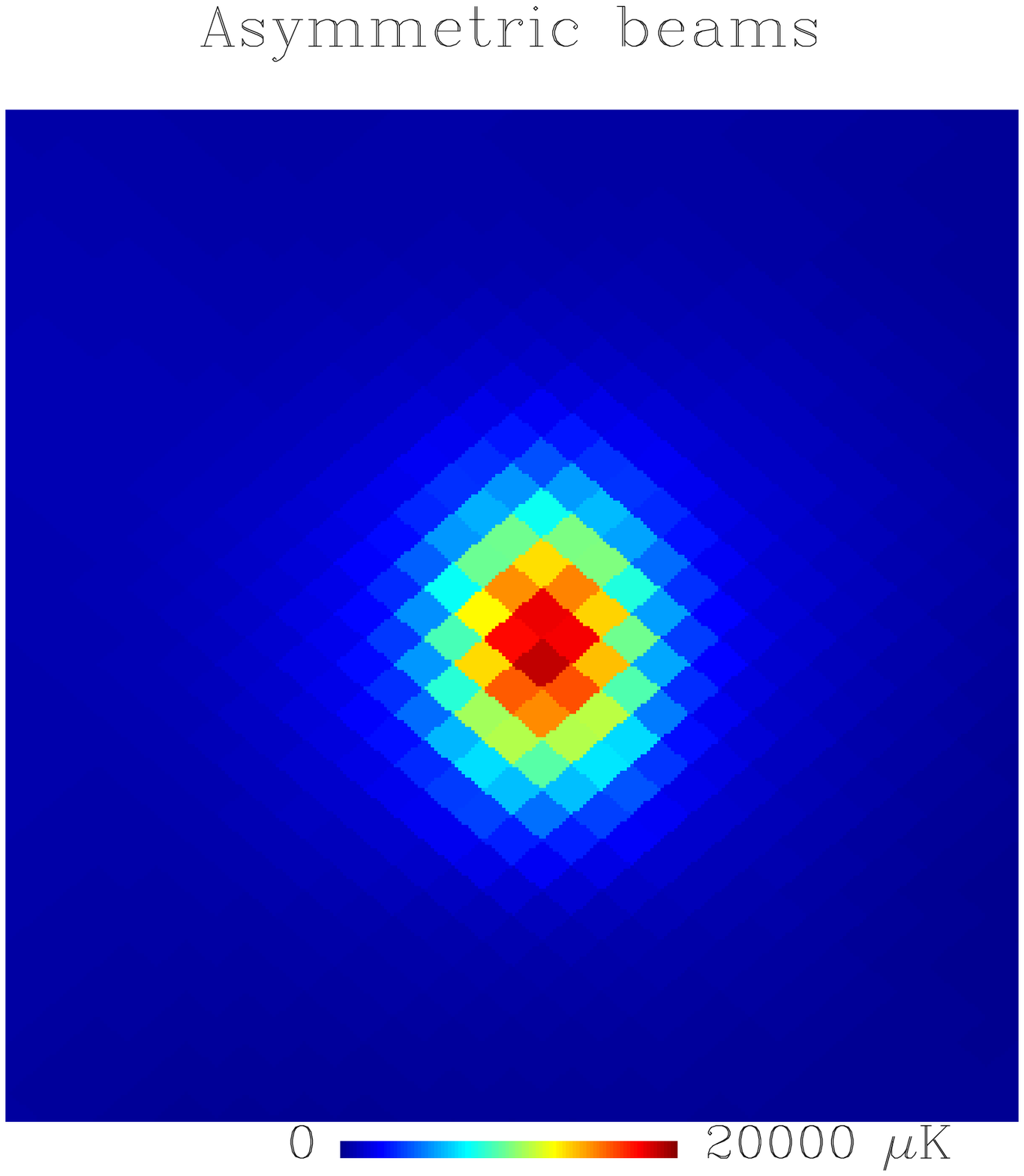}
\end{tabular}}
\end{center}
\caption{Effect of beams in the point source observations. These
plots show $3^{\circ} \times3^{\circ}$ patches of the noiseless
Madam temperature maps (from Fig.~\ref{fig:maps}), zoomed to the
vicinity of the ecliptic plane. The location of the patch is given
as a square box in the top left map of Fig.~\ref{fig:maps}.}
\label{fig:maps1}
\end{figure}

\begin{figure} [!tbp]
\begin{center}
\resizebox{\hsize}{!}{
\begin{tabular}{cc}
\includegraphics{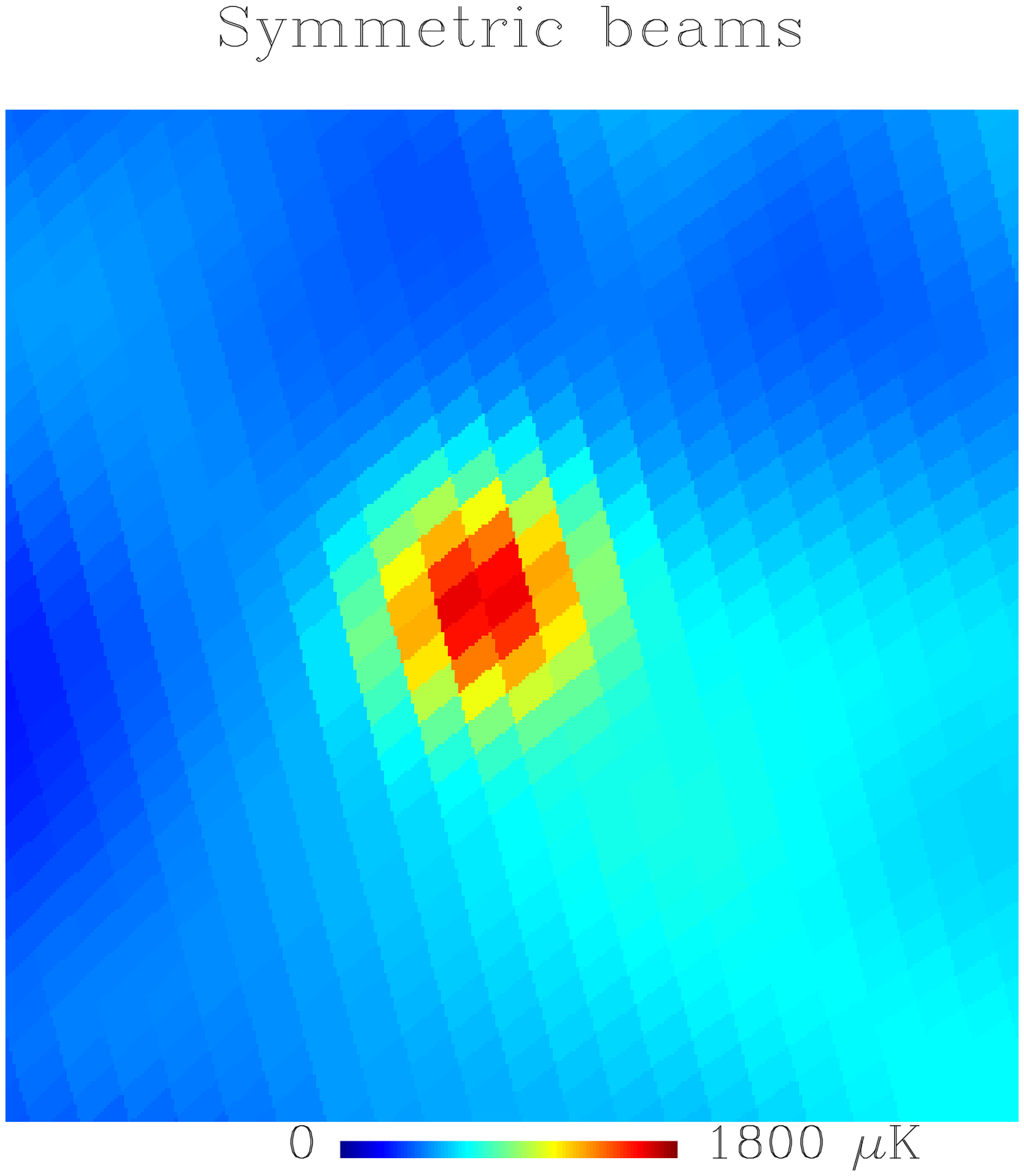}&
\includegraphics{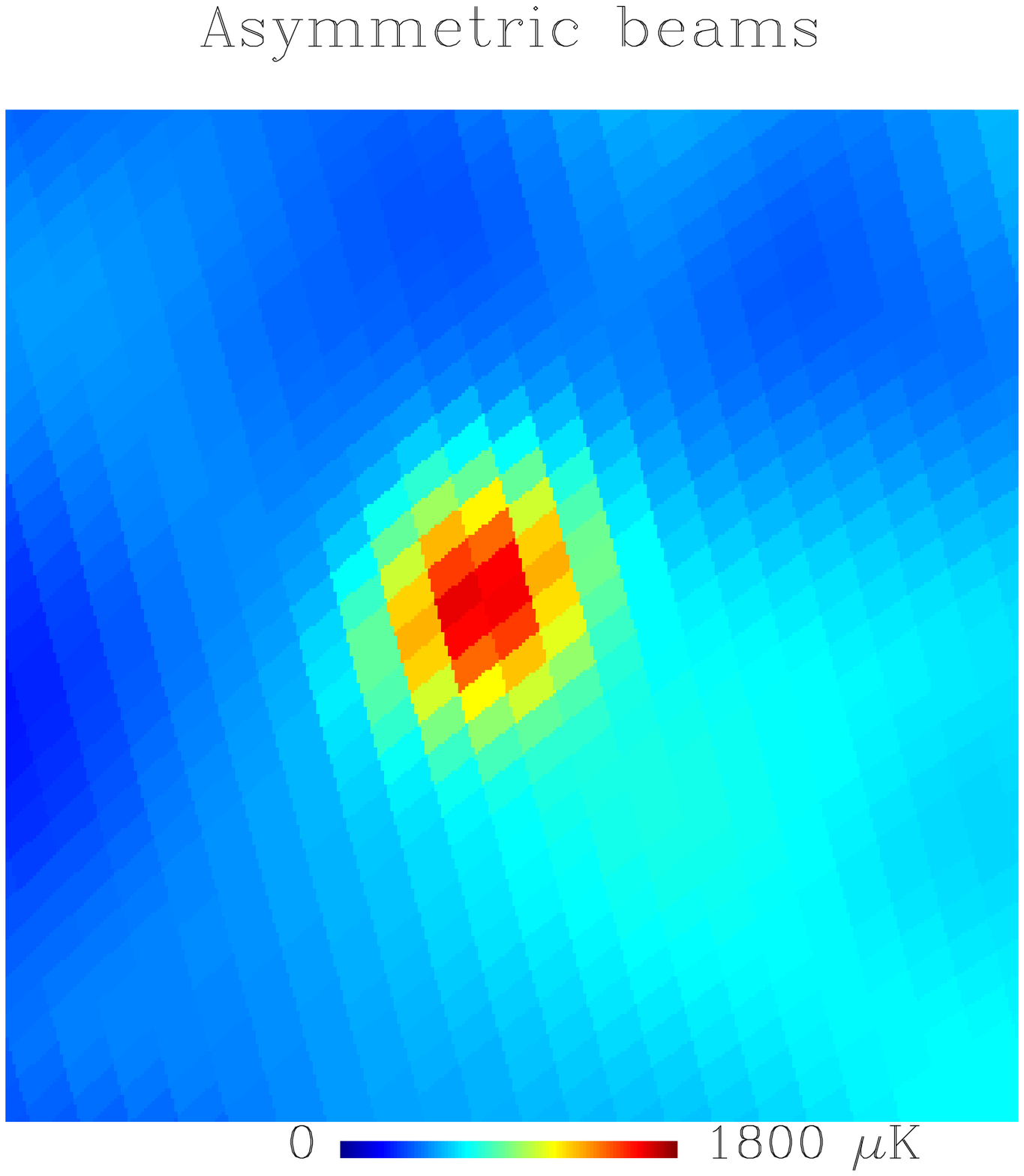}
\end{tabular}}
\end{center}
\caption{Similar plots as in Fig.~\ref{fig:maps1}, but now we have
zoomed to a point source near the south ecliptic pole (in
$(\lambda,\beta) = (255^{\circ},-81^{\circ})$ of ecliptic longitude
and latitude). The size of the patch is $3^{\circ}
\times3^{\circ}$.} \label{fig:maps2}
\end{figure}

Another way to see the effects of the beams in the maps is to
examine how point sources show up in the maps. The image of a point
source in the map shows the effective beam at that location of the
sky.  Figs.~\ref{fig:maps1} and \ref{fig:maps2} show two such point
sources of the noiseless Madam temperature maps.
Fig.~\ref{fig:maps1} is a patch from the vicinity of the ecliptic
plane. There the scanning is mainly in one direction and therefore
the difference in ellipticities of the effective beams (of the
symmetric and asymmetric cases) is clearly visible.
Fig.~\ref{fig:maps2} shows a similar comparison near the south
ecliptic pole. Here the wide range of the scanning directions makes
the effective beams more symmetric. The effective beams of the
symmetric and asymmetric cases are now more alike, but we can still
detect some difference in their ellipticities.

In our simulations beams and sample integration distort the TOD
before the instrument noise is added. Independent of the noise,
these effects are best explored in the binned noiseless maps,
independent of any particular mapmaking algorithm.  We examine the
effects of the beams and sample integration on the binned noiseless
maps in Sect.~\ref{subsec:binned_noiseless_maps}.

Because the beams and sample integration affect the signal gradients
of the observations, they have an impact on the signal error too. We
examine these effects in Sect.~\ref{subsec:residual_maps}, and
compare differences between mapmaking codes.  It is only through the
signal error that the effects of beams and sample integration show
up differently in the maps.  The binned noiseless map contribution
that is also affected by beams and sampling stays the same in all
maps.  In our simulations beams should have no effect in the
residual noise maps. Because there is a half a sample timing offset
between the detector pointings of sampling on and off cases, the hit
count maps of these two cases differ slightly, which shows up as
small differences in the residual noise maps.

Finally we discuss sorption cooler fluctuations and detector
pointing errors and assess their impacts in the maps.

\subsection{Binned noiseless maps}
\label{subsec:binned_noiseless_maps}

\begin{figure*} [!tbp]
\begin{center}
\includegraphics[scale=0.4]{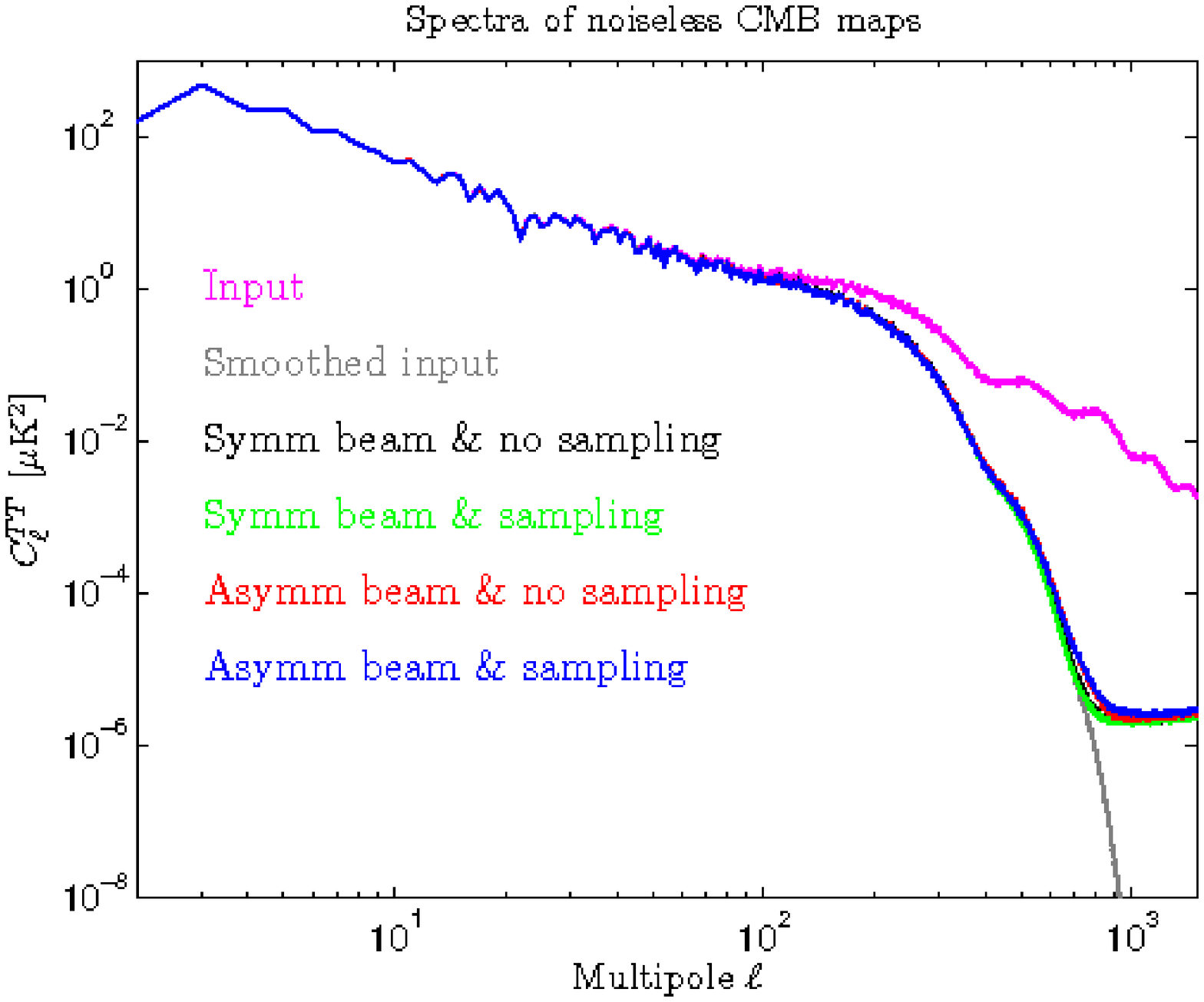}
\includegraphics[scale=0.4]{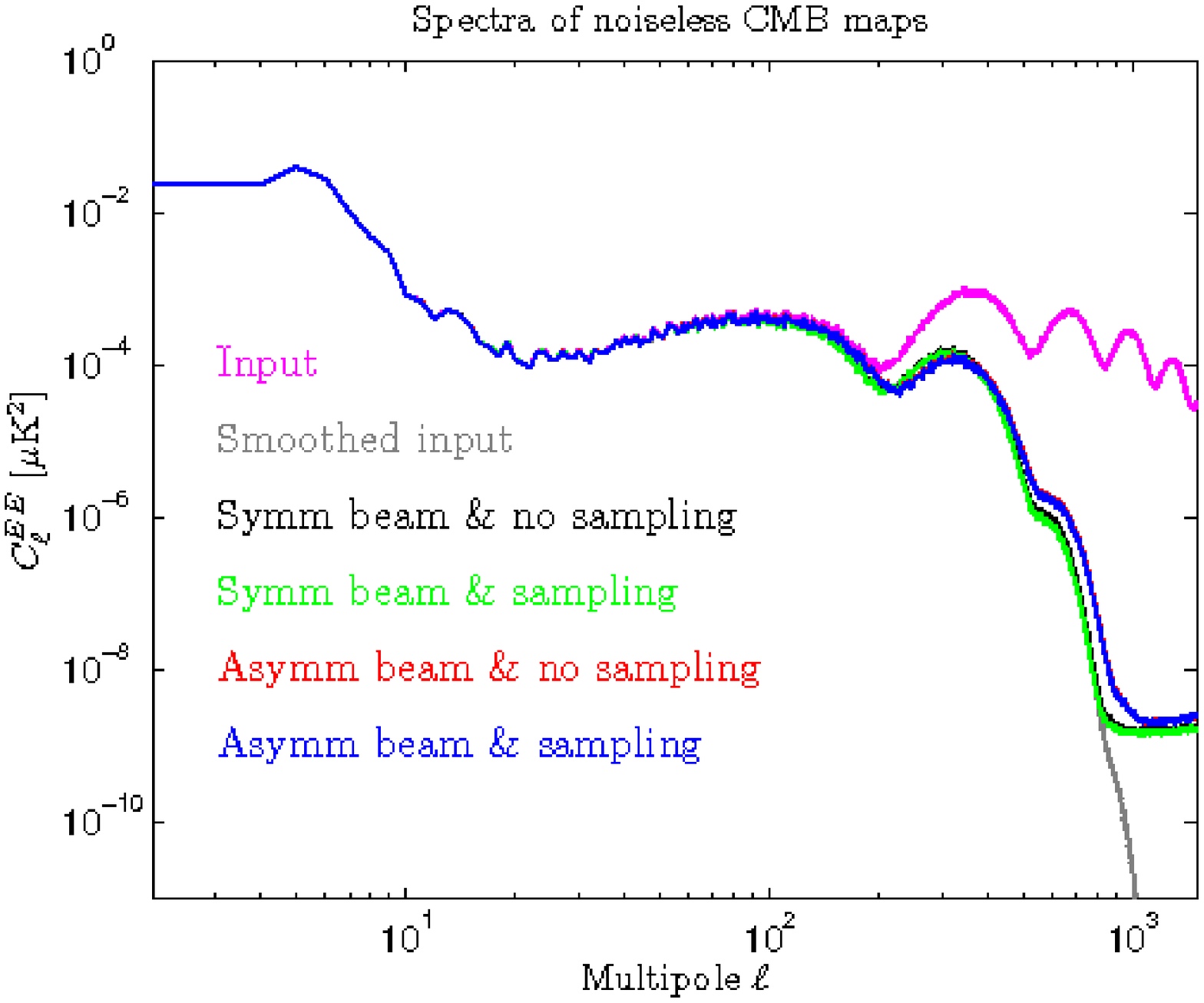}
\includegraphics[scale=0.4]{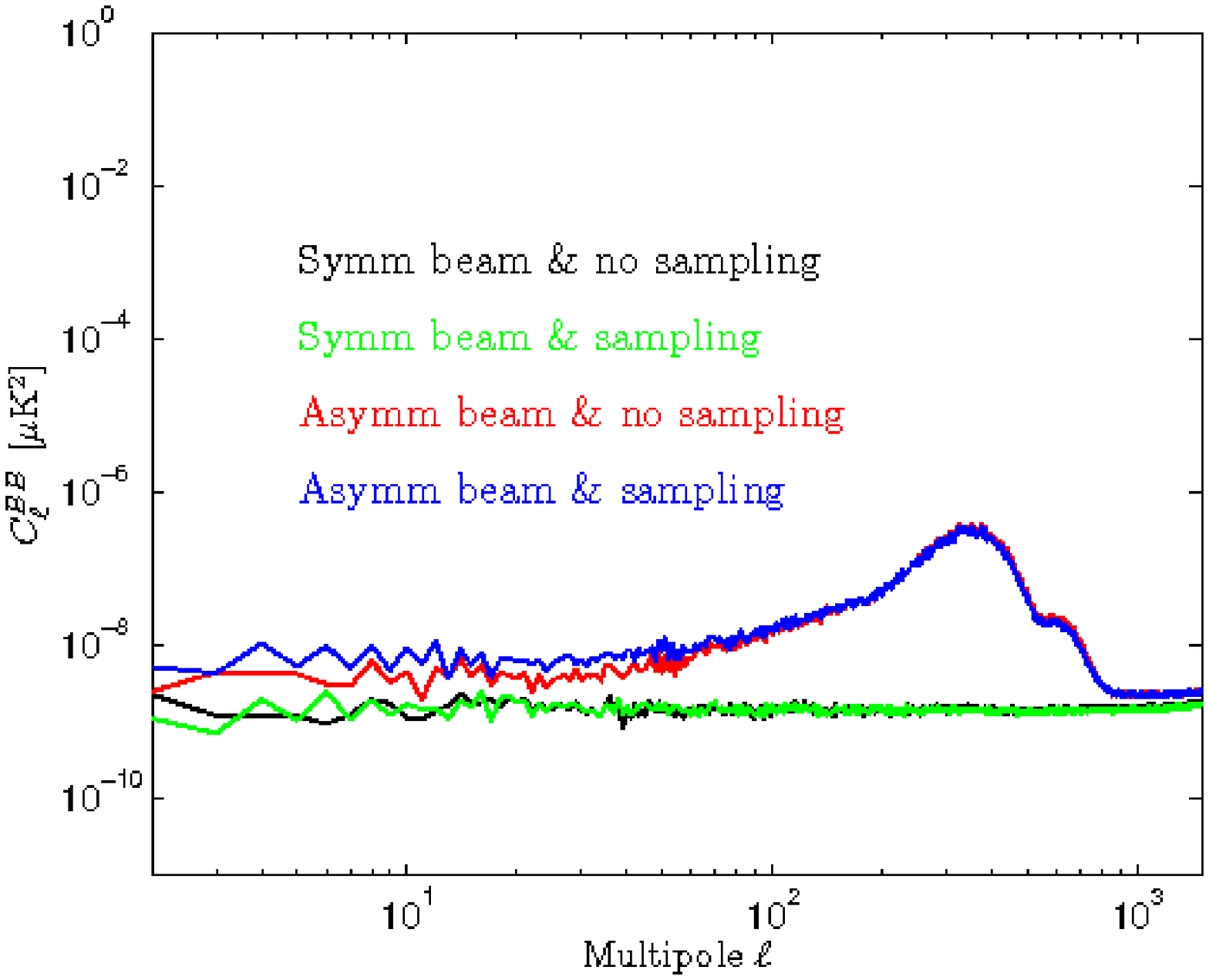}
\includegraphics[scale=0.4]{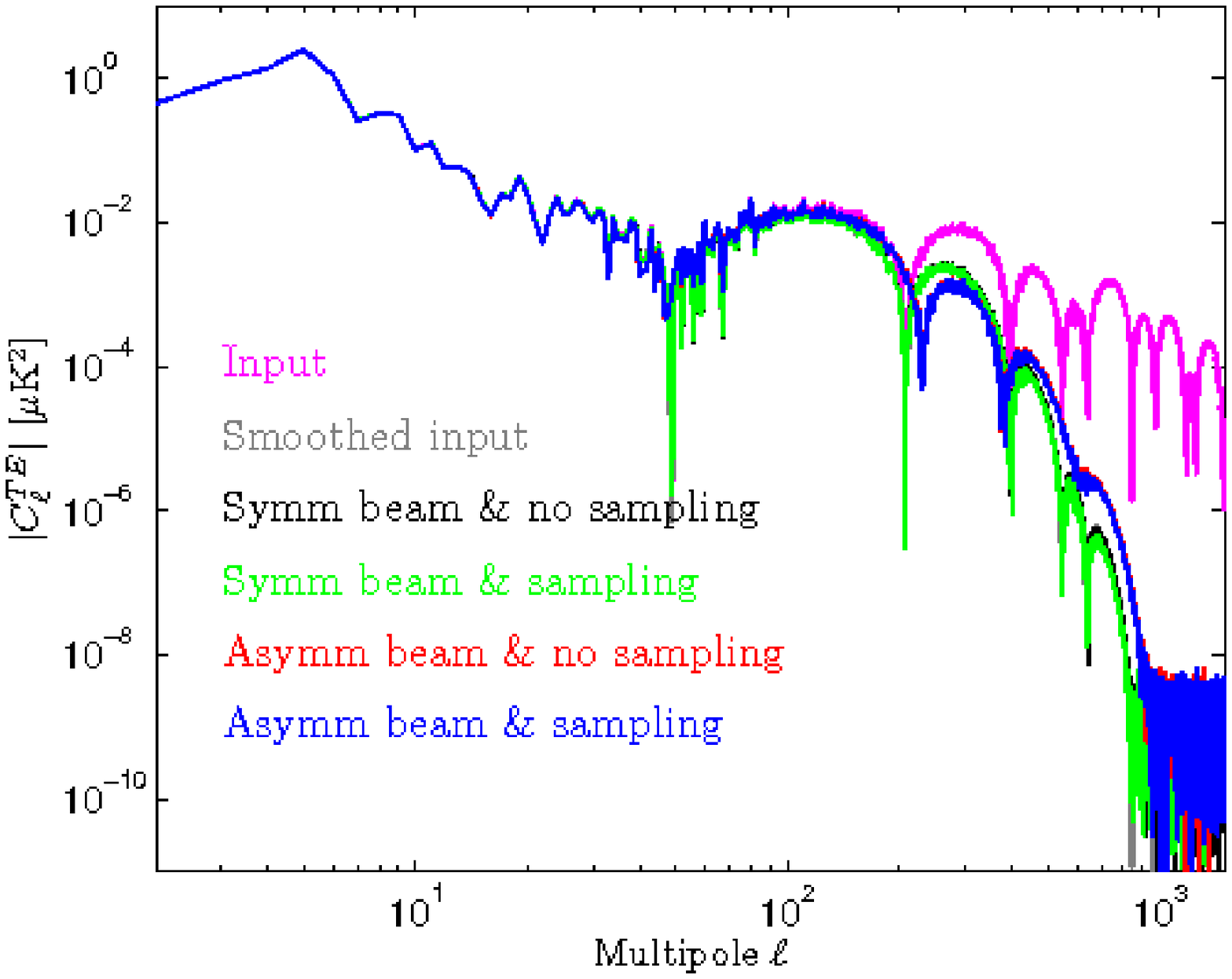}
\includegraphics[scale=0.4]{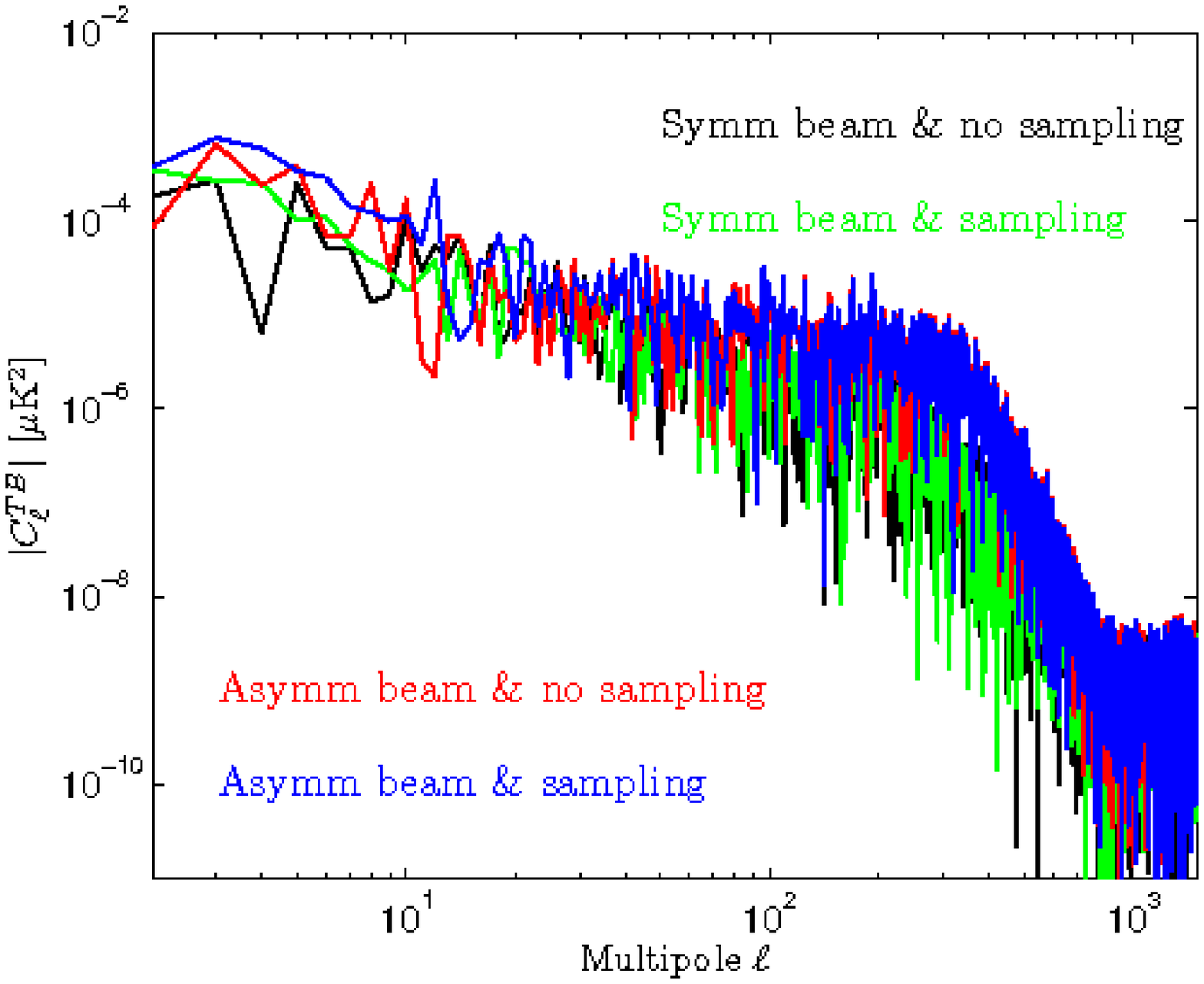}
\includegraphics[scale=0.4]{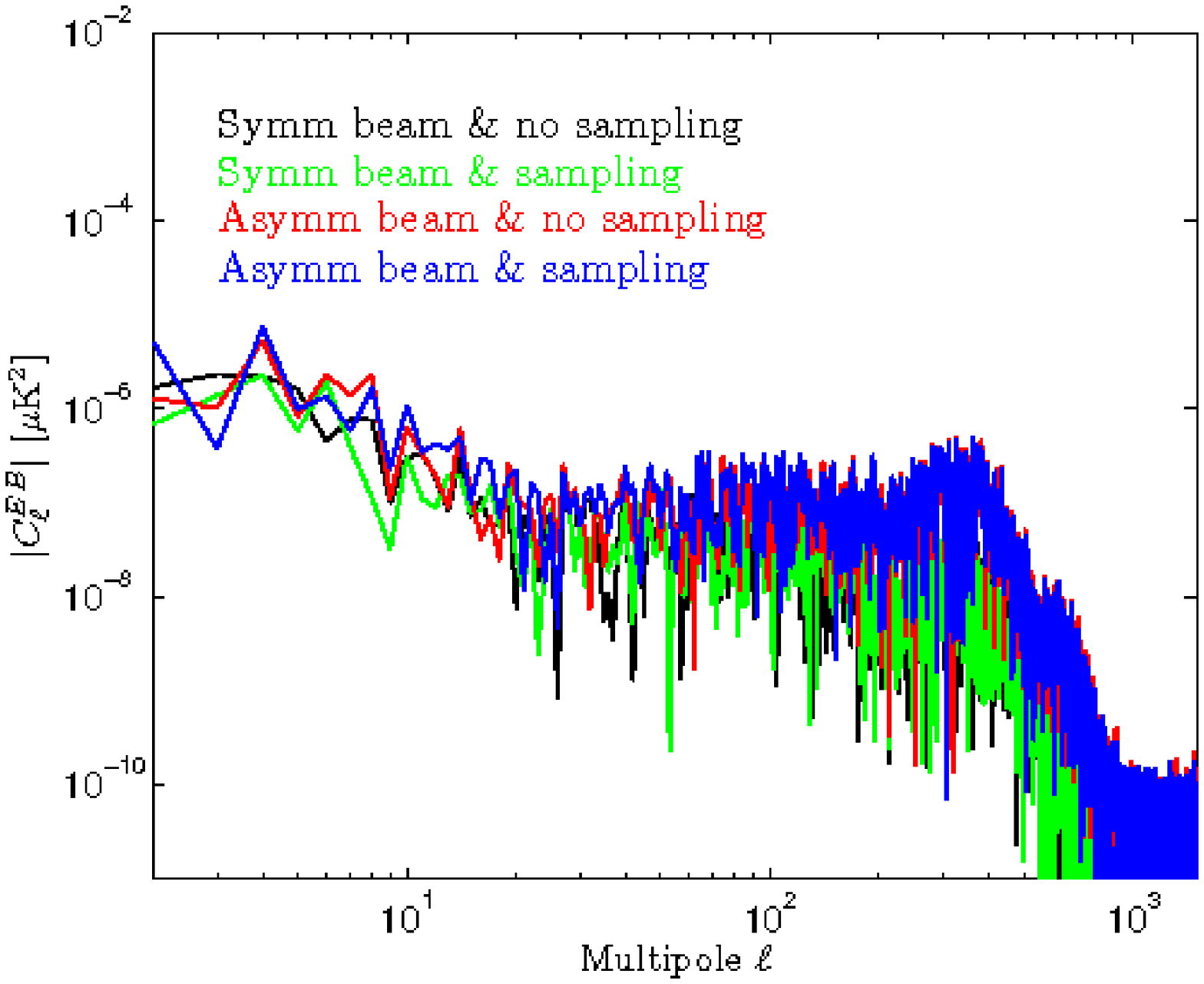}
\end{center}
\caption{Overview of the effects of beams and sample integration on
the angular power spectra of the CMB maps. We computed the spectra
of the noiseless CMB maps that were binned from the four simulated
LFI 30\,GHz TODs. The TODs represent different beam and sample
integration cases. The CMB input sky contains no $B$ mode power.
Note that the plots of the cross-spectra show their absolute values.
In this paper we define $C_{\ell}^{XY} \equiv
\frac{1}{2\ell+1}\sum_{m=-\ell}^{\ell}{a_{\ell m}^{X}(a_{\ell
m}^{Y})^{\star}}$. Here $X,Y = T,E,B$ and $Z^{\star}$ is the complex
conjugate of $Z$.} \label{fig:binned_spectra}
\end{figure*}

The effects of beams and sample integration were examined from the
binned ($I,Q,U$) maps made from the four noiseless TODs of different
beam and sampling cases. The TODs contained only \hbox{CMB}. The
angular power spectra of the four maps are shown in
Fig.~\ref{fig:binned_spectra}.  The TT angular power spectra of the
symmetric and asymmetric beams differ mainly at large multipoles
($\ell \gtrsim 600$). This is a result of the different effective
beam window functions of these two cases.

The EE spectra of symmetric and asymmetric beams are different at
both intermediate ($\ell \gtrsim 100$) and high multipoles.  Due to
the beam mismatch the EE spectrum of the asymmetric beam case is
influenced by the cross-coupling from the temperature. A similar
cross-coupling does not occur in the symmetric beam case (no beam
mismatch). Therefore the EE spectra behave differently than the TT
spectra. We discuss these issues more in
Sects.~\ref{subsubsec:beam_mismatch} and ~\ref{subsubsec:window} and
give there the explanation of the behaviors of the TT and EE
spectra.

Sample integration is effectively a low-pass filter in the TOD
domain. Therefore it introduces an extra spectral smoothing that
removes some small-scale signal power from the map. This effect is
just barely visible in Fig.~\ref{fig:binned_spectra}, where the TT
and EE spectra with ``sampling'' are slightly suppressed as compared
to their ``no sampling'' counterparts. We discuss these effects more
in Sect.~\ref{subsubsec:sampling}.

The TT and EE spectra of the binned maps become flat at high $\ell$
($\gtrsim 1000$). This is a result of spectral aliasing ($\ell$ mode
coupling) that arises from the non-uniform sampling of the pixel
areas. The aliasing couples power from low-$\ell$ to high-$\ell$. We
see this effect in the maps too, where we called it the pixelization
error.

Because the CMB input sky contained no $B$-mode power, the signals
in the BB spectra of the binned maps (see
Fig.~\ref{fig:binned_spectra}) must arise from temperature and
$E$-mode polarization signals leaking to the $B$-mode. The
magnitudes of the symmetric beam BB spectra are comparable to the
magnitude of the high-$\ell$ flat part of the corresponding EE
spectra. This observation and the fact that the symmetric beams were
all identical (no beam mismatch) suggest that the BB spectrum of the
symmetric beam case mainly arises from the spectral aliasing from
the E mode (due to pixelization error). The BB spectrum of the
asymmetric beam case shows a distinct signal at multipoles between
100 and 800. We expect this signal to originate from temperature and
$E$-mode polarization signals that cross-couple to BB due to beam
mismatch.  We discuss these cross-couplings more in
Sect.~\ref{subsubsec:beam_mismatch}.

\subsubsection{Beam mismatch, cross-couplings of $I$, $Q$, and $U$} \label{subsubsec:beam_mismatch}

We can quantify the effects of beam mismatch on the maps by
calculating power spectra. We consider the asymmetric beam case with
no sampling\footnote{Note that for this analysis, the smoothing
effect from detector sample integration could be subsumed into the
asymmetric beams, but we decided to ignore it for simplicity.}. A
number of authors have worked on beam mismatch systematics and their
impacts on maps and angular power spectra (Hu et al.~\cite{Hu03},
Rosset et al.~\cite{Ros07}, O'Dea et al.~\cite{Ode07}, Shimon et
al.~\cite{Shi08}).

The LFI main beams vary in width and orientation from detector to
detector. These variations are fully represented in our asymmetric
beams.  (In contrast, the widths of our symmetric beams were the
same in all detectors.)  Due to beam mismatch, the detectors of a
horn see different Stokes I; this difference appears as an artifact
in the polarization map. This is a potentially serious issue for a
CMB experiment such as {\sc Planck}, because the fraction of the
strong temperature signal that pollutes the polarization map
(denoted $T\rightarrow P$) may be fairly large compared to the weak
CMB polarization signal itself.

Strong foreground emission dominates the $T\rightarrow P$ signal in
a full sky 30~GHz map. This map is useful for CMB power spectrum
estimation only if the pixels with strong foreground contribution
are removed from the map (galactic cut); however, the $T\rightarrow
P$ signal will depend on the mask used in the cut.  To avoid this,
and to study the beam mismatch effects in a more general case for
which our results would not depend much on the details of the data
processing (e.g., masks), we decided to continue to work with the
full sky noiseless maps that we binned from the CMB \hbox{TOD}.

\begin{figure*} [!tbp]
\begin{center}
\includegraphics[scale=0.4]{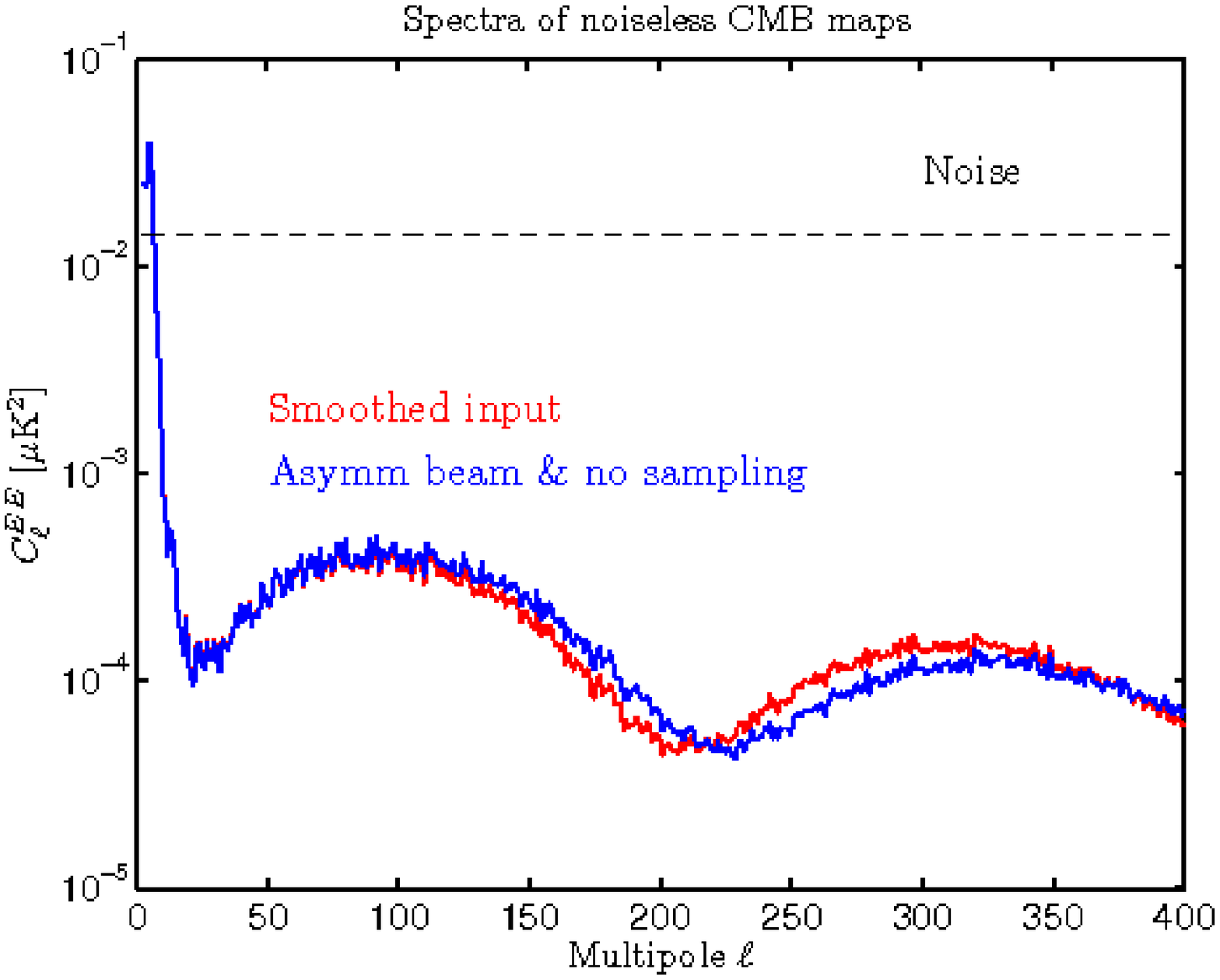}
\includegraphics[scale=0.4]{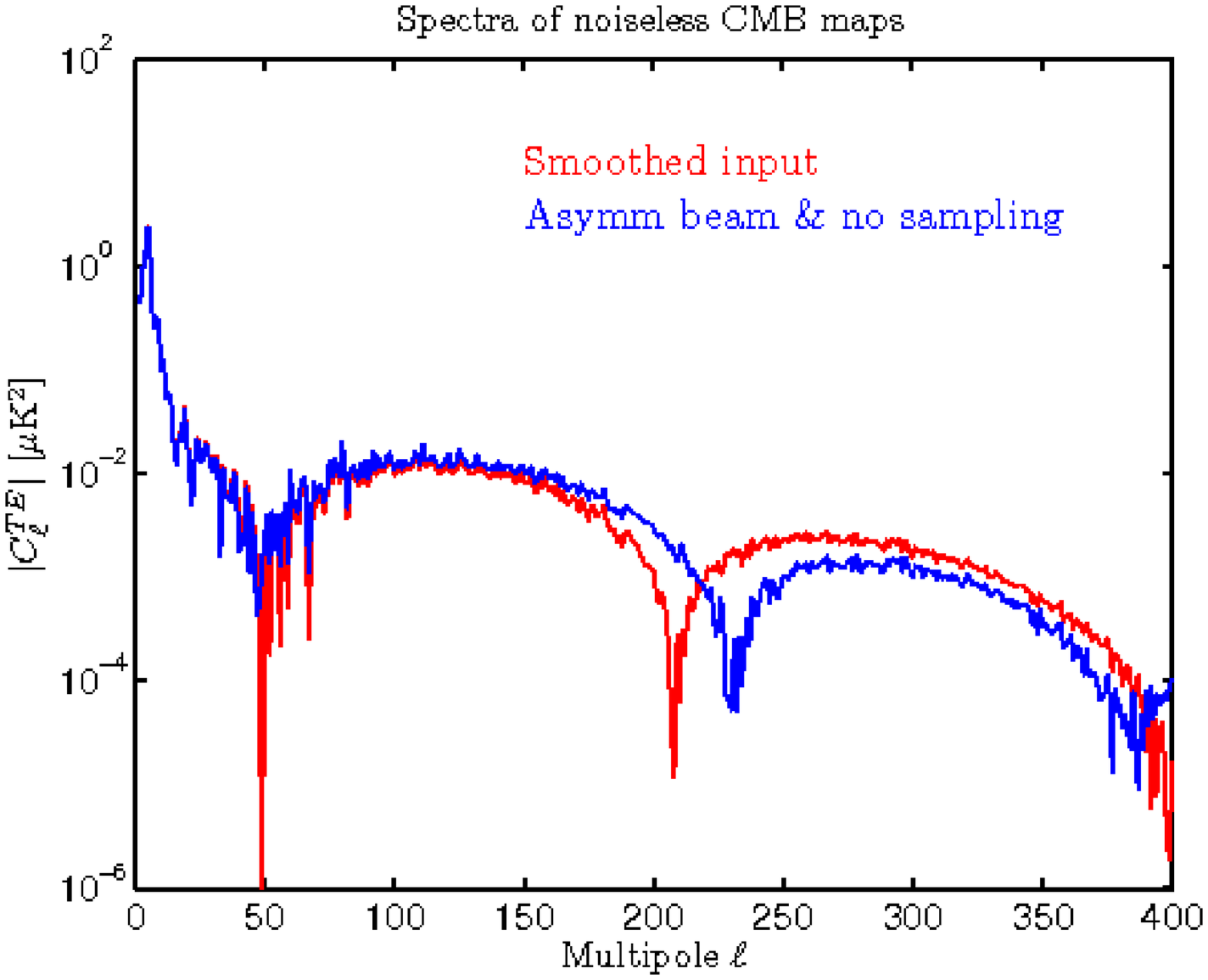}
\end{center}
\caption{Effects of $T\rightarrow P$ cross-coupling arising from
beam mismatches. \textit{Left panel:} Replot of the EE spectra from
Fig.~\ref{fig:binned_spectra}, showing spectra of a binned noiseless
CMB map (blue curve) and smoothed input map (red curve). The
expected EE spectrum of the white noise map is indicated by the
horizontal dashed line. \textit{Right panel:} TE spectra of the same
maps. We show the absolute values of the spectra.  The TE spectrum
does not have a noise bias, because white noise is uncorrelated in
the temperature and polarization maps. Therefore the dashed line of
noise bias is missing from this plot.} \label{fig:demo}
\end{figure*}

To give low-$\ell$ details we replotted the EE and TE spectra (from
Fig.~\ref{fig:binned_spectra}) of the binned noiseless CMB map
(asymmetric beams \& no sampling) and the smoothed input map. The
replotted spectra are shown in Fig.~\ref{fig:demo}. The
$T\rightarrow P$ cross-coupling is a significant systematic for the
power spectrum measurement, as Fig.~\ref{fig:demo} shows. The
acoustic peaks and valleys of the binned map are shifted towards
higher multipoles. In the EE spectrum the effect is smaller than the
noise, and may not be detectable.  For the TE spectrum the error of
the E mode polarization gets amplified by the large temperature
signal. This results in a larger artifact that is visible even in
the noisy TE spectrum (see Fig.~\ref{fig:te_3} and the discussion in
Sect.~\ref{subsec:residual_maps}).

We designed a simple analytic model that gives a reasonably good
description of the effects of the beam mismatch in the angular power
spectra of the ($I,Q,U$) signal maps (Appendix~\ref{sec:model}). Our
model shows that the mismatch of the beams of a horn is relevant for
$T\rightarrow P$.  The beams can be different from horn to horn, but
only a weak $T\rightarrow P$ may arise if the two beams of a horn
are identical.

\begin{figure*}[!tbp]
\begin{center}
\includegraphics[scale=0.4]{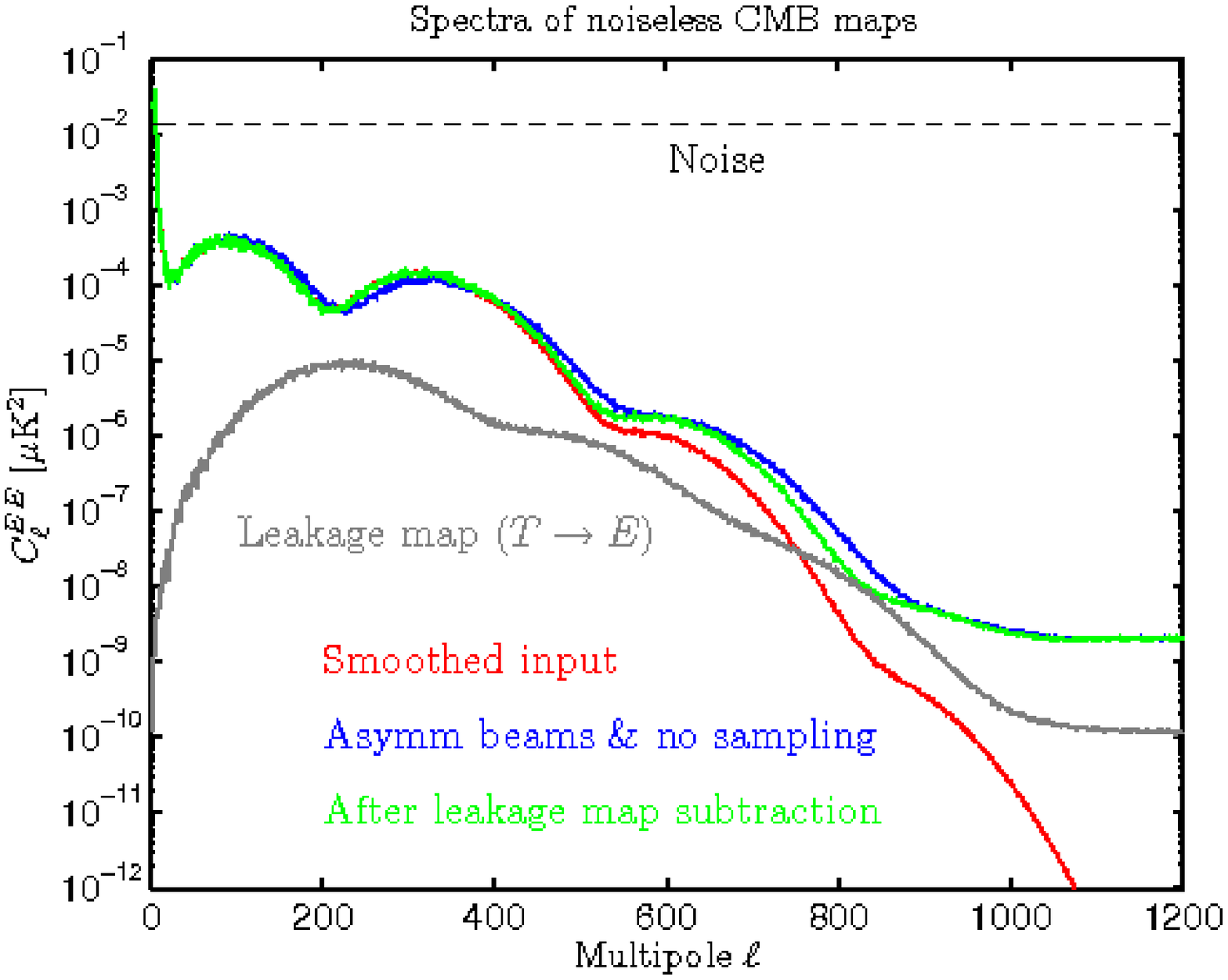}
\includegraphics[scale=0.4]{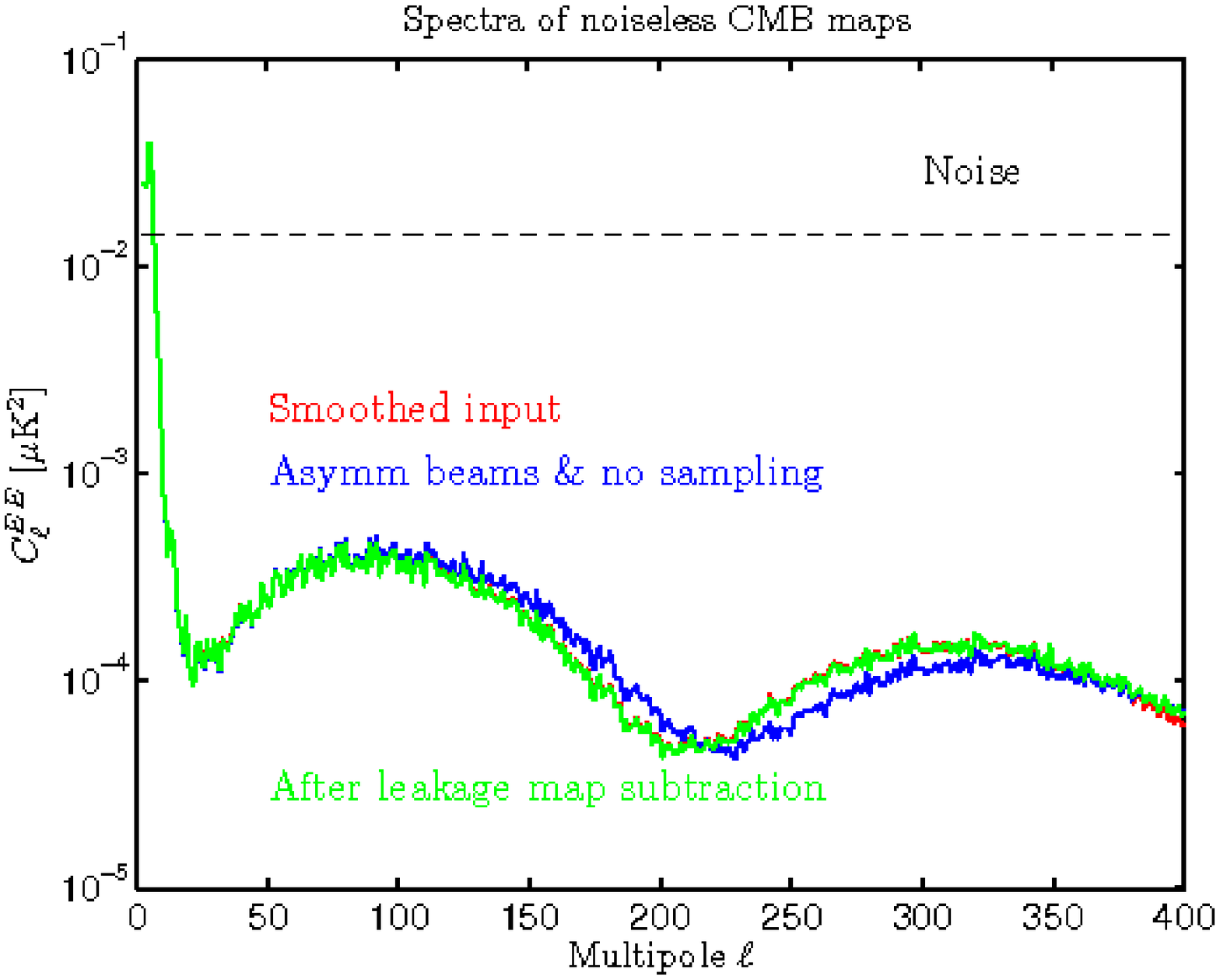}
\end{center}
\caption{Demonstration of how removing the $T\rightarrow P$
cross-coupling restores the acoustic peaks and valleys of the CMB in
their original positions. \textit{Left hand panel:} We show here a
number of EE spectra. The gray curve shows the EE spectrum of the
leakage map. The red and blue curves are as in Fig.~\ref{fig:demo}.
The green curve shows the spectrum of the binned noiseless map,
where the leakage map has first been subtracted. Note that the green
curve is nearly on top of the red curve at low multipoles ($\ell <
400$). The red and green curves differ at $\ell> 450$ mainly due to
differences in the beams: the red spectrum is smoothed with
symmetric beams while the green spectrum is smoothed with asymmetric
beams. At $\ell> 1050$ the green and blue spectra become flat due to
spectral aliasing arising from pixelisation errors. $T\rightarrow P$
becomes flat too for the same reason. The expected spectrum of the
white noise map is indicated by the horizontal dashed line.
\textit{Right hand panel:} Same as left hand panel but zoomed to
lower $\ell$. The red and green curves are nearly on top of each
other.} \label{fig:ee_1}
\end{figure*}

\begin{figure*}[!tbp]
\begin{center}
\includegraphics[scale=0.4]{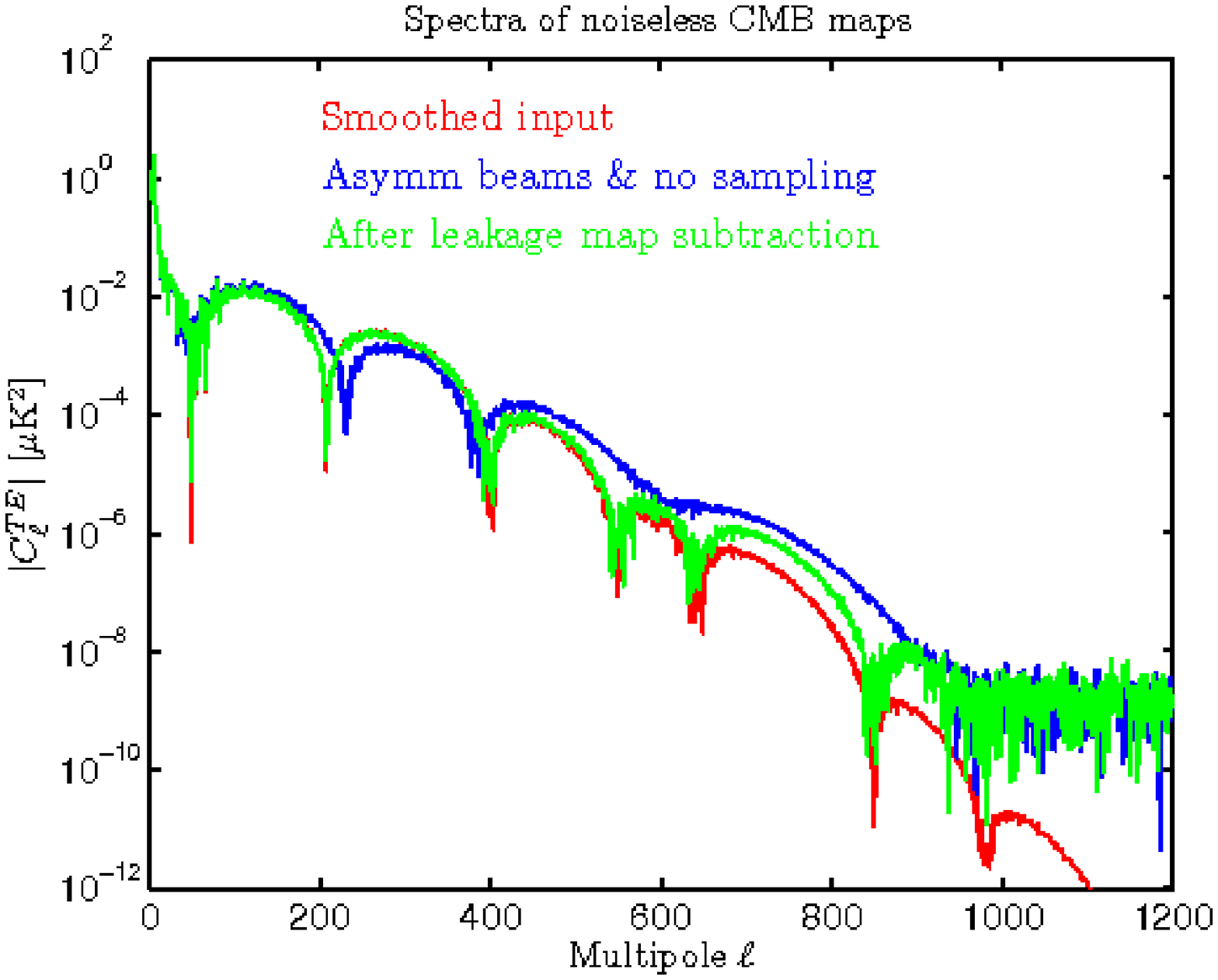}
\includegraphics[scale=0.4]{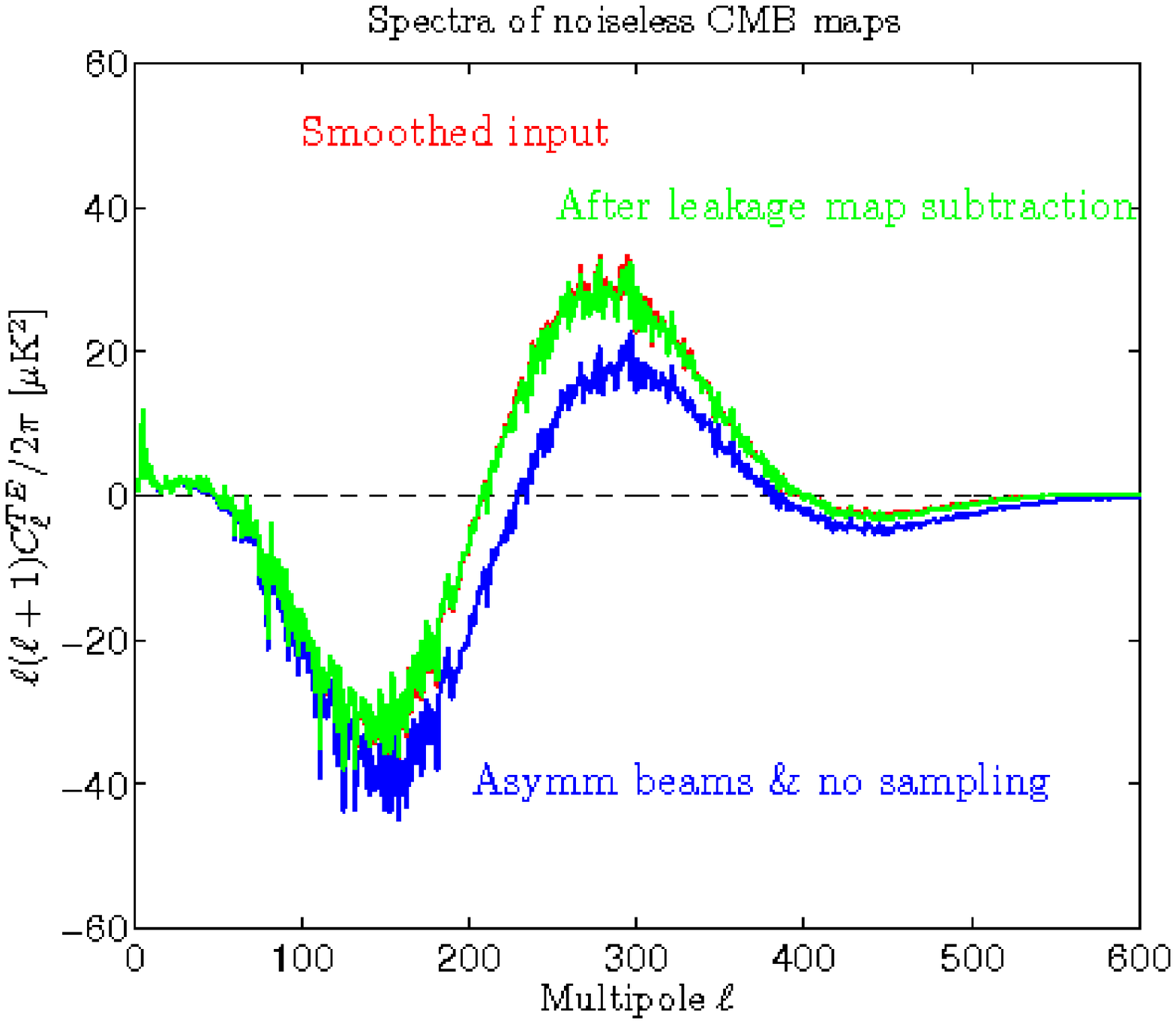}
\end{center}
\caption{Restoration of the acoustic peaks and valleys in the TE
spectrum. \textit{Left hand panel:} Similar to the left hand panel
of Fig.~\ref{fig:ee_1}, but for TE spectra. Note that we plot the
absolute value of the spectra. \textit{Right hand side:} Same as the
left hand panel, but we plot now $\ell(\ell+1)C_{\ell}^{TE}/2\pi$
and zoom to lower multipoles. The red and green curves are nearly on top
of each other.} \label{fig:te_1}
\end{figure*}

To isolate the effect of $T\rightarrow P$, we generated test TODs
for all four 30\,GHz radiometers. The TODs contained Stokes $I$ of
the CMB only. We binned a new $(I,Q,U)$ map from these TODs. The
polarization part of the map should contain the temperature leakage
only. We therefore call this map the \textit{leakage map}. The power
spectra of the leakage map and the original binned map with the
leakage map subtracted are shown in Figs.~\ref{fig:ee_1} and
\ref{fig:te_1}. Removing $T\rightarrow P$ restores the acoustic
peaks and valleys of the EE spectrum in their original positions.
The mismatch of the beams is important at or below the beam scale.
Therefore the magnitude of the $T\rightarrow P$ leakage, and the
corresponding error in the EE spectrum, are small at low $\ell$.

Beam mismatches cause cross-coupling in the opposite
direction too, namely $E\rightarrow T$ and $B\rightarrow T$; however,
because the E and B mode powers are small compared to the power of
the temperature signal, these couplings have an insignificant effect
in the maps and angular power spectra.

\begin{figure*}[!tbp]
\begin{center}
\includegraphics[scale=0.4]{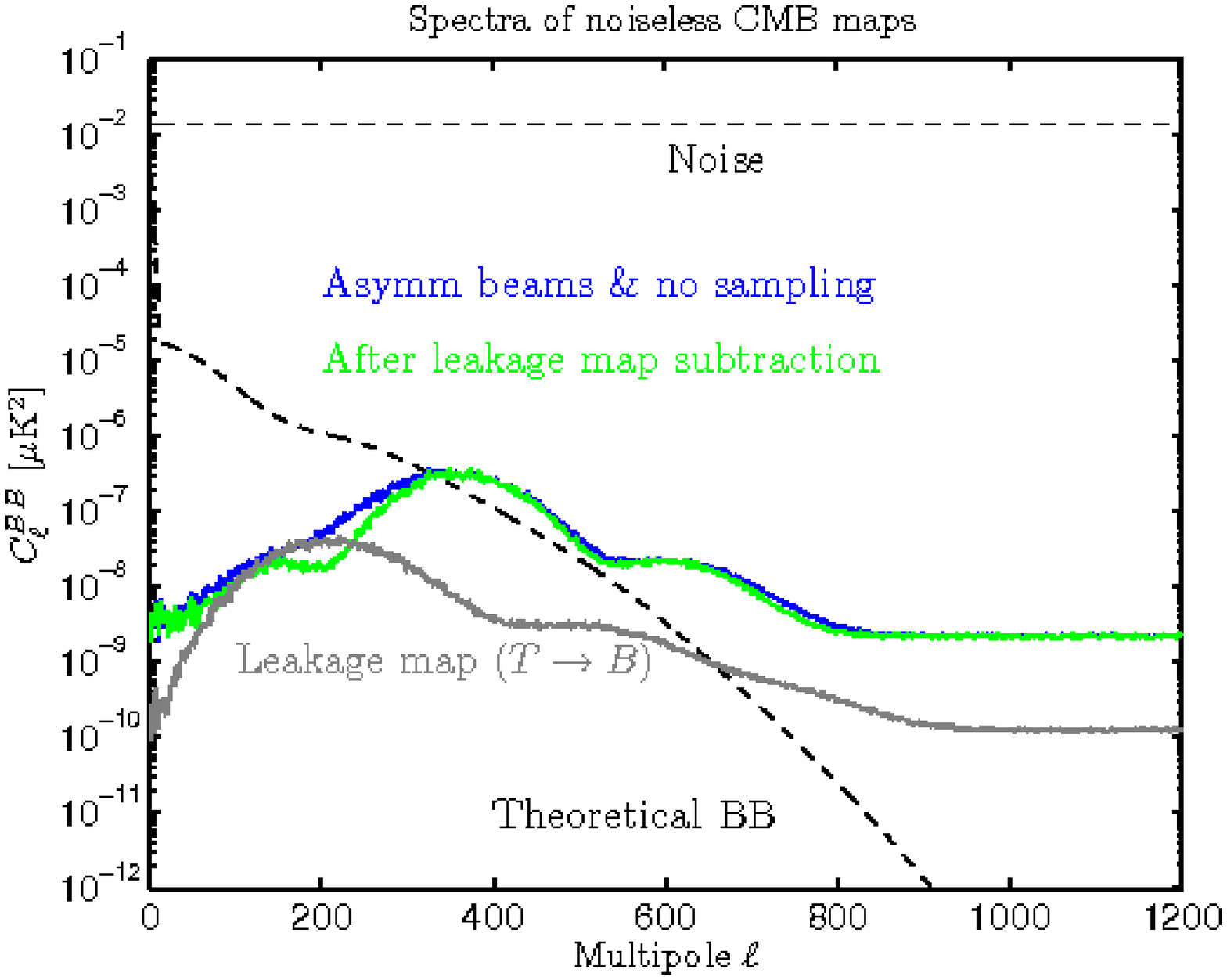}
\includegraphics[scale=0.4]{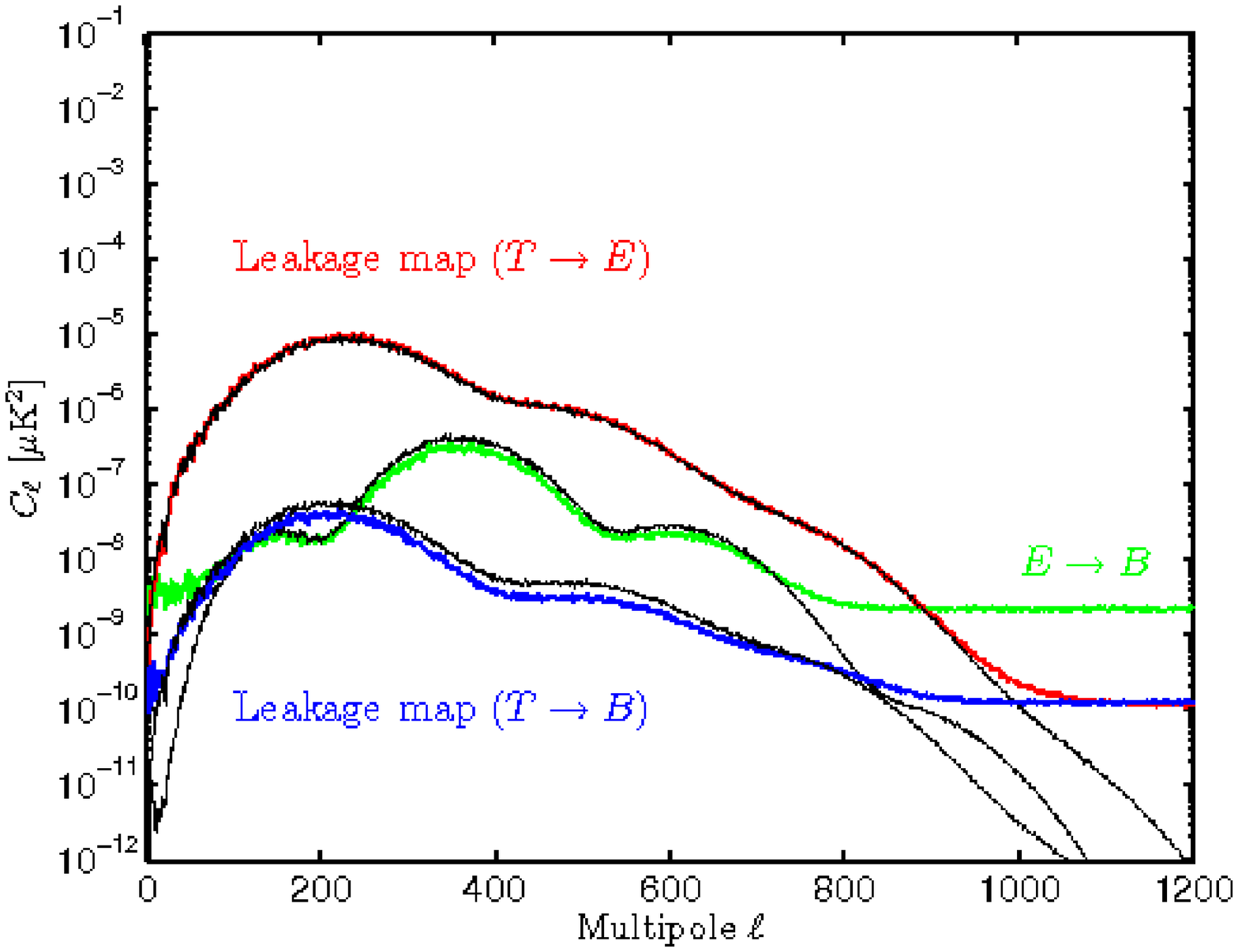}
\end{center}
\caption{\textit{Left hand panel:} Effects of $T\rightarrow P$ and
$E\rightarrow B$ cross-couplings on the BB spectra. The blue curve
is the spectrum of the binned noiseless CMB map. The gray curve is
the BB spectrum of the $T\rightarrow P$ leakage map. Because the
input CMB sky had zero $B$-mode power, the remaining BB power after
the subtraction of the leakage map is mostly cross-coupling from the
$E$-mode polarization. This is shown by the green $E\rightarrow B$
curve. The expected spectrum of a white noise map is indicated by
the horizontal dashed line. For comparison, we show a theoretical
$B$-mode spectrum corresponding to a 10\% tensor-to-scalar ratio and
including lensing from $E$. \textit{Right hand panel:} EE and BB
spectra of the $T\rightarrow P$ leakage map (red and blue curves,
respectively) and the BB spectrum of the $E\rightarrow B$ coupling
(green curve). They are the same as the gray and green curves of the
left-hand panel of this figure and the gray curve of the left-hand
panel of Fig.~\ref{fig:ee_1}. The thin black lines are the
predictions of these signals computed from the analytical model
(Appendix~\ref{sec:model}).   Flattening at high $\ell$ results from
non-uniform pixel sampling. This effect is not included in our
model, however,  so the model spectra do not become flat.}
\label{fig:bb}
\end{figure*}

The $B$-mode spectra (left-hand panel of Fig.~\ref{fig:bb}) show two
notable effects. First, the cross-coupling from the temperature to
the $E$- and $B$-mode spectra are not the same.  Though similar in
shape, the $T \rightarrow E$ power is  2--3 orders of magnitude
larger than the $T \rightarrow B$ power (at $\ell \backsim$ 200).
This discrepancy is rooted in the beam widths and orientations
coupled with the scan pattern. Our analytic model (of
Appendix~\ref{sec:model}) is able to predict this behavior (see the
right-hand panel of Fig.~\ref{fig:bb}). The analytic model also
shows that if the detector beams were axially symmetric but
different in widths, $E$ and $B$ would show the same amount of
leakage from temperature.

Second, the $B$-mode spectrum shows more power than can be explained by
$T\rightarrow P$ alone (see the green curve of
Fig.~\ref{fig:bb}).  Because the CMB sky did not contain $B$-mode
polarization, the $B$-mode power remaining after the removal of the
$T\rightarrow P$ must result from the $E\rightarrow B$
cross-coupling. The source of the distinct $E\rightarrow B$ signal
at $\ell < 800$ is the spin-flip coupling. In
Appendix~\ref{sec:model} we show that the relevant quantity in
$E\rightarrow B$ is the sum of the beam responses of the pair of
detectors sharing a horn. If there is a mismatch of these sums
(mismatch between the horns), the two polarization fields of
opposite spins ($Q + iU$ and $Q - iU$) get mixed. This is the
spin-flip coupling (Hu et al.~\cite{Hu03}). Appendix~\ref{sec:model}
shows how the spin-flip coupling arises from the beam mismatch and
how it creates \hbox{$E\rightarrow B$}.  In these simulations the source of
the mismatch of the sum responses is not the widths of the beams,
but their orientations (see Sect.~\ref{subsec:beams}). The widths of
the pair of beams of a horn are different, but the pairs of beams
have these same values in both horns. The orientations of these
pairs are, however, different in the two horns.

In spite of the fact that the CMB $E$-mode polarization signal is
significantly weaker than the temperature signal, the magnitude of
the $E\rightarrow B$ signal is larger than the magnitude of
$T\rightarrow B$ signal. It seems that the difference of the
orientations of the pairs of beams produces an $E\rightarrow B$
signal that is stronger than the $T\rightarrow B$ signal produced by
the mismatch of widths of the two beams of a horn. The power
transfer between the polarization modes operates equally in both
directions. $B\rightarrow E$ cross-coupling occurs too (same
coupling transfer function as in $E\rightarrow B$), but it has no
effect in $E$, because the $B$-mode power is zero. At large
multipoles (flat part of the BB spectrum at $\ell > 800$) the main
source of $E\rightarrow B$ is the $\ell$ mode coupling of the
non-uniform sampling of the pixels (pixelization error).

Finally, the $T\rightarrow B$ and $E\rightarrow B$ signals could be
compared to the magnitude of the CMB $B$-mode signal that we might
expect to detect with LFI 30~GHz detectors. In the right-hand panel
of Fig.~\ref{fig:bb} the leakage spectra are compared to a
theoretical BB spectrum of the CMB (including lensing from $E$ mode
and corresponding to a 10\% tensor-to-scalar ratio). It can be seen
that the cross-coupling to $B$ is small compared to this signal for
$\ell \leq 300$.

\subsubsection{Effective window functions}\label{subsubsec:window}

The effects of instrument response and data processing on the map
can be described in terms of an effective window function, which we
compute as the ratio of the map power spectrum to the input
spectrum. In our simulations, beams, sample integration, and
sampling of the pixel area (pixel window function) are the main
contributors to the effective window function. In some cases, to
reveal more details, we compute the effective window functions
relative to the smoothed input spectrum.

\begin{figure*}[!tbp]
\begin{center}
\includegraphics[scale=0.4]{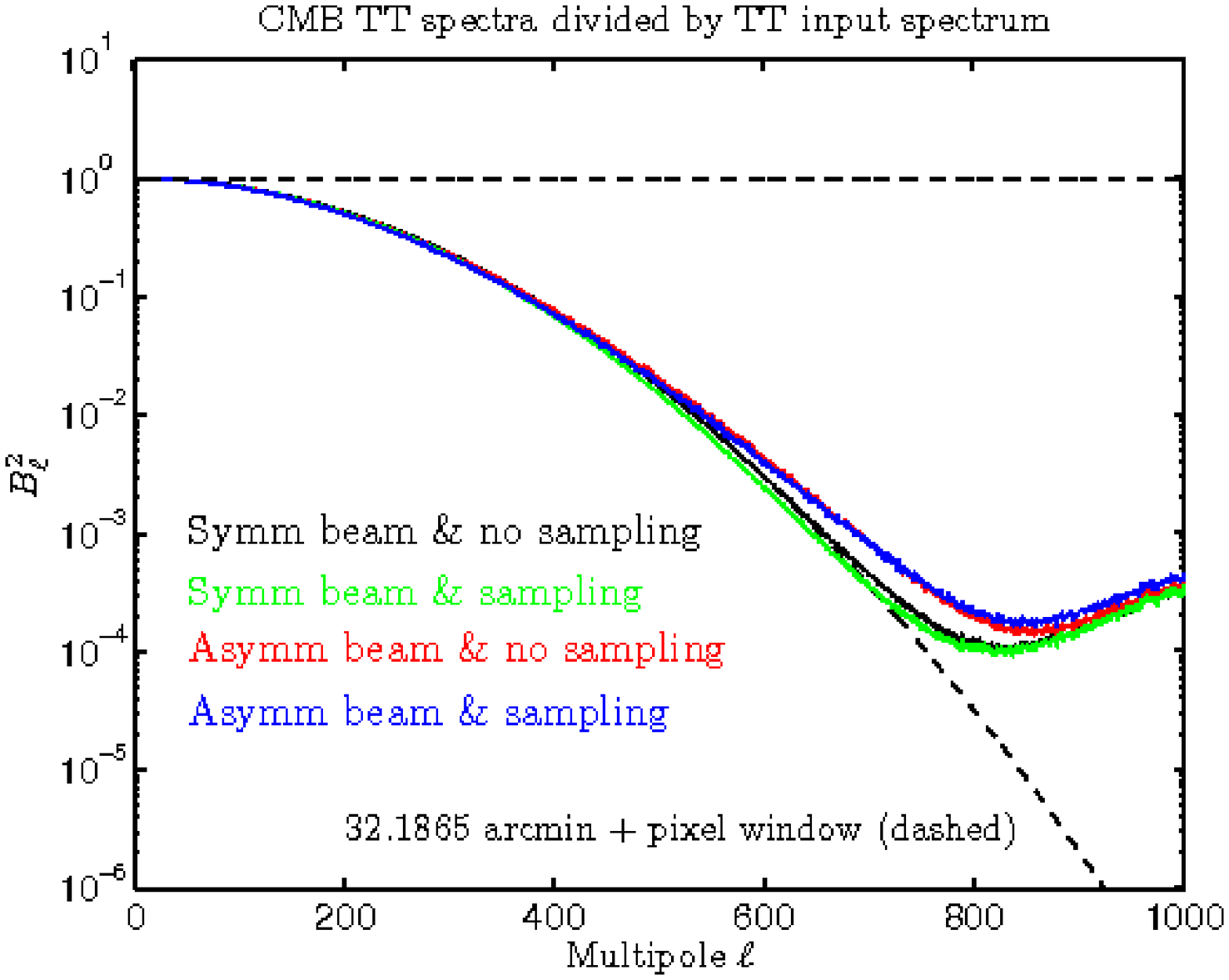}
\includegraphics[scale=0.4]{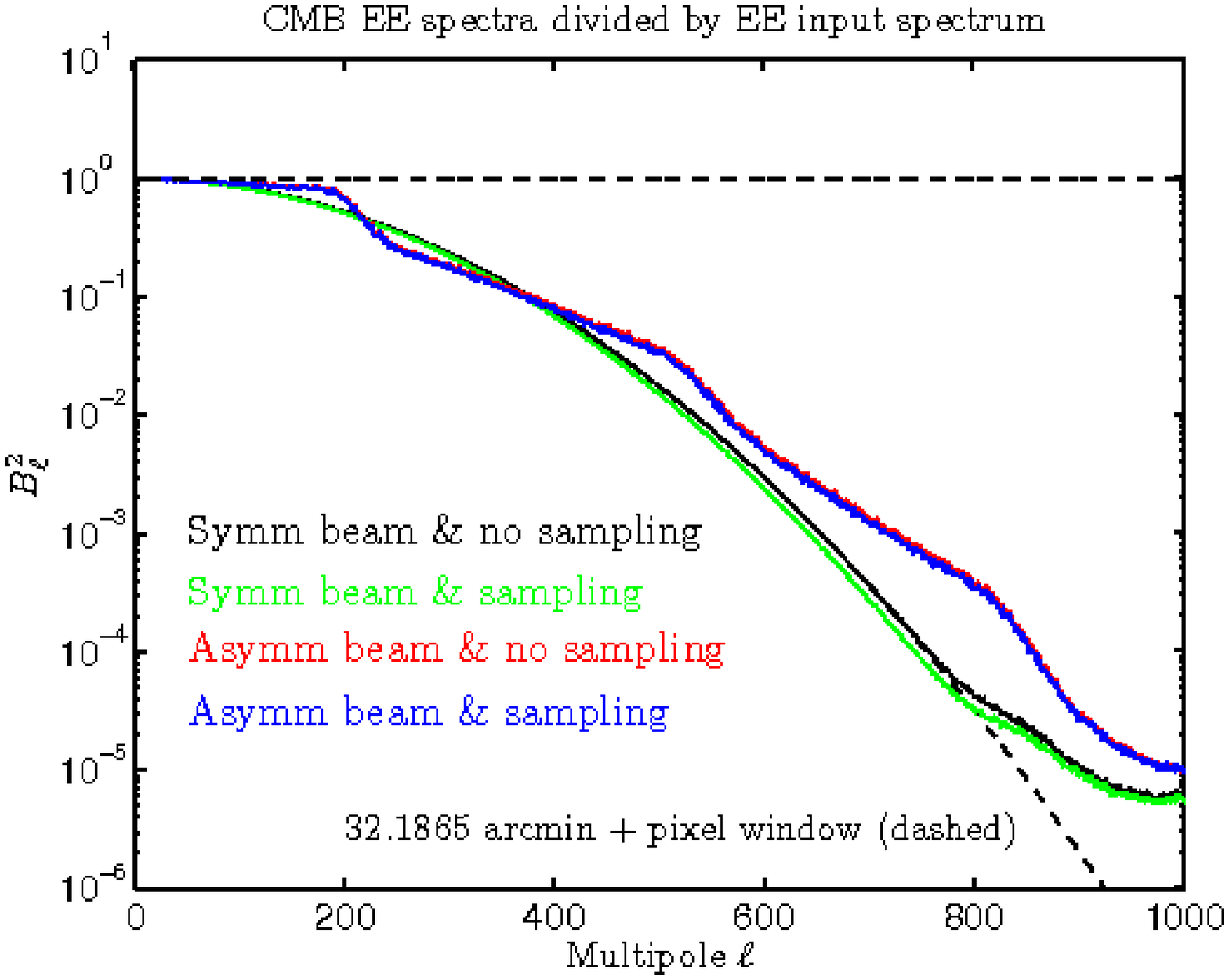}
\includegraphics[scale=0.4]{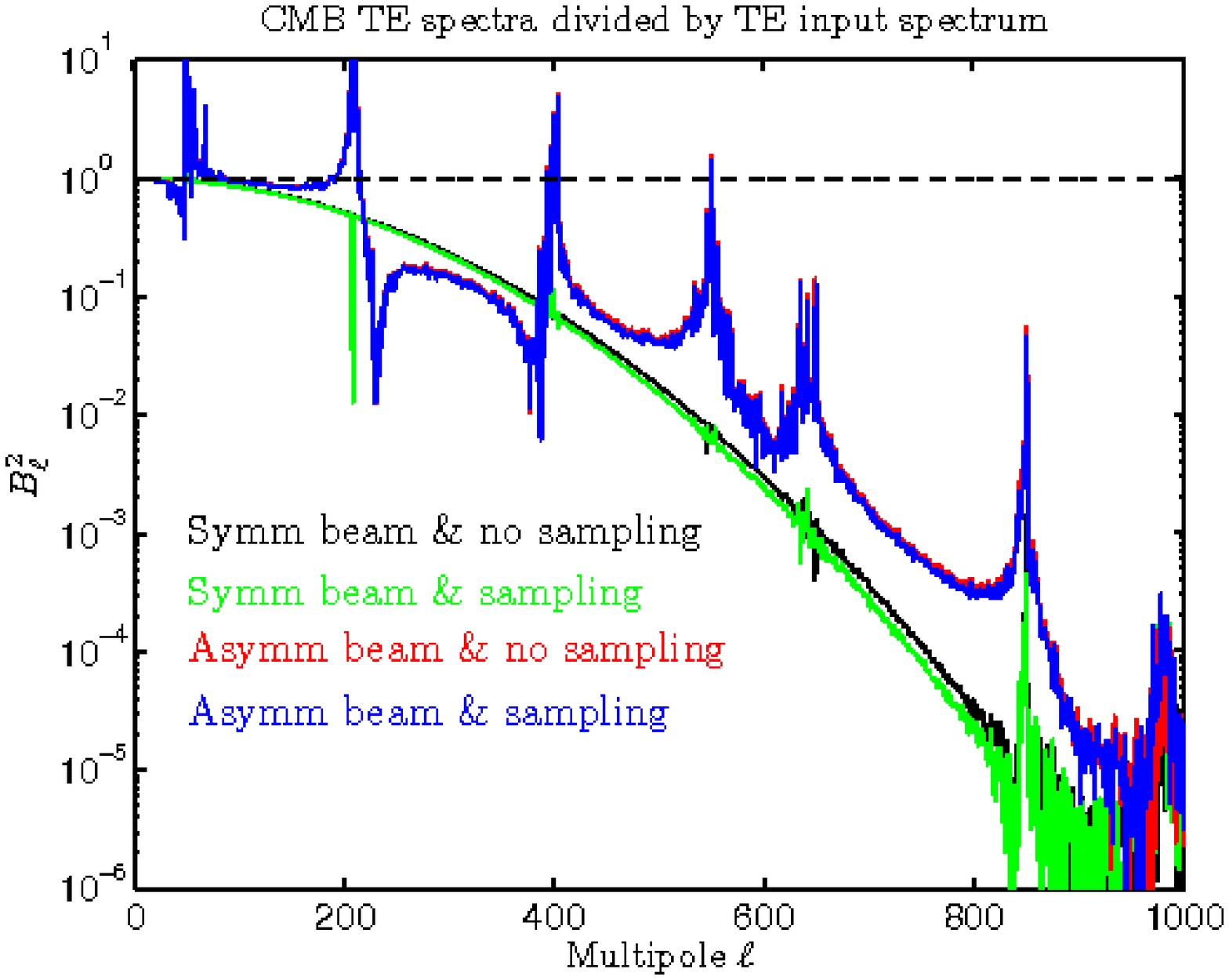}
\end{center}
\caption{TT, EE, and TE effective window functions, computed as
ratios of the angular power spectra of the binned noiseless CMB maps
and the input \hbox{CMB}.  Because the $B$-mode power of the CMB was
zero, no BB window function could be computed.  For comparison we
show the total response of the axially symmetric Gaussian beam
(dashed curve; FWHM = $32\farcm1865$ and $N_{\rm side}$ = 512
HEALPix pixel window function). This is the window function that we
applied to the input spectrum to obtain the smoothed input
spectrum.} \label{fig:raw_window}
\end{figure*}

Figure~\ref{fig:raw_window} shows the TT, EE, and TE effective
window functions of the binned noiseless CMB maps that we introduced
earlier in this section.  The Gaussian window function approximation
breaks down at high $\ell$ and the function becomes flat or starts
to increase (as in the TT window functions). This is an effect of
the pixelization error in the binned noiseless maps. Polarity
changes of the TE spectra cause the sudden jumps seen in the TE
window functions. The EE and TE window functions of the asymmetric
beams show another non-regular characteristic. They deviate
significantly from the regular Gaussian response. The source of this
effect is the $T\rightarrow P$ cross-coupling. The magnitudes of the
window functions with ``sampling on'' are systematically smaller
than their ``sampling off'' counterparts. This is more clearly
visible in the window functions of the symmetric beams. This
difference in window functions comes from the extra spectral
smoothing of the sample integration.

\begin{figure*}[!tbp]
\begin{center}
\includegraphics[scale=0.4]{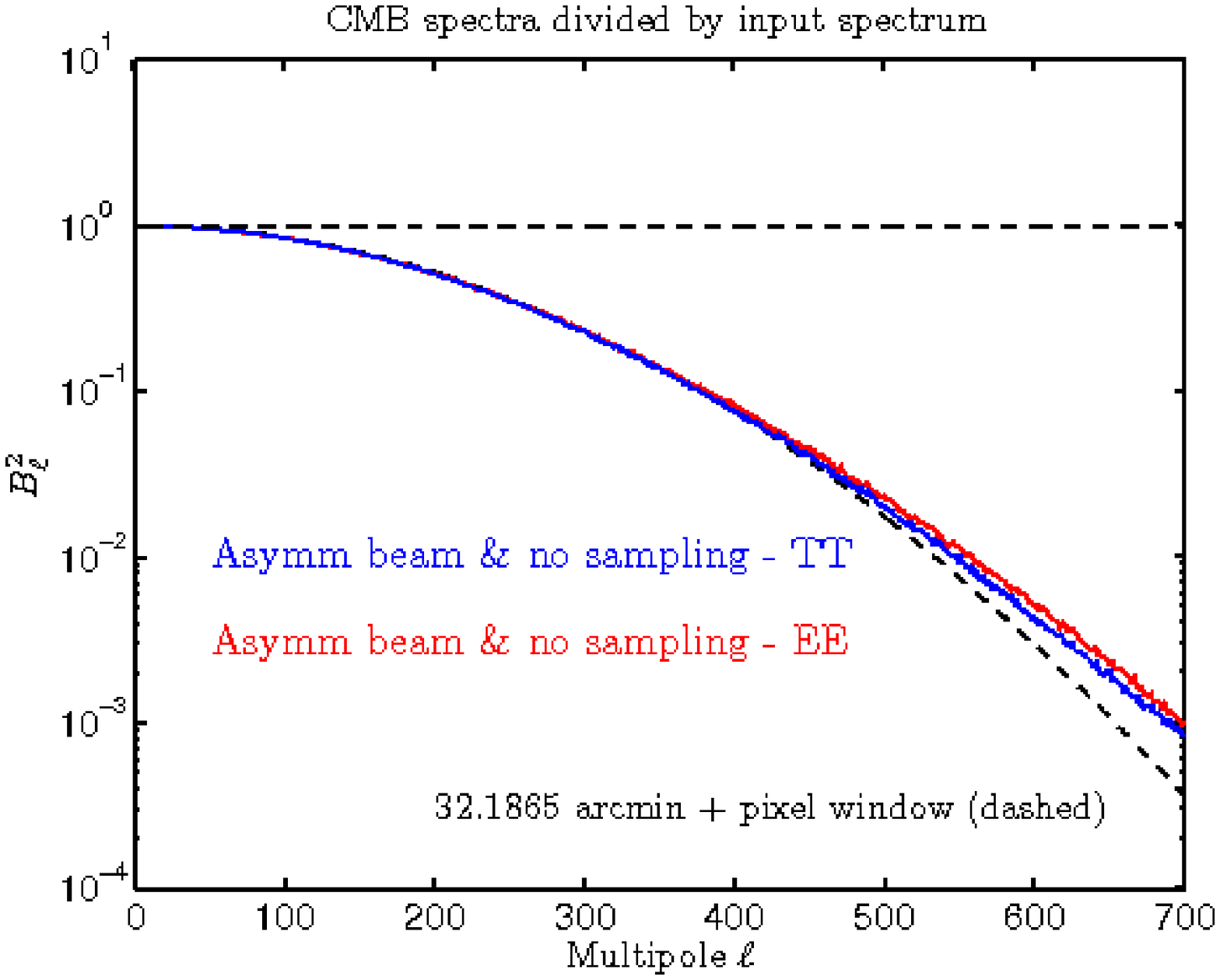}
\includegraphics[scale=0.4]{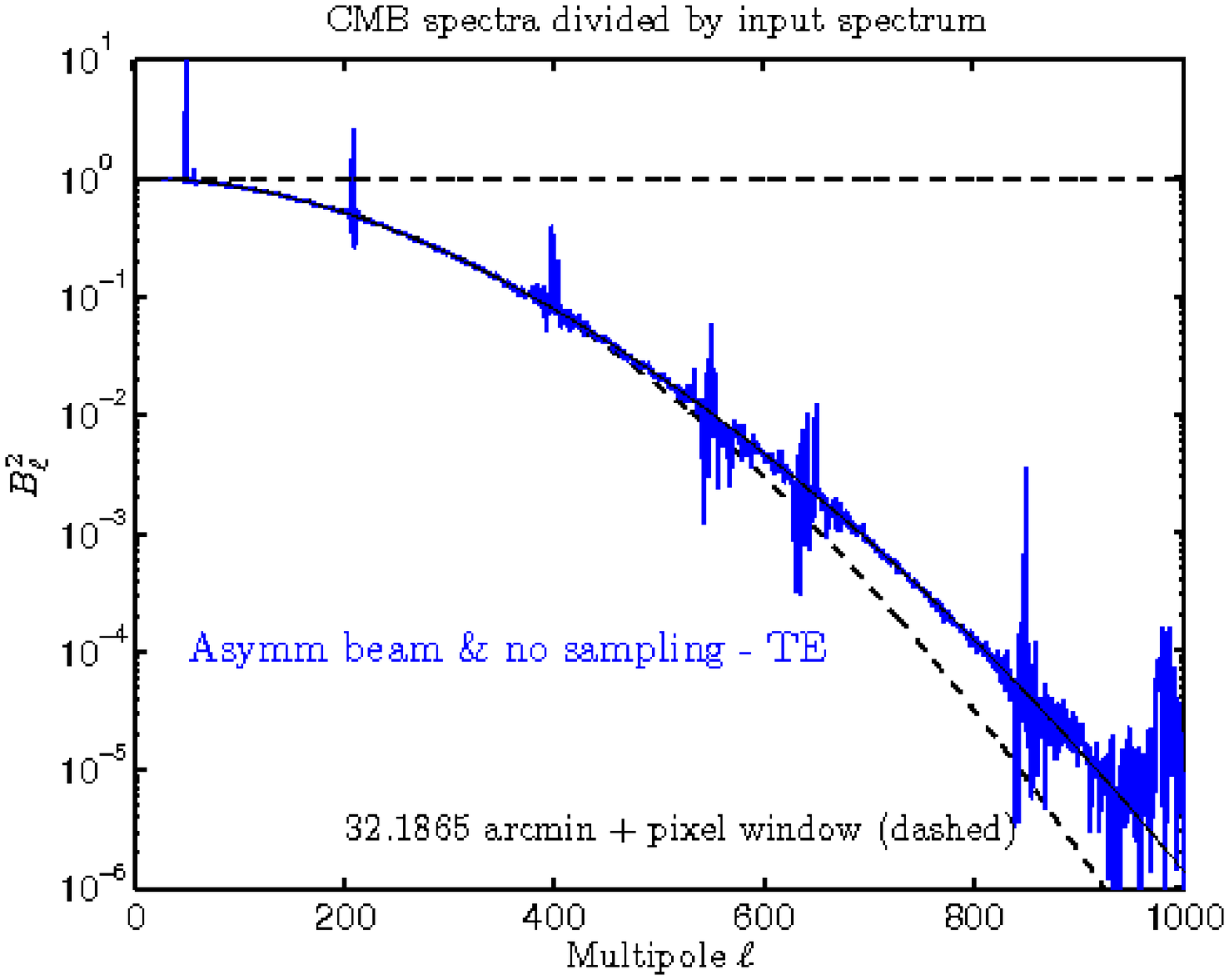}
\includegraphics[scale=0.4]{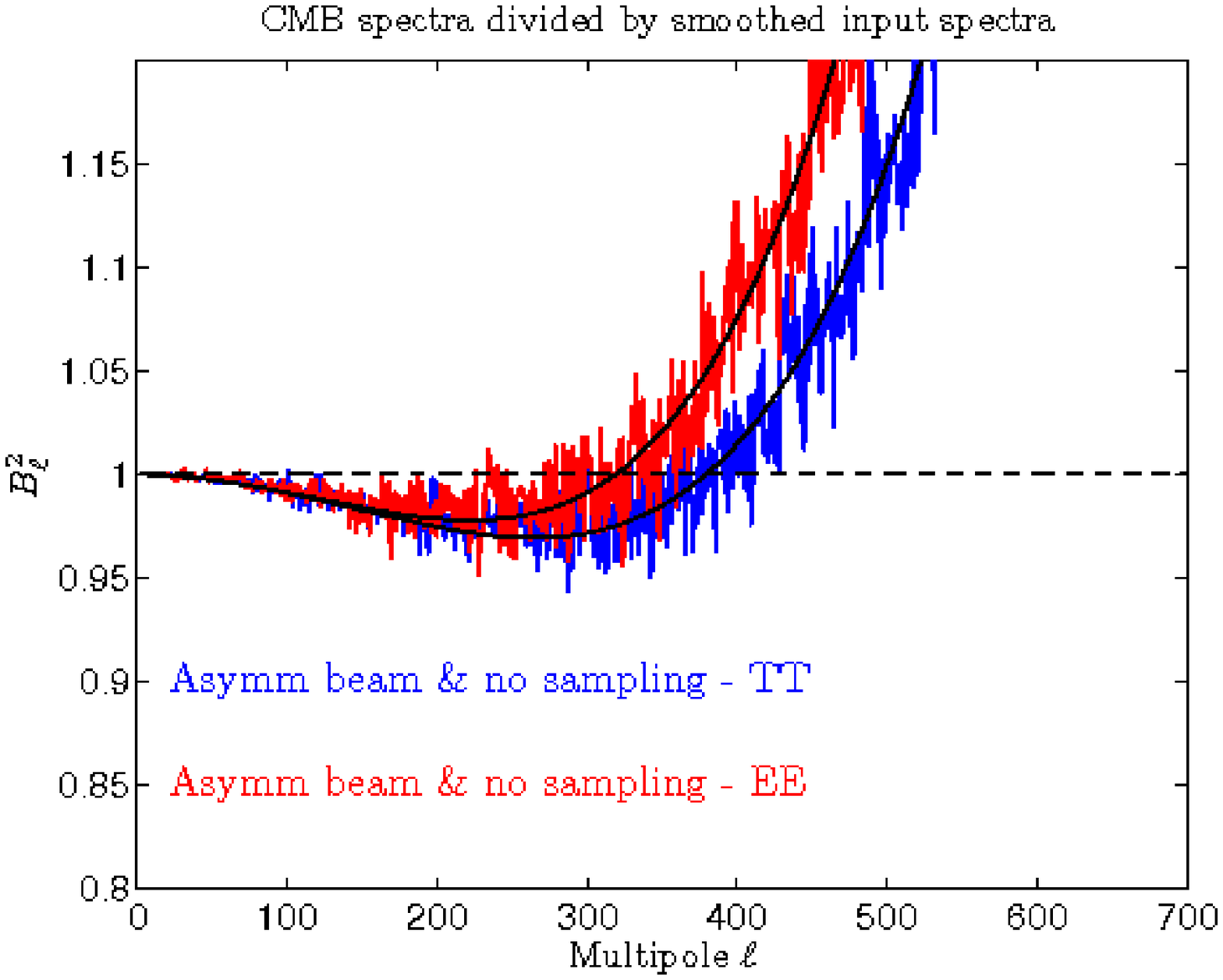}
\end{center}
\caption{TT, EE, and TE effective window functions after the
$T\rightarrow P$ leakage map has been removed from the binned
noiseless CMB map (asymmetric beam \& no sampling). \textit{Top
row:} The TT window function is the ratio of the power spectra of
the binned noiseless CMB map and the CMB input map. The EE and TE
window functions were calculated similarly, except that $Q$ and $U$
leakage maps were first subtracted from the original binned
noiseless map. Symmetric Gaussian responses were separately fitted
to the TT and EE window functions (in range $\ell = 0 \ldots 500$).
The fit FWHM were 32\parcm32 (TT) and 31\parcm94 (EE).
\textit{Bottom panel:} The TT and EE window functions of the top
left panel were divided by the dashed theoretical window function.
The resulting ratios are shown in this plot. Thin black curves are
window functions that we computed from our analytical model for this
case (see Appendix~\ref{sec:model}).} \label{fig:window}
\end{figure*}

To examine the window functions without the effects of the
$T\rightarrow P$ cross-coupling seen above, we subtracted the
$T\rightarrow P$ leakage map from the polarization part of the
binned noiseless CMB map and recomputed the  window functions, as
shown in Fig.~\ref{fig:window}.  We did this for the asymmetric
beams/no sampling case, because we had the leakage map for this case
only. The window functions have now more regular shapes. The TT and
EE window functions are not identical; the TT function is slightly
steeper than the EE function. In the case of perfectly matched beams
the TT and EE window functions would be identical\footnote{The
off-diagonals of the Mueller matrix would be zero and all diagonals
would be identical (see Eq. (\ref{eq:A1_7})).}.

The subtraction of the $T\rightarrow P$ leakage map from the binned
map does not influence the $E\rightarrow B$ cross-coupling. In
addition to creating a $B$-mode polarization signal from the
$E$-mode signal, this coupling influences the original $E$-mode
signal (see Sect.~\ref{subsec:spin_flip} of
Appendix~\ref{sec:model}), so that the TT and EE window functions
become different. Fig.~\ref{fig:window} demonstrates that our
analytical model is able to explain these window functions.

The effective window function of the symmetric beams  with no
sampling is simple to compute because all detectors have the same
beamwidths. If we assume that the scanning is simply from pole to
pole and all detector beams have fixed orientations relative to the
local meridian, the effective TT window function of the asymmetric
beams (no sampling) can be estimated in the following way:

\begin{enumerate}
\item Denote by $B_k(\theta,\varphi)$ the beam response of
the radiometer $k$ of the LFI focalplane (see
Fig.~\ref{fig:focalplane}). The beams are normalized to
$\int_{4\pi}{B_k(\theta,\varphi)d\Omega} = 1$.
\item Compute the coefficients ($b_{\ell m}^{k}$) of the spherical harmonic
expansion of the beam $B_k(\theta,\varphi)$.
\item Compute the mean $b_{\ell m}$ over the radiometer beams:\\
$b_{\ell m} = \frac{1}{N_{\rm det}}\cdot\sum_{k=1}^{N_{\rm
det}}{b_{\ell m}^{k}}$. For LFI 30~GHz $N_{\rm det}$ = 4.
\item Compute the effective beam window function as\\
$B_l^2 = \frac{4\pi}{2\ell+1}\cdot\sum_{m}|b_{lm}|^2$.
\end{enumerate}

If this window function is convolved with the HEALPix pixel window
function, the result explains well the TT window function of
Fig.~\ref{fig:window}.

\subsubsection{Correction of beam mismatch effects} \label{subsubsec:correction}

For temperature observations, the $a_{lm}^{T}$ of the sky get
convolved with a beam, which is fully described by one complex
number for every $\ell$ and $m$. For polarization, where we need
three complex quantities to describe the sky signal
($a_{lm}^{T,E,B}$), the beam in general is a complex $3\times 3$
matrix (for every $\ell$ and $m$). Its non-diagonal elements, which
arise from the beam mismatch, are responsible for the
cross-couplings of temperature and polarization. The leakage map
approach that we used earlier to remove the $T\rightarrow P$ effects
is not applicable in real experiments where Stokes
$I$-only timelines cannot be constructed independently. For real
experiments, more practical methods to correct beam effects are
required.

Mapmaking methods that address beam convolution properly
have been proposed.  Deconvolution mapmaking with a proper treatment
of the detector beams is a method to produce maps free from
cross-couplings of $T$, $E$, and $B$. Two implementions of this
method have been introduced (Armitage and Wandelt~\cite{Arm04};
Harrison et al.~\cite{Har08}). Both have shown results indicating
that they may be computationally practical for the lower-resolution
{\sc Planck} channels (in the LFI).  Another map-domain method able
to correct for beam effects is the FICSBell approach (Hivon et.
al.~\cite{Hiv08}), in which asymmetries of the main beam are
treated as small perturbations from an axially symmetric Gaussian
beam, and are averaged over each pixel taking into account the
orientation of the detector beams at each visit of that pixel.

In Appendix~\ref{sec:model} we developed an analytical model to
predict the effects of the beams in the angular power spectra of the
CMB maps. This model was inverted (also in Appendix~\ref{sec:model})
to turn it to a correction method.  It can deconvolve the effects of
the asymmetric beams from the angular power spectrum and return a
spectrum that is an approximation of the spectrum of the input sky.
Our method is based on a number of simplifying assumptions that
limit its accuracy in real experiments; however, we can use it to
compute coarse corrections that can be improved with, e.g., Monte
Carlo simulations. The correction capability of our method is
demonstrated in Fig.~\ref{fig:te_3}.

\subsubsection{Sample integration effects} \label{subsubsec:sampling}

\begin{figure}[!tbp]
\begin{center}
\resizebox{\hsize}{!}{\includegraphics[width=0.8\textwidth]{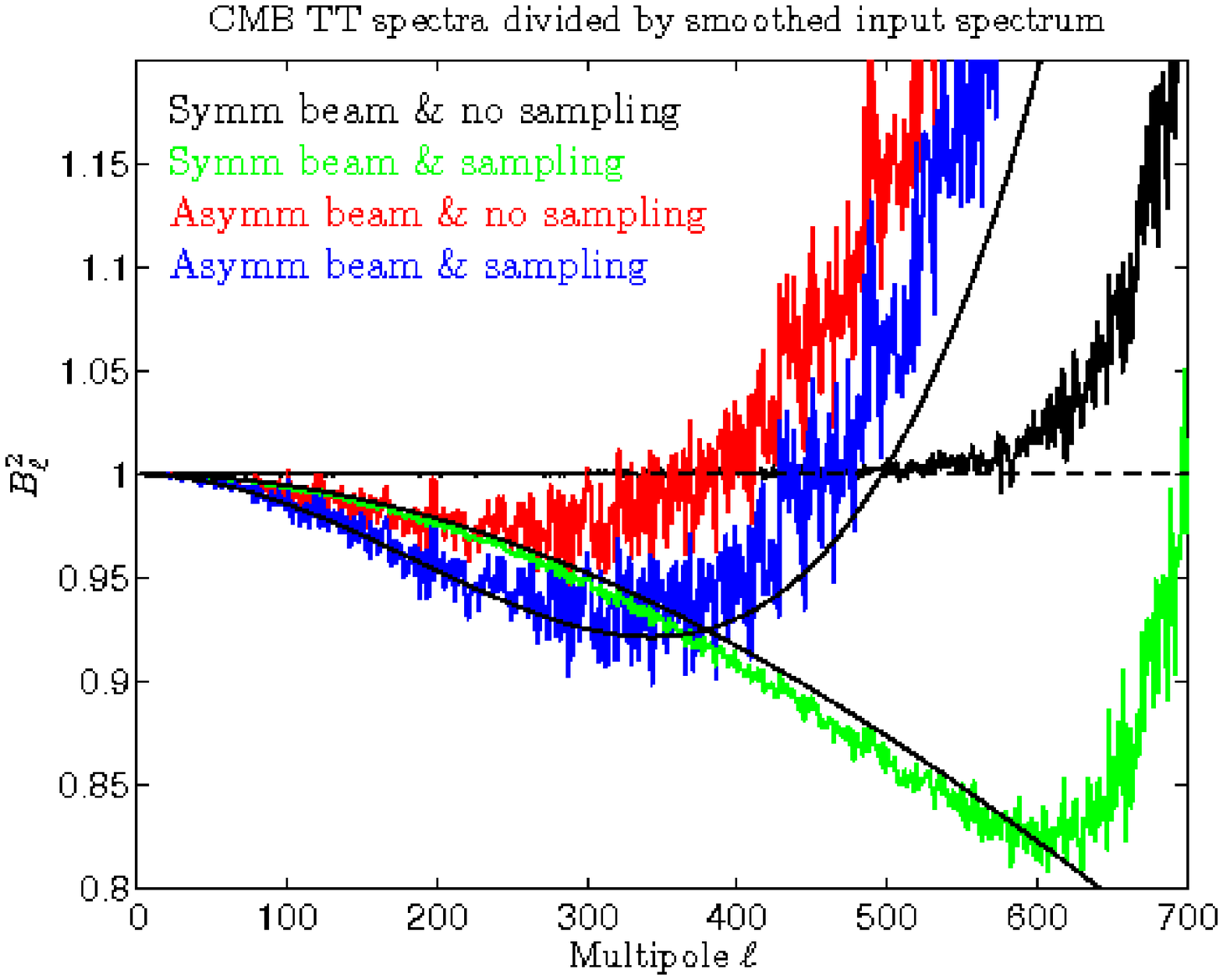}}
\end{center}
\caption{Sample integration effects in the effective TT window
functions. These window functions are the ratios of the TT angular
spectra of the binned noiseless CMB maps and the smoothed CMB input
spectrum. The red curve is the same as the red curve of
Fig.~\ref{fig:window}. The window functions of the symmetric beams
(black and green curves) blow up at $\ell \approx 600$ due to
pixelization error. The corresponding blow up of the other window
functions occurs outside the scales of this plot. The thin black
curve that tracks the green curve was computed from
Eq.~(\ref{eq:beam4}). The other black curve (that tracks the blue
curve) is the window function prediction of our analytical model
multiplied with Eq. (\ref{eq:beam4}).} \label{fig:sampling_window}
\end{figure}

We recomputed the effective TT window functions of
Fig.~\ref{fig:raw_window}, but this time we divided the angular
power spectra of the binned noiseless CMB maps with the smoothed
input spectrum (instead of the input spectrum). The recomputed
effective TT window functions are shown in
Fig.~\ref{fig:sampling_window}. A pairwise comparison of the
``sampling'' and ``no sampling'' window functions (of the same beam
type) reveals the effect of the sample integration in the angular
power spectrum of a CMB map. The extra spectral smoothing due to the
sample integration can be clearly seen.

\begin{figure}[!tbp]
  \begin{center}
    \resizebox{\hsize}{!}{\includegraphics{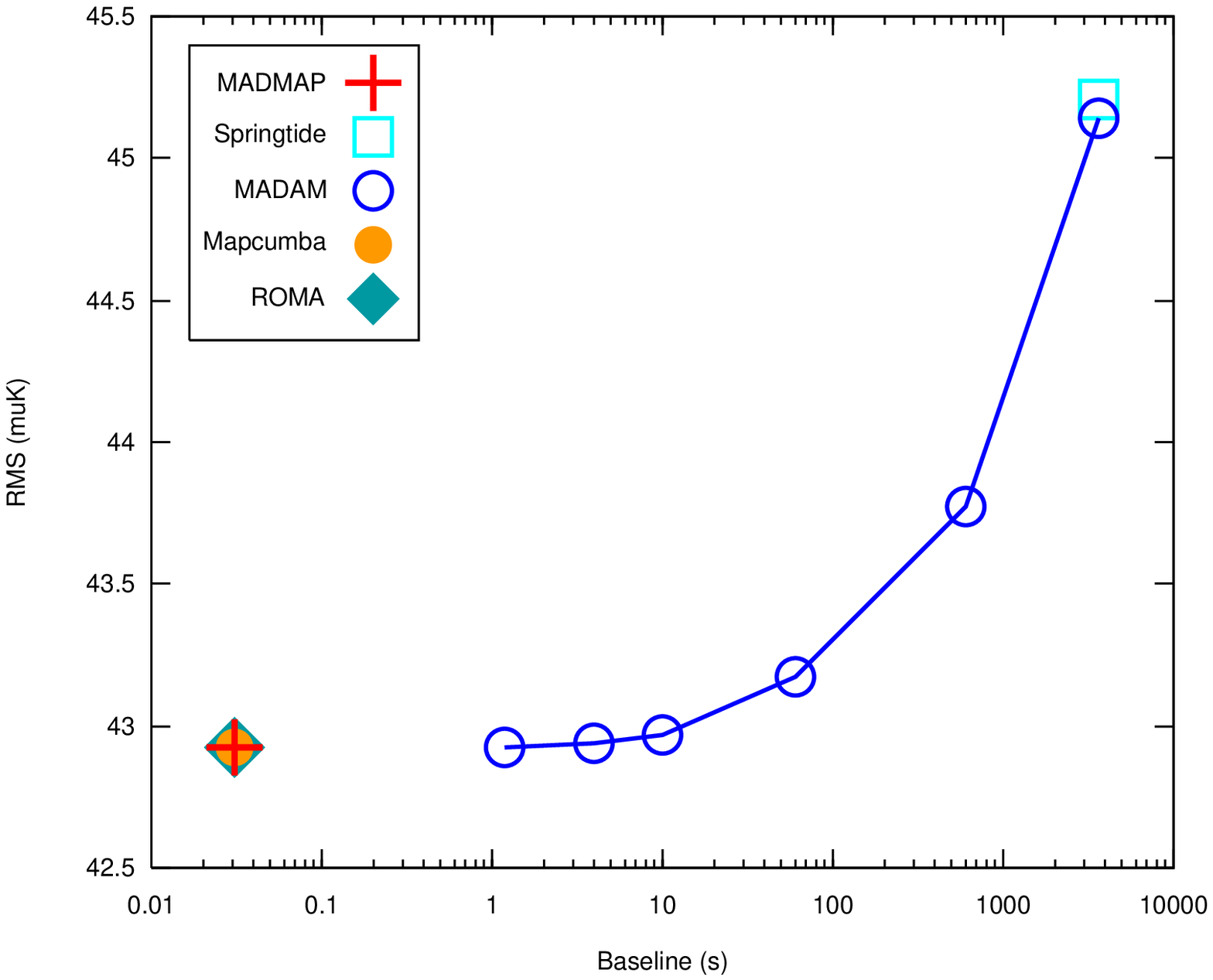}}
  \end{center}
\caption{RMS of the residual temperature maps for LFI 30\,GHz
observations. The maps were $N_{\rm side} = 512$ full sky maps and
they contained CMB, instrument noise, and the effects of the
sorption cooler temperature fluctuations. Asymmetric beams were
used, and sample integration was turned on. The y-axis units are
thermodynamic microkelvins. The angular power spectra of some of the
residual maps are shown in Fig.~\ref{fig:cnrdiff}.}
\label{fig:kevin}
\end{figure}

Let us consider an axially symmetric Gaussian beam and its
elongation in the direction of the scan. The window function of the
initial beam (before elongation) is
 \beq
 B_{\ell}^2 = e^{-\ell(\ell + 1)\sigma_0^2},
 \label{eq:beam1}
 \eeq
where $\sigma_0 = \rm{FWHM}/\sqrt{8\ln2}$.\footnote{In this study
the FWHM of the axially symmetric LFI 30\,GHz beams is
$32\farcm1865$.} This window function operates in the angular power
spectrum domain. The beam $\sigma$ stays in its original value
($\sigma = \sigma_0$) in the perpendicular direction of the scan.
Along the scan the $\sigma$ gets modified to (Burigana et
al.~\cite{Bur01})
 \beq
 \sigma_{\rm eff}^2 = \sigma_0^2 + \frac{\Delta\theta_{\rm s}^2}{12}.
 \label{eq:beam2}
 \eeq
Here $\Delta\theta_{\rm s} = 2\pi f_{\rm spin}/f_{\rm sample}$, the
angle through which the beam center pointing rotates during a
detector sample time. For the nominal satellite spin rate ($f_{\rm
spin}$ = 1~rpm) the ellipticity ($\sigma_{\rm eff}/\sigma_0$) of the
elongated beam is 1.027. The ellipticity produced by the scanning is
significantly smaller than the ellipticities of our asymmetric beams
($\backsim$1.35).

The geometric mean of the $\sigma$'s of the elongated beam is
 \beq
 \sigma_{\rm e}^2 = \sigma_0\sigma_{\rm eff} = \sigma_0\sqrt{\sigma_0^2 + \frac{\Delta\theta_{\rm s}^2}{12}}
 \approx \sigma_0^2 + \frac{\Delta\theta_{\rm s}^2}{24}.
 \label{eq:beam3}
 \eeq
The last form is an approximation that could be made because
$\frac{\Delta\theta_{\rm s}^2}{12} \ll \sigma_0^2$.
Eq.~(\ref{eq:beam3}) suggests that the effect of the sample
integration in the angular power spectra of the maps could be
approximated by a symmetric Gaussian window function \beq
 E_{\ell}^2 = e^{-\ell(\ell + 1)\sigma_{\rm s}^2},
 \label{eq:beam4}
 \eeq
where $\sigma_{\rm s} = \Delta\theta_{\rm s}/\sqrt{24}$. In our
simulations this $\sigma_{\rm s}$ corresponds to FWHM = $5\farcm32$.
We compare the predictions of this model to the actual window
functions of our simulation in Fig.~\ref{fig:sampling_window}. The
comparison shows that the accuracy of our simple model is good in
the symmetric beam case but somewhat worse for the asymmetric beam
case.

\begin{figure*}
  \begin{center}
    \includegraphics[scale=0.4]{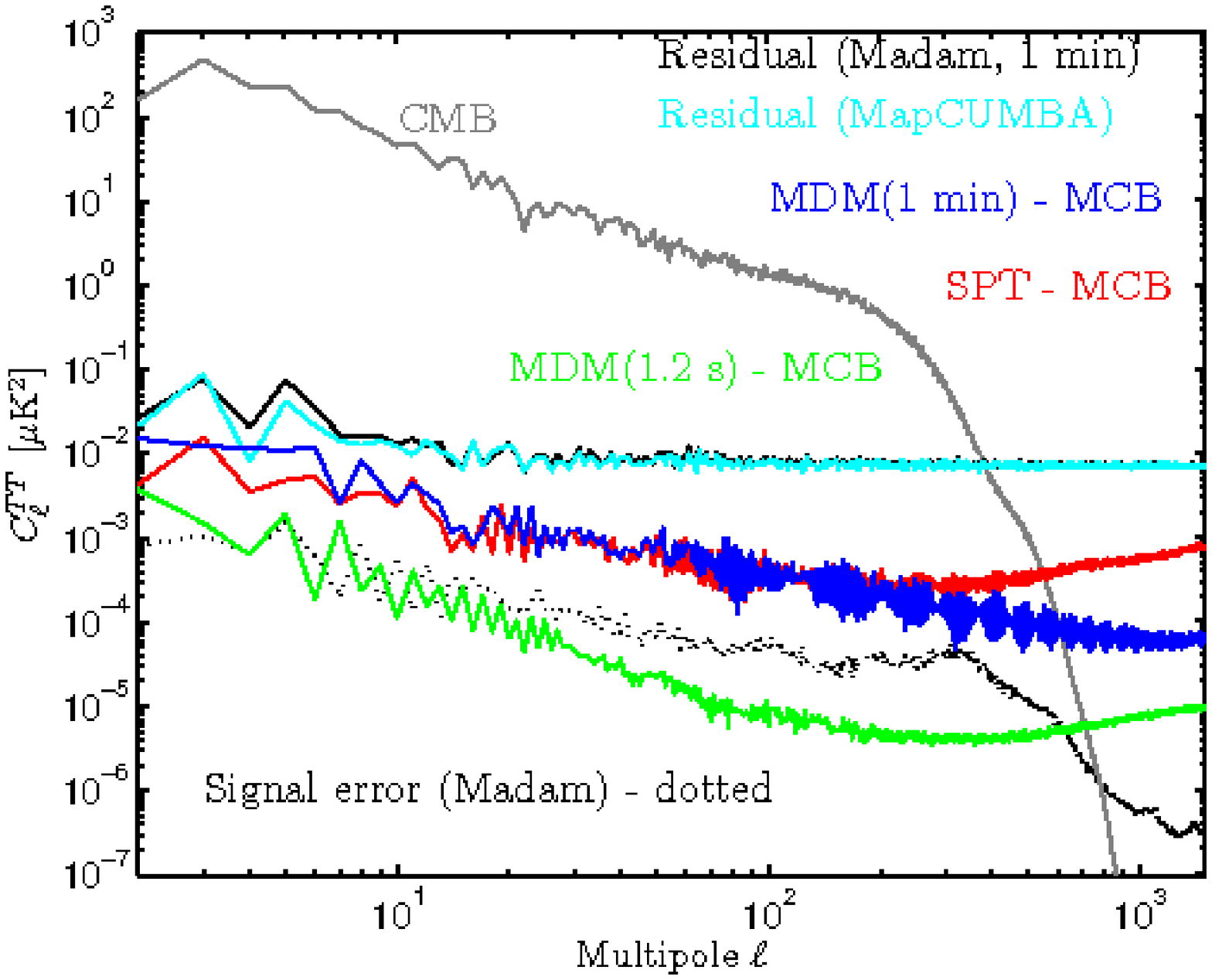}
    \includegraphics[scale=0.4]{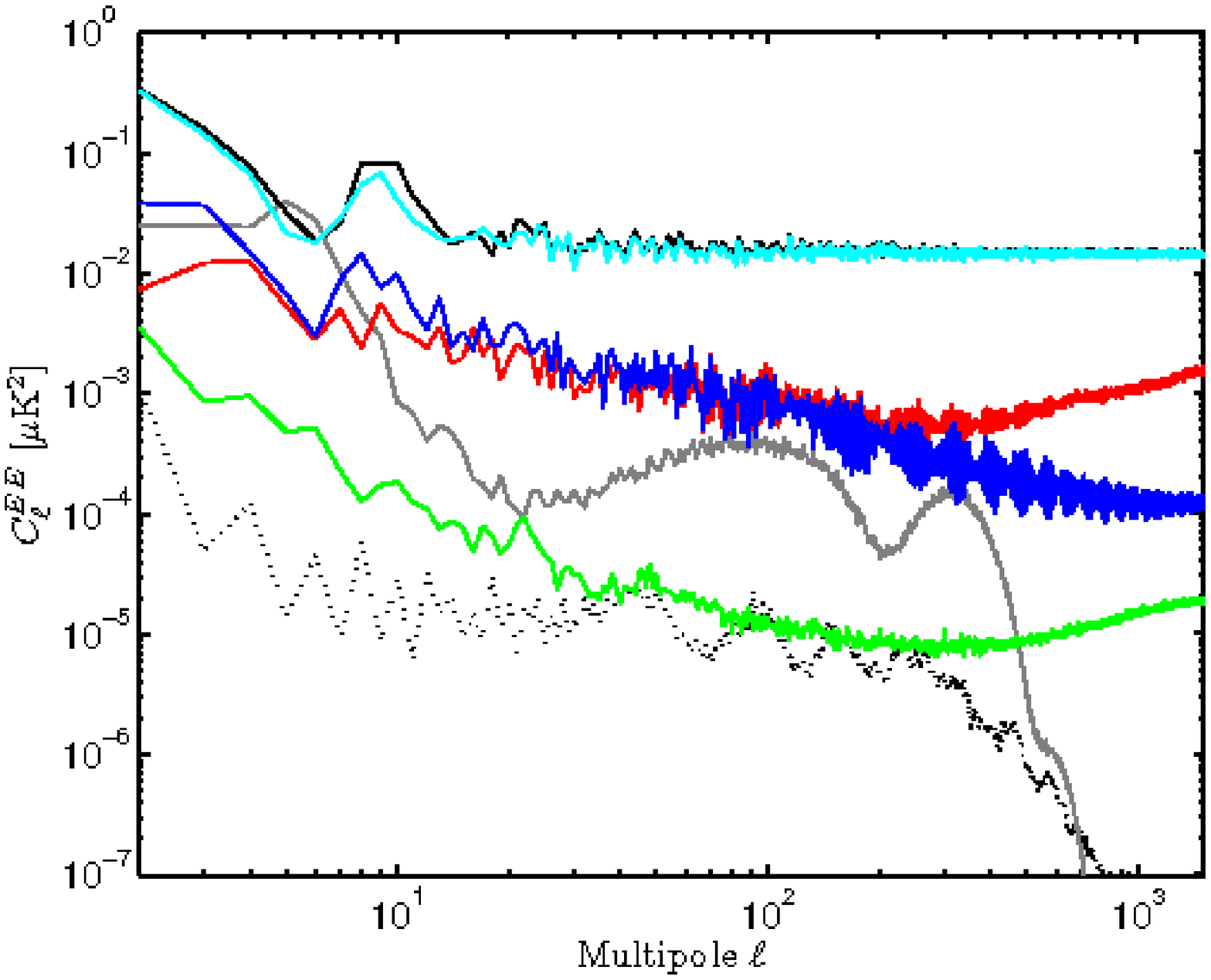}
    \includegraphics[scale=0.4]{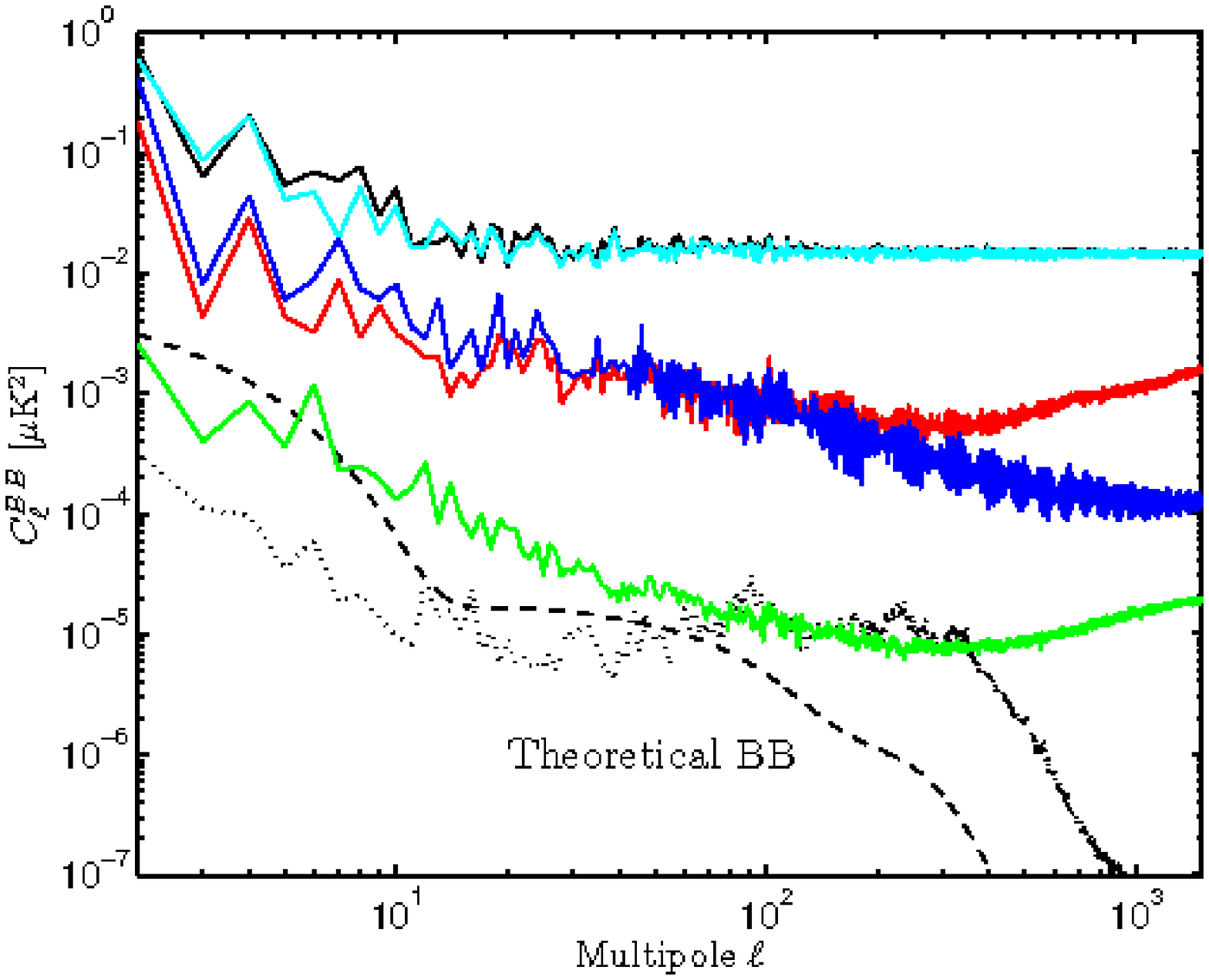}
  \end{center}
\caption{TT, EE, and BB angular power spectra of Madam (with
1-minute baselines and without the noise filter) and MapCUMBA
residual maps (black and light blue curves, respectively). Their RMS
values were shown in Fig.~\ref{fig:kevin}. The maps represent the
observations of four LFI 30\,GHz detectors (CMB, instrument noise,
cooler fluctuations; asymmetric beams; and sample integration on).
The red curve is for the difference of the Springtide (SPT) and
MapCUMBA (MCB) residual maps.  The blue curve is the same for Madam
of 1-minute baselines (MDM (1 min)) and MapCUMBA, and the green
curve is for Madam of 1.2 s baselines (MDM (1.2 s)) and
\hbox{MapCUMBA}. Madam used either 1.2\,s or 1-minute baselines,
whereas Springtide used long 1\,hour baselines. For comparison we
also show the CMB spectrum of our simulations (gray curve), the
spectrum of the Madam signal error map (including all sky emissions,
asymmetric beams, and sample integration, dotted curve), and a
theoretical CMB $B$-mode spectrum (dashed curve, see also
Fig.~\ref{fig:bb}).} \label{fig:cnrdiff}
\end{figure*}

\subsection{Residual maps} \label{subsec:residual_maps}

To compare mapmaking codes, we constructed residual maps,
specifically the difference between the output map and the binned
noiseless map. Smaller residuals imply smaller mapmaking errors. We
examined the RMS of the residuals, and computed power spectra to
study scale dependence.  We are interested in anisotropy; the mean
sky temperature is irrelevant. Therefore, whenever we calculated a
map RMS, we subtracted the mean of the observed pixels from the map
before squaring. The RMS of a map was always calculated over the
observed pixels.

Fig.~\ref{fig:kevin} shows the RMS of the residual temperature maps
for some of our mapmaking codes and for a number of destriper
(Madam) baseline lengths. The data of Fig.~\ref{fig:kevin} were
derived from HEALPix $N_{\rm side} = 512$ maps.  At that pixel size
the instrument noise is the dominant contributor in the residual
maps\footnote{Mapmaking errors arising from the subpixel signal
structure typically show up in maps with considerably larger pixel
size.}(Poutanen et al.~\cite{Pou06}, Ashdown et al.~\cite{Ash07b}).
The optimal codes consistently deliver output maps with the smallest
residuals, which were nearly the same between the several codes.
Madam with short uniform baselines (e.g., 1.2\,s) produces maps with
residuals essentially the same as those of the optimal codes, at
considerably lower computational cost.  Destriper maps with longer
baselines showed larger residuals. The RMS of the other Stokes
parameters ($Q$ and $U$) were larger as expected, but behaved
similarly as a function of baseline length.

\begin{figure*} [!tbp]
\begin{center}
\includegraphics[scale=0.3,angle=90]{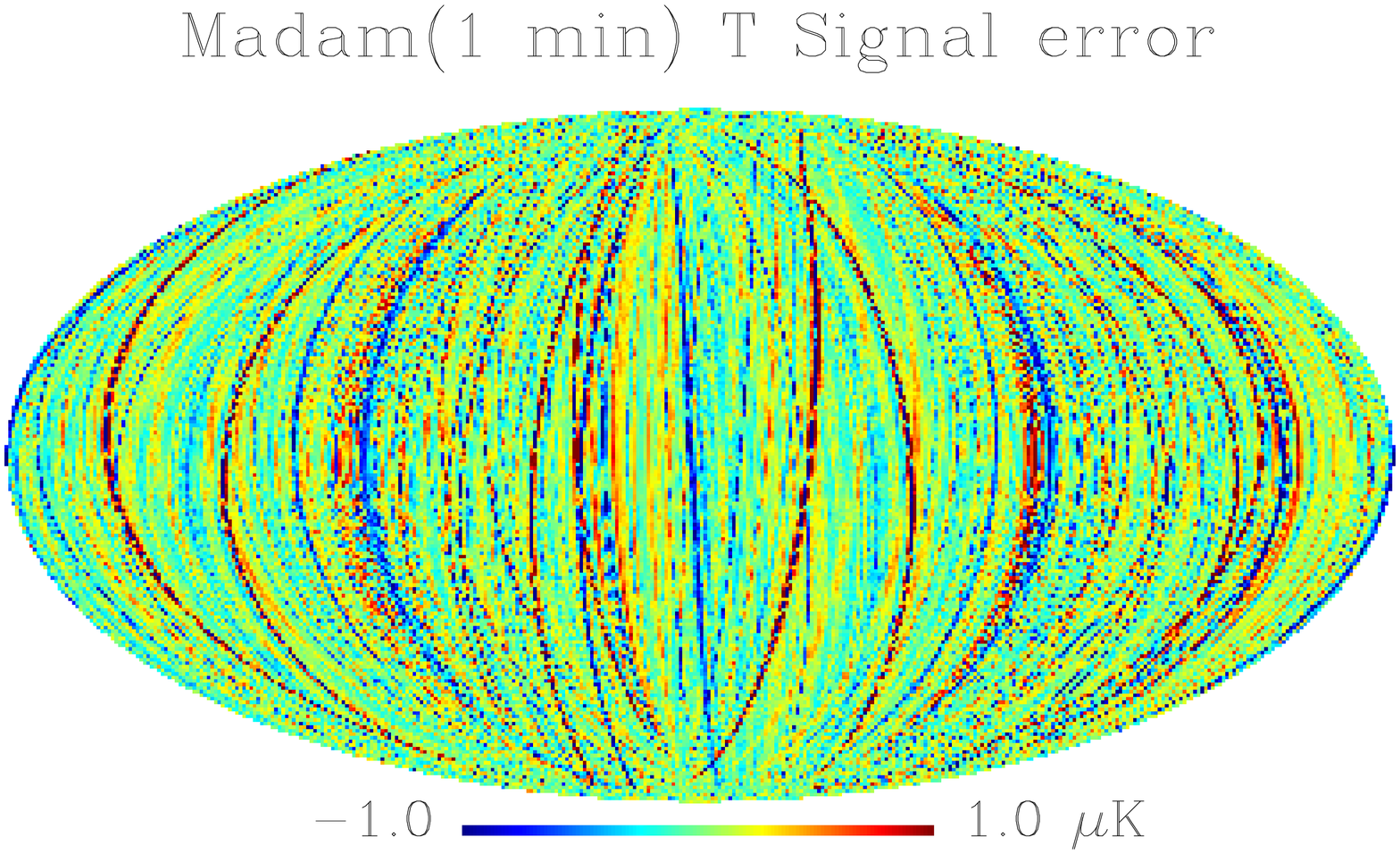}
\includegraphics[scale=0.3,angle=90]{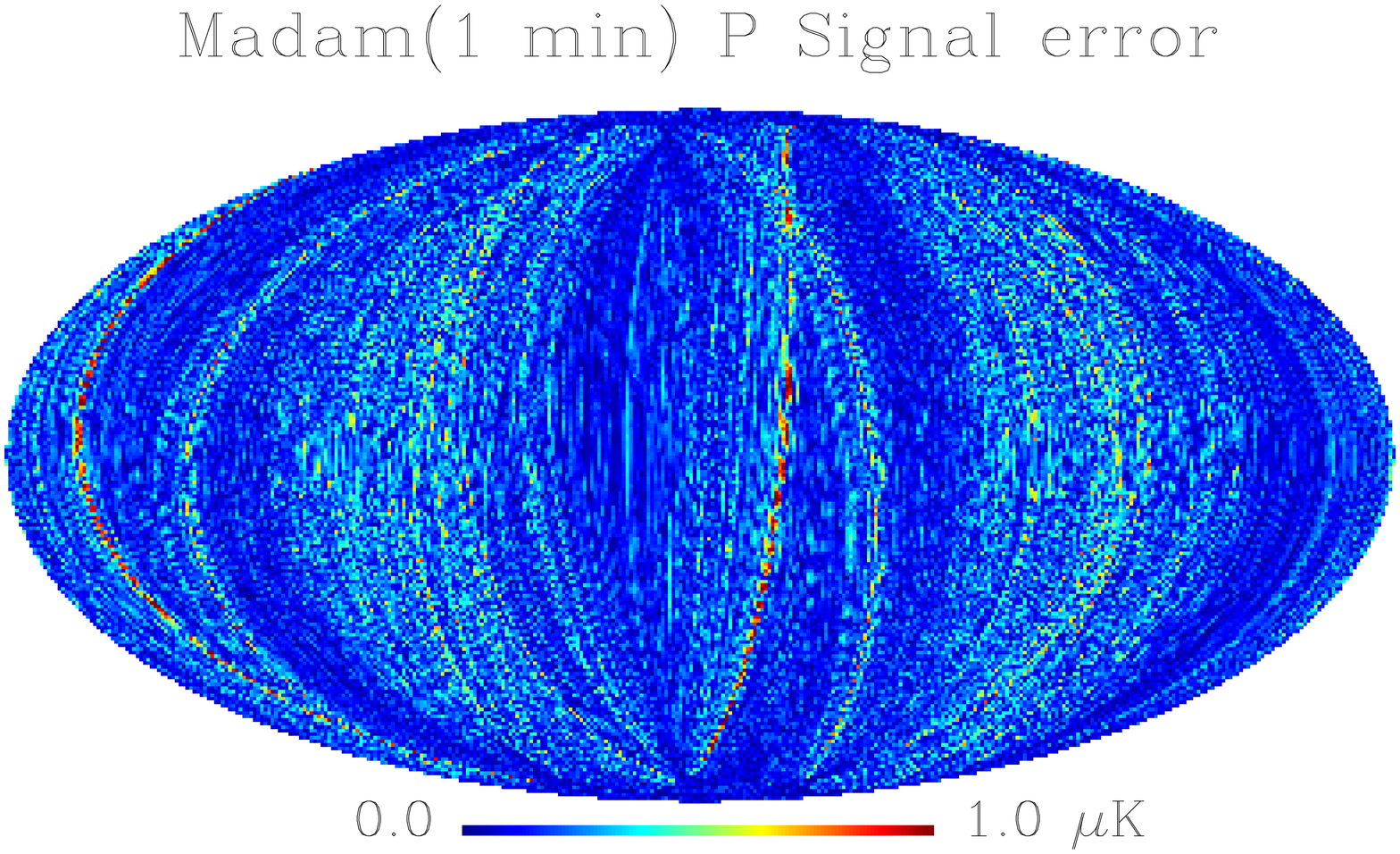}
\includegraphics[scale=0.3,angle=90]{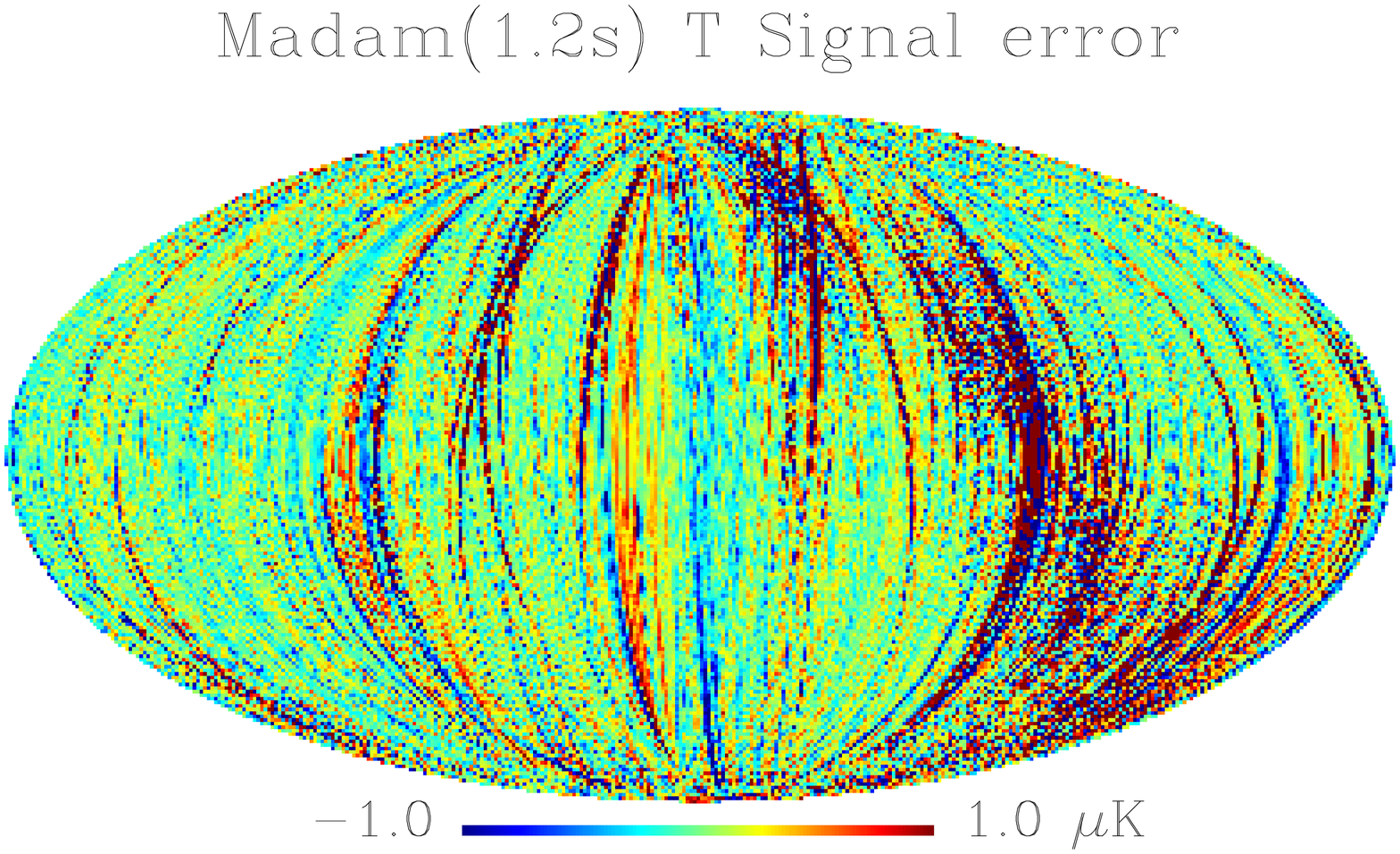}
\includegraphics[scale=0.3,angle=90]{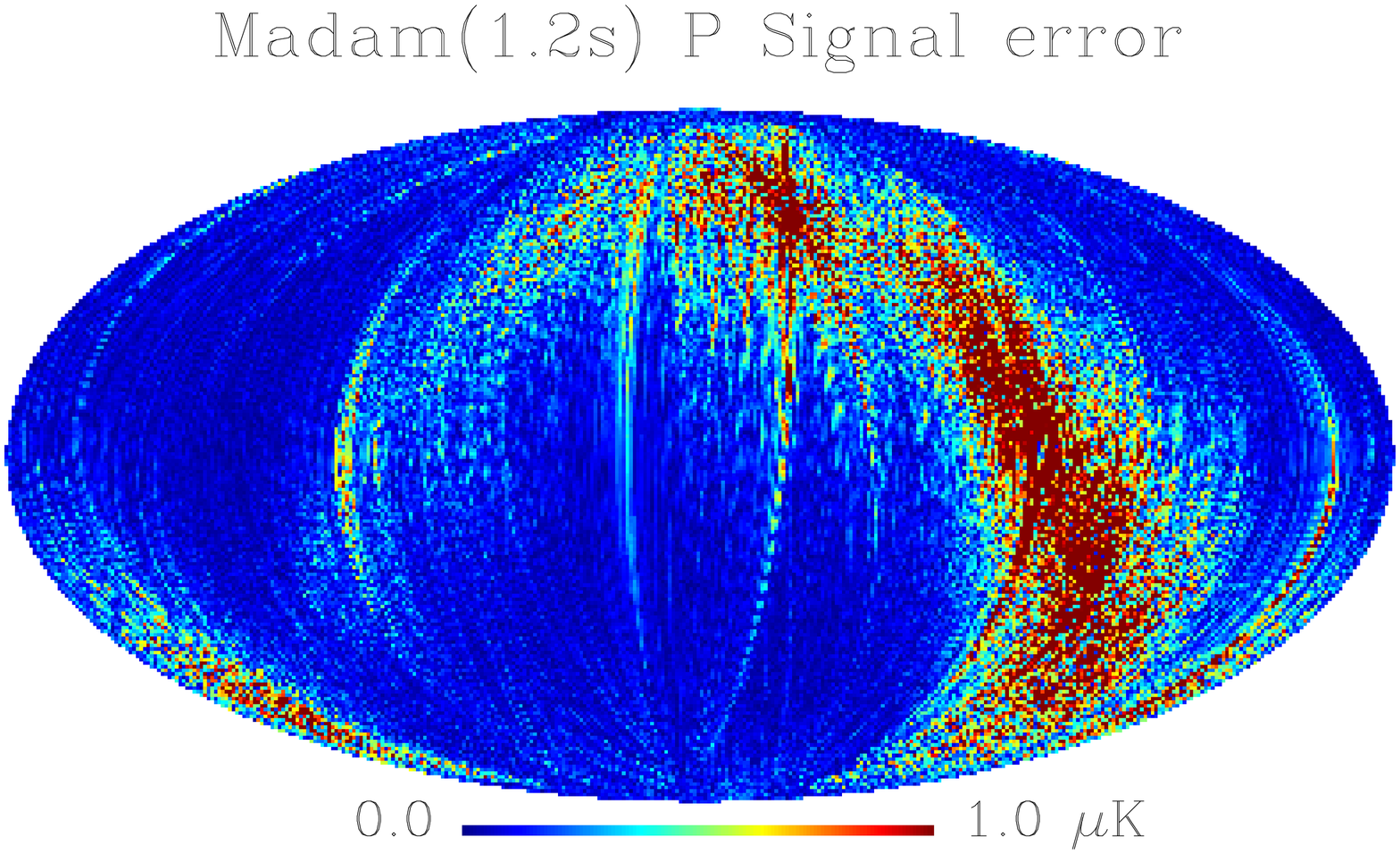}
\includegraphics[scale=0.3,angle=90]{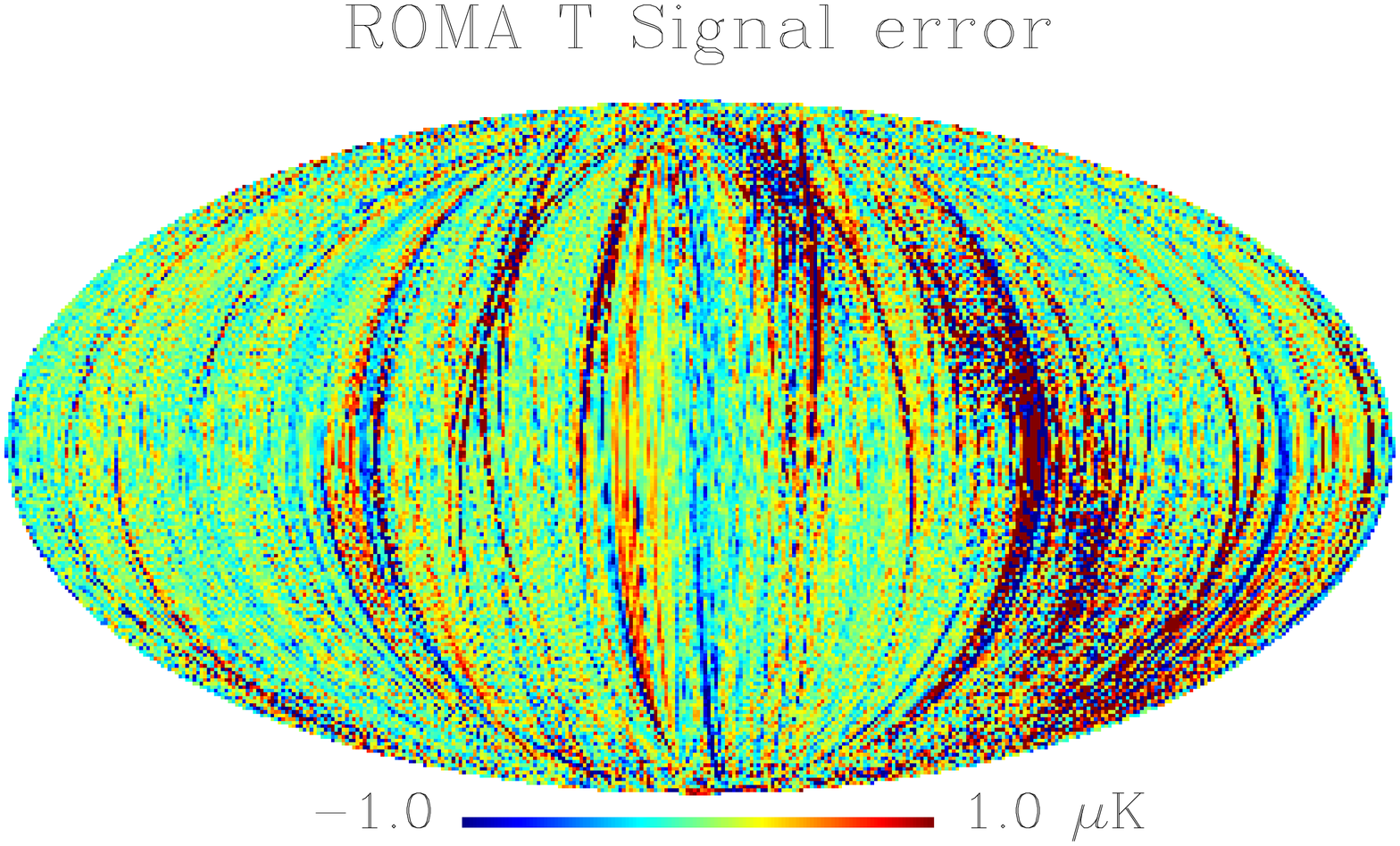}
\includegraphics[scale=0.3,angle=90]{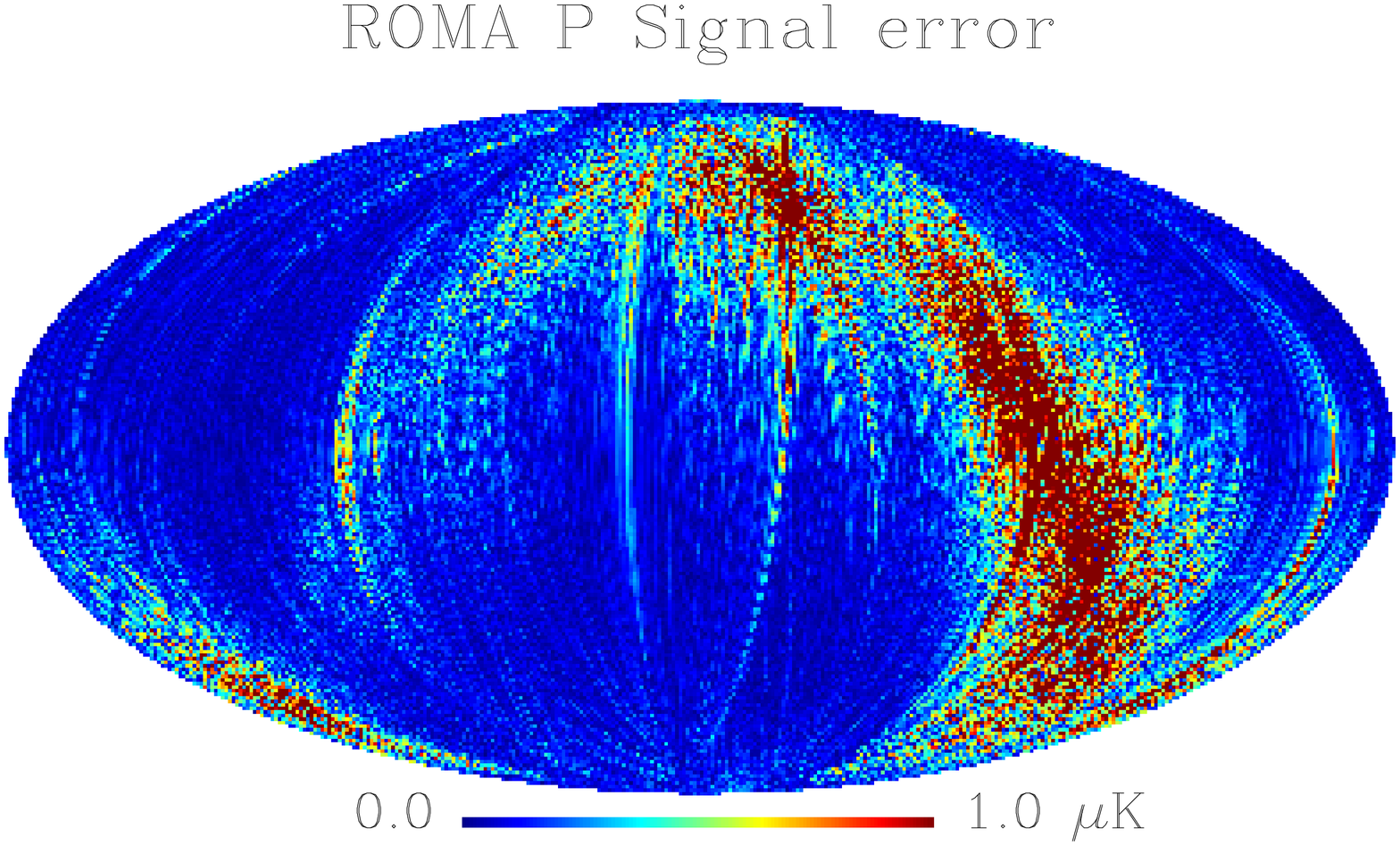}
\includegraphics[scale=0.3,angle=90]{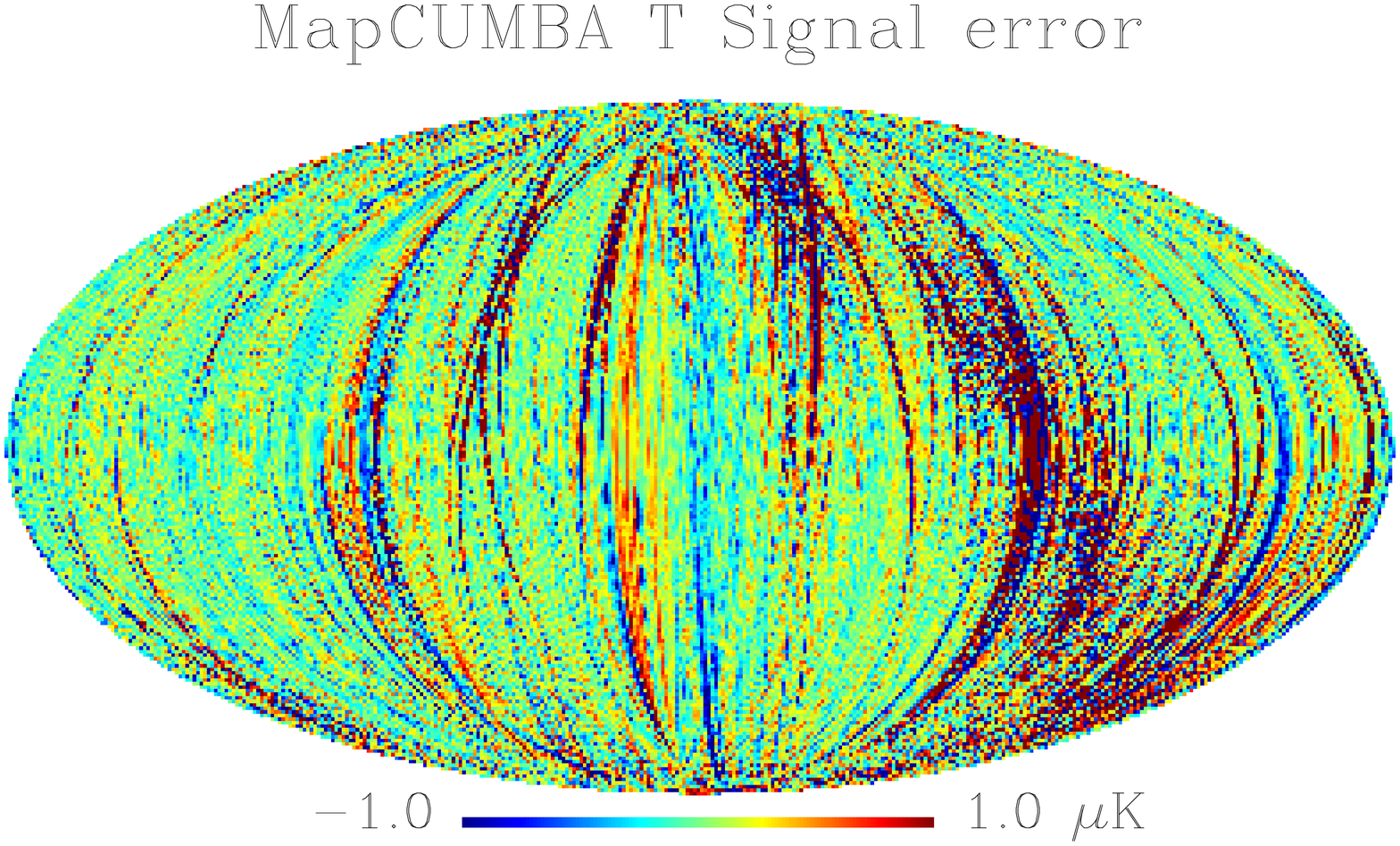}
\includegraphics[scale=0.3,angle=90]{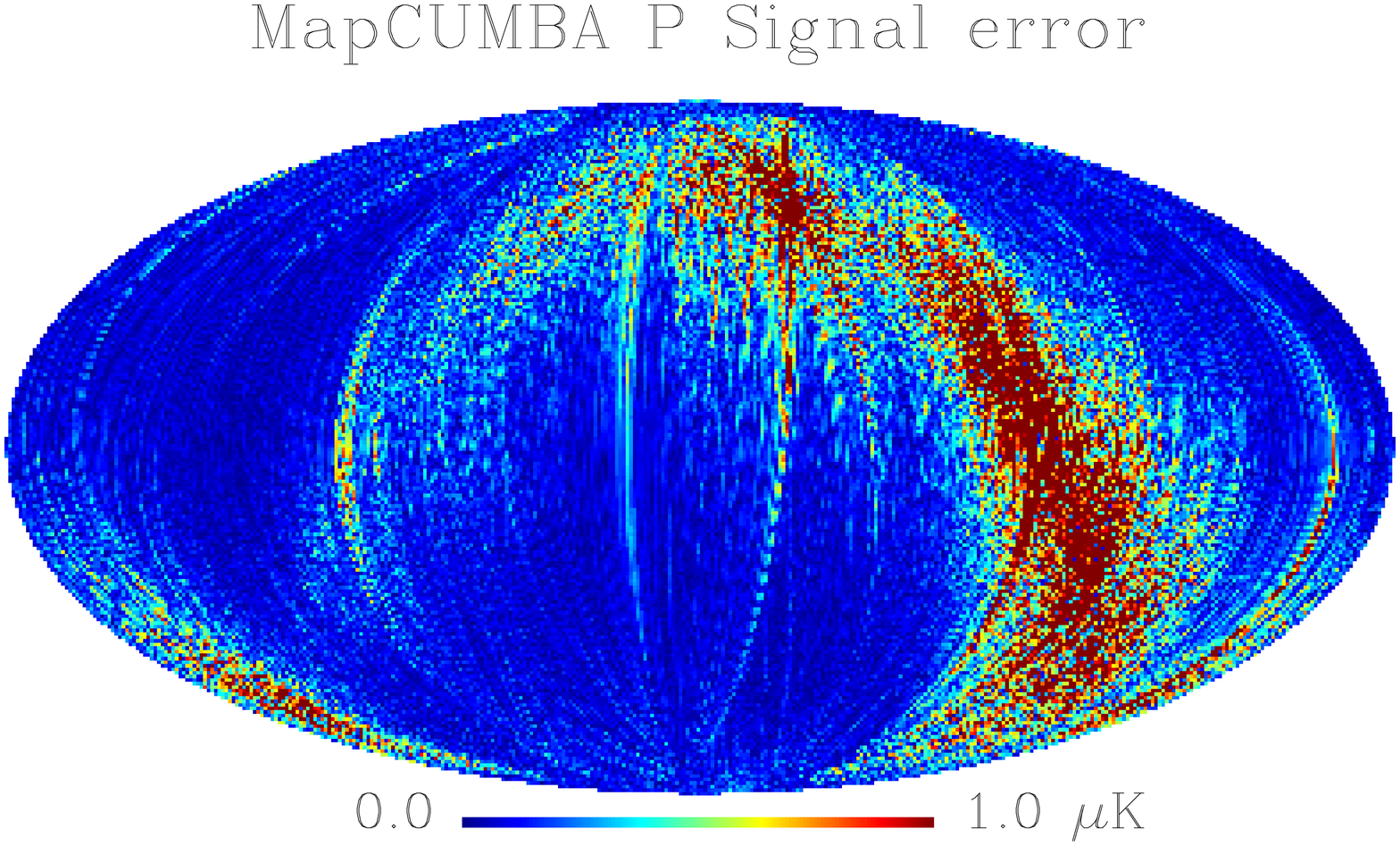}
\end{center}
\caption{Signal error maps of different mapmaking codes. The signal
error is the difference between the noiseless output map and the
binned noiseless map. These maps contain all sky emissions (CMB,
dipole, and foregrounds) and they are in ecliptic coordinates. The
beams were asymmetric and sample integration was off. The left
column is for the temperature; the right column is for polarization
magnitude. The statistics of these maps are given in Table~3. An
example of the effects of beams and sample integration in the signal
error maps is shown in Fig.~\ref{fig:signal_error_madam}.}
\label{fig:signal_error}
\end{figure*}

\begin{figure*} [!tbp]
\begin{center}
\includegraphics[scale=0.3,angle=90]{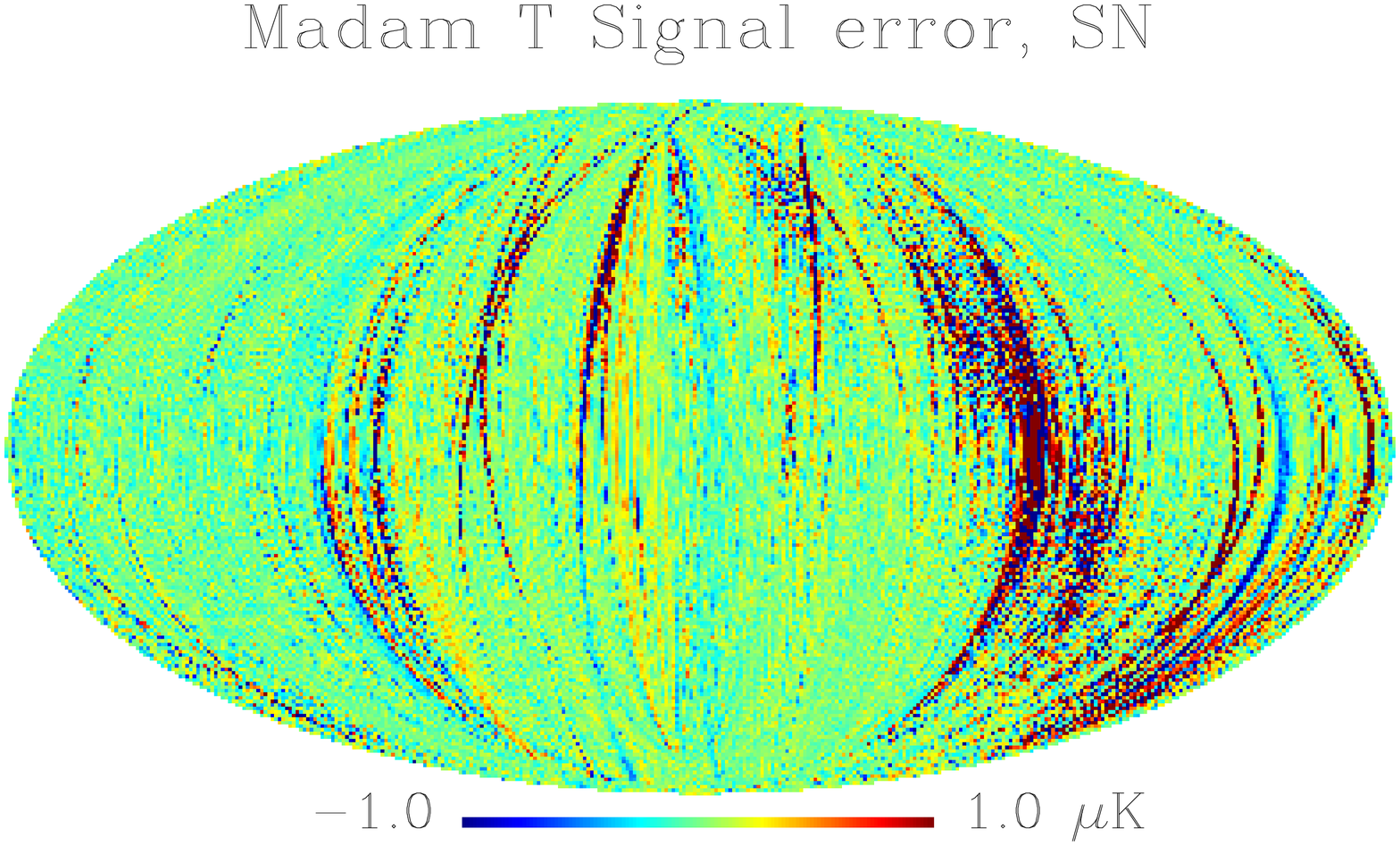}
\includegraphics[scale=0.3,angle=90]{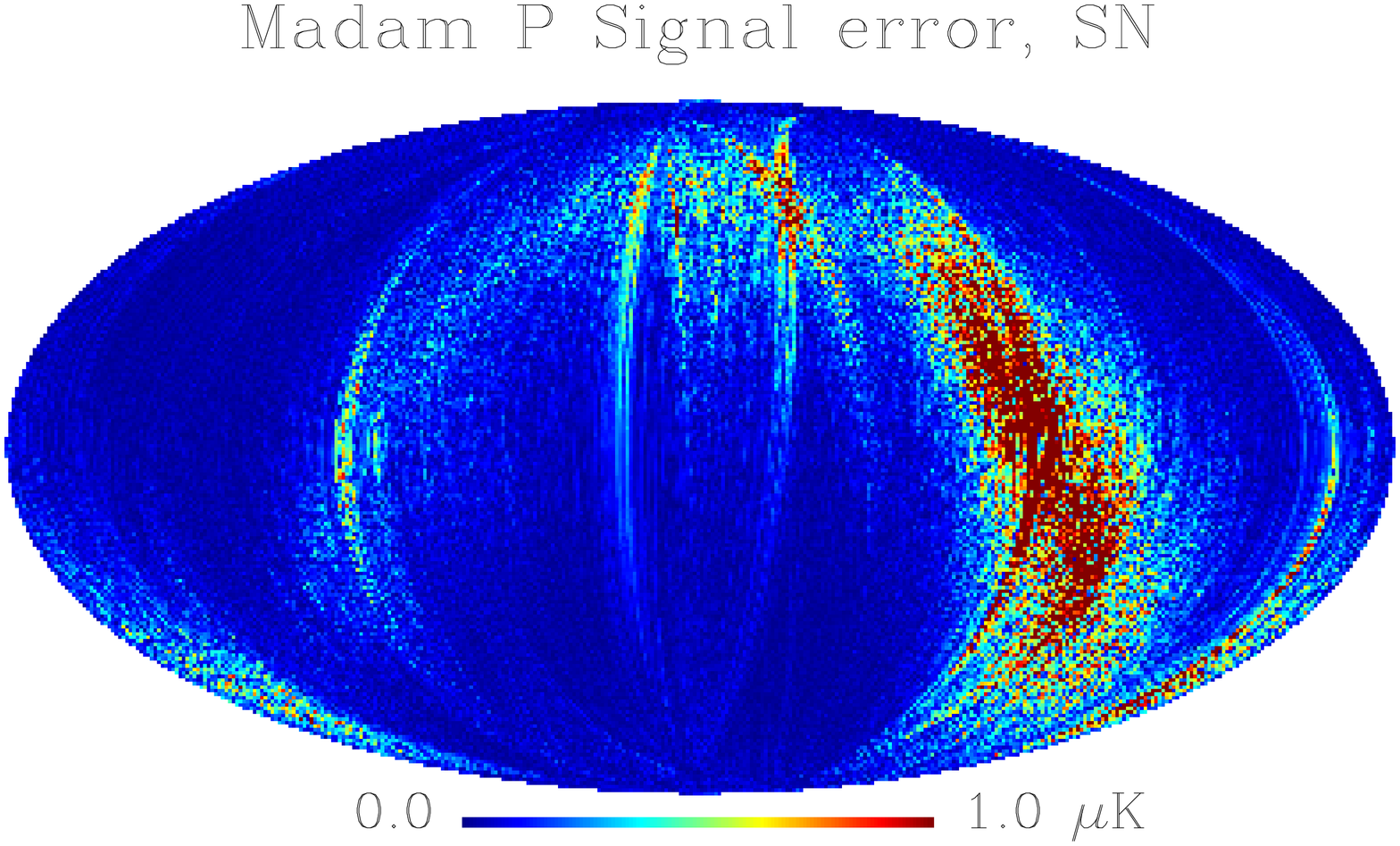}
\includegraphics[scale=0.3,angle=90]{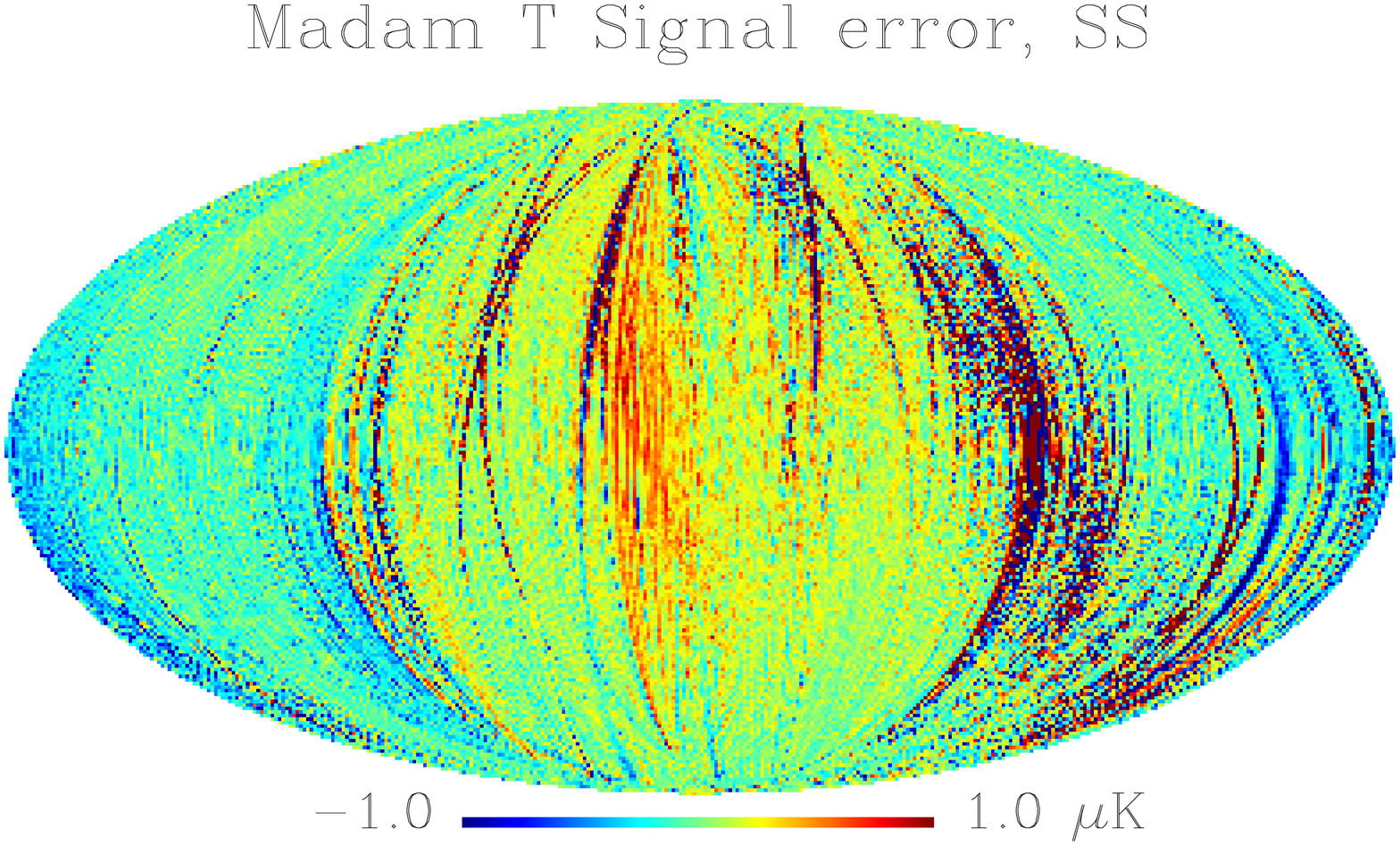}
\includegraphics[scale=0.3,angle=90]{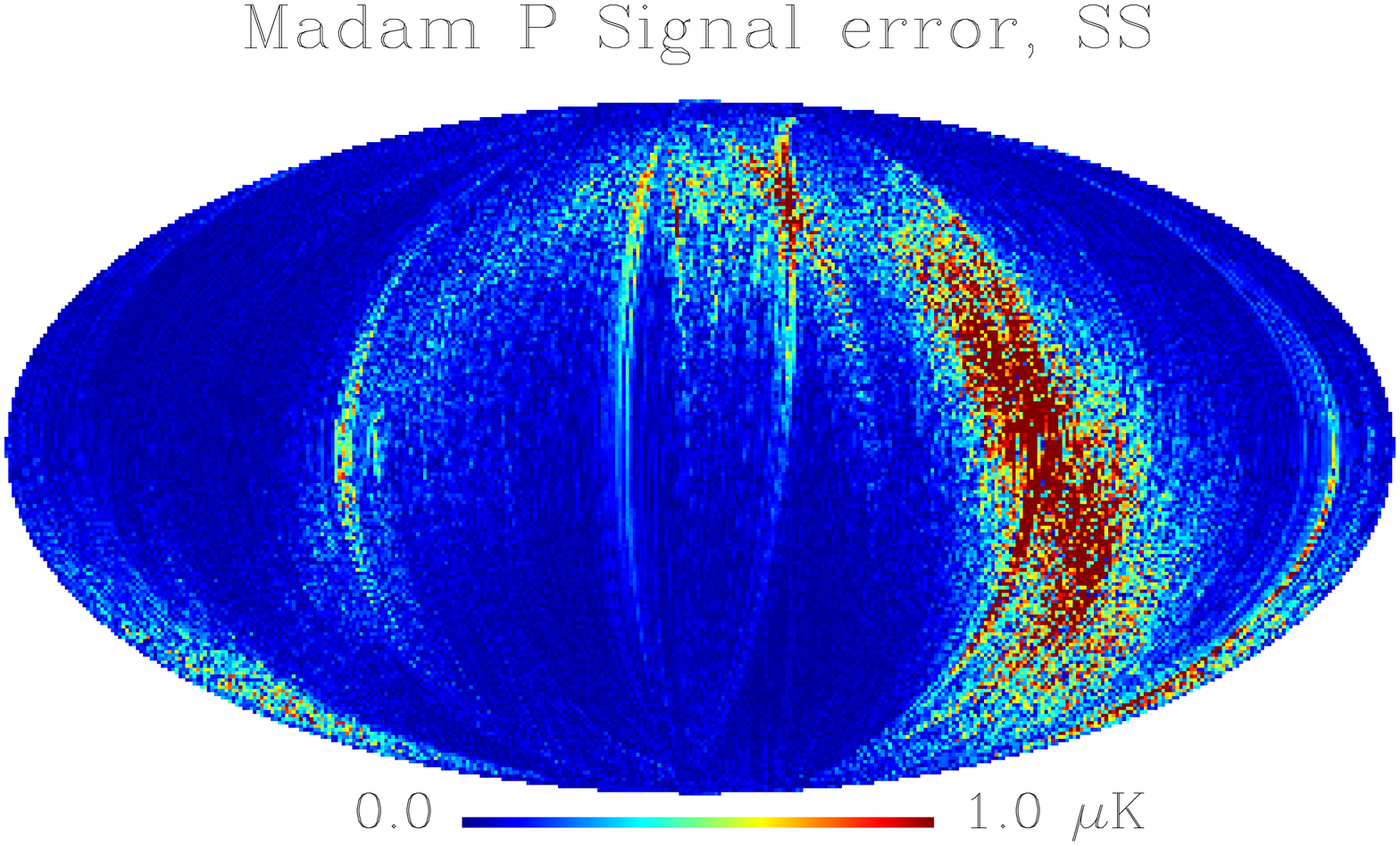}
\includegraphics[scale=0.3,angle=90]{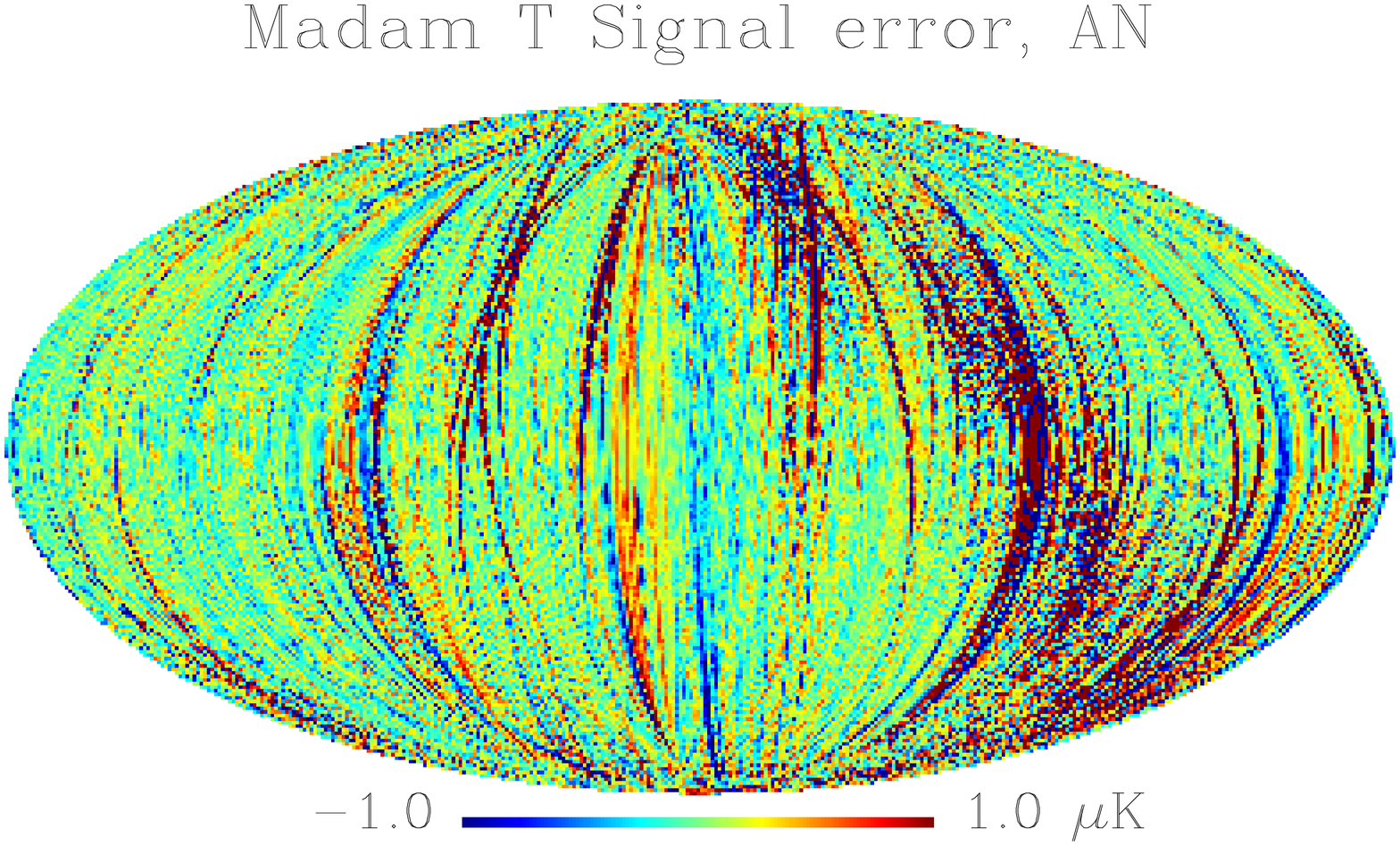}
\includegraphics[scale=0.3,angle=90]{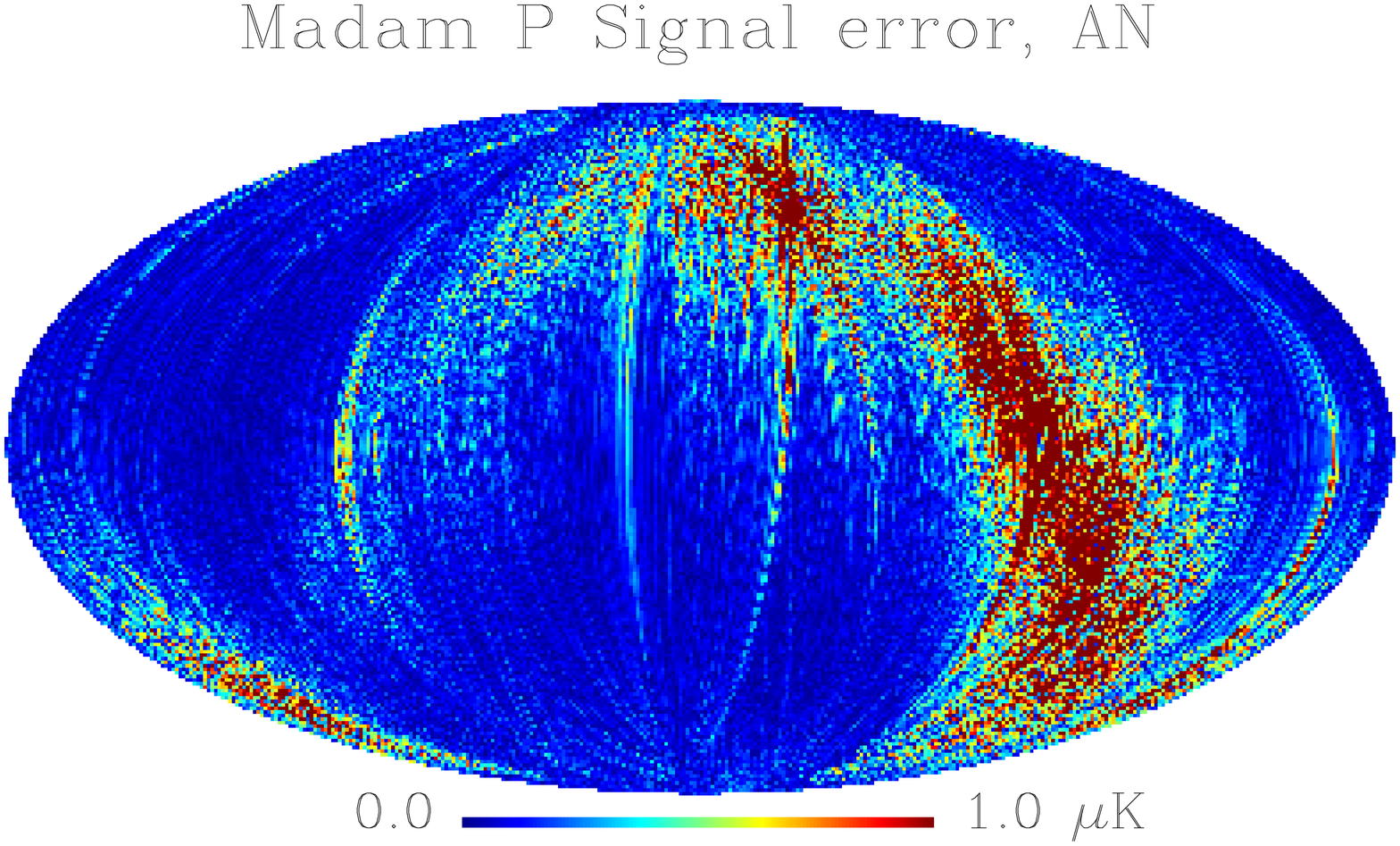}
\includegraphics[scale=0.3,angle=90]{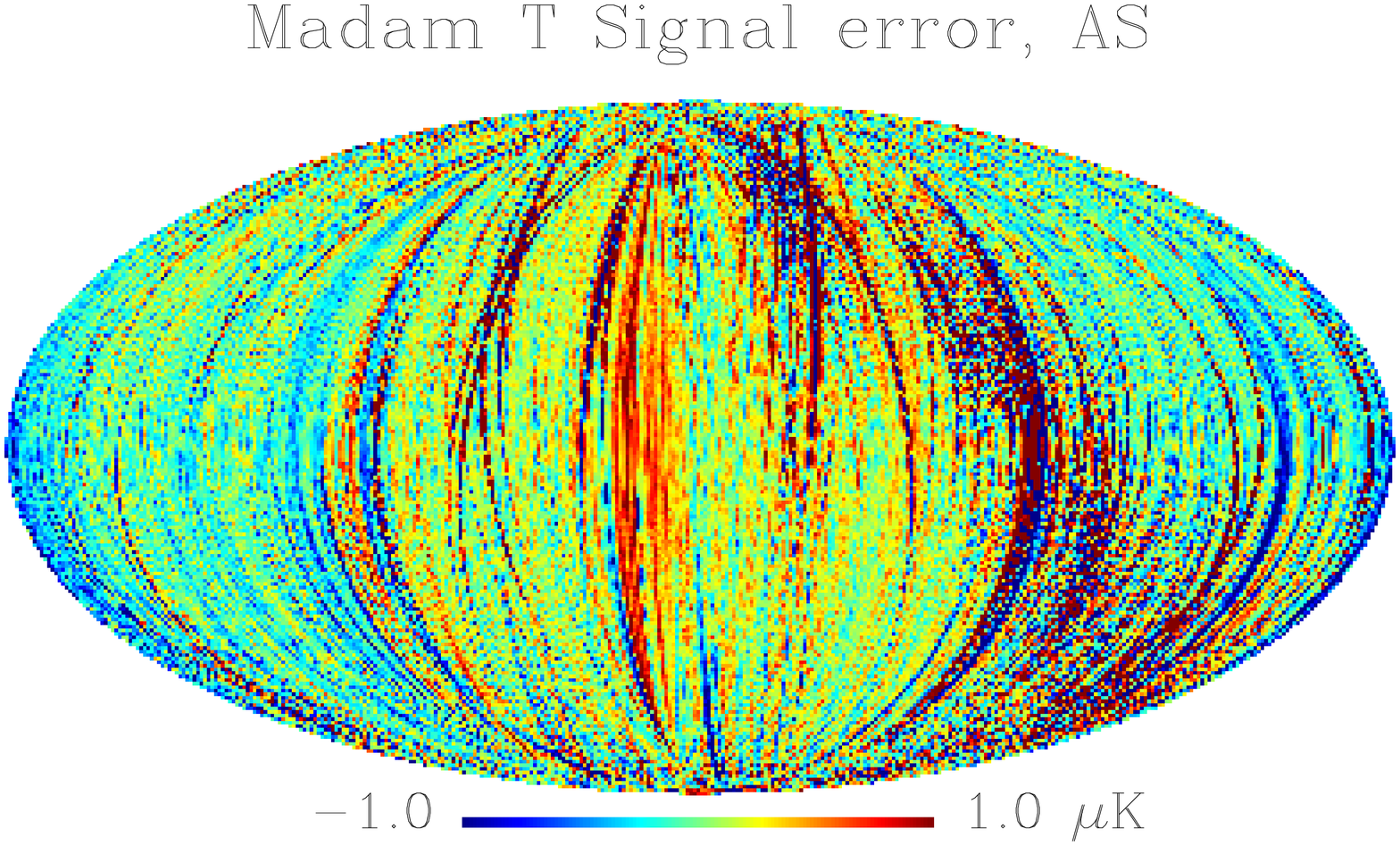}
\includegraphics[scale=0.3,angle=90]{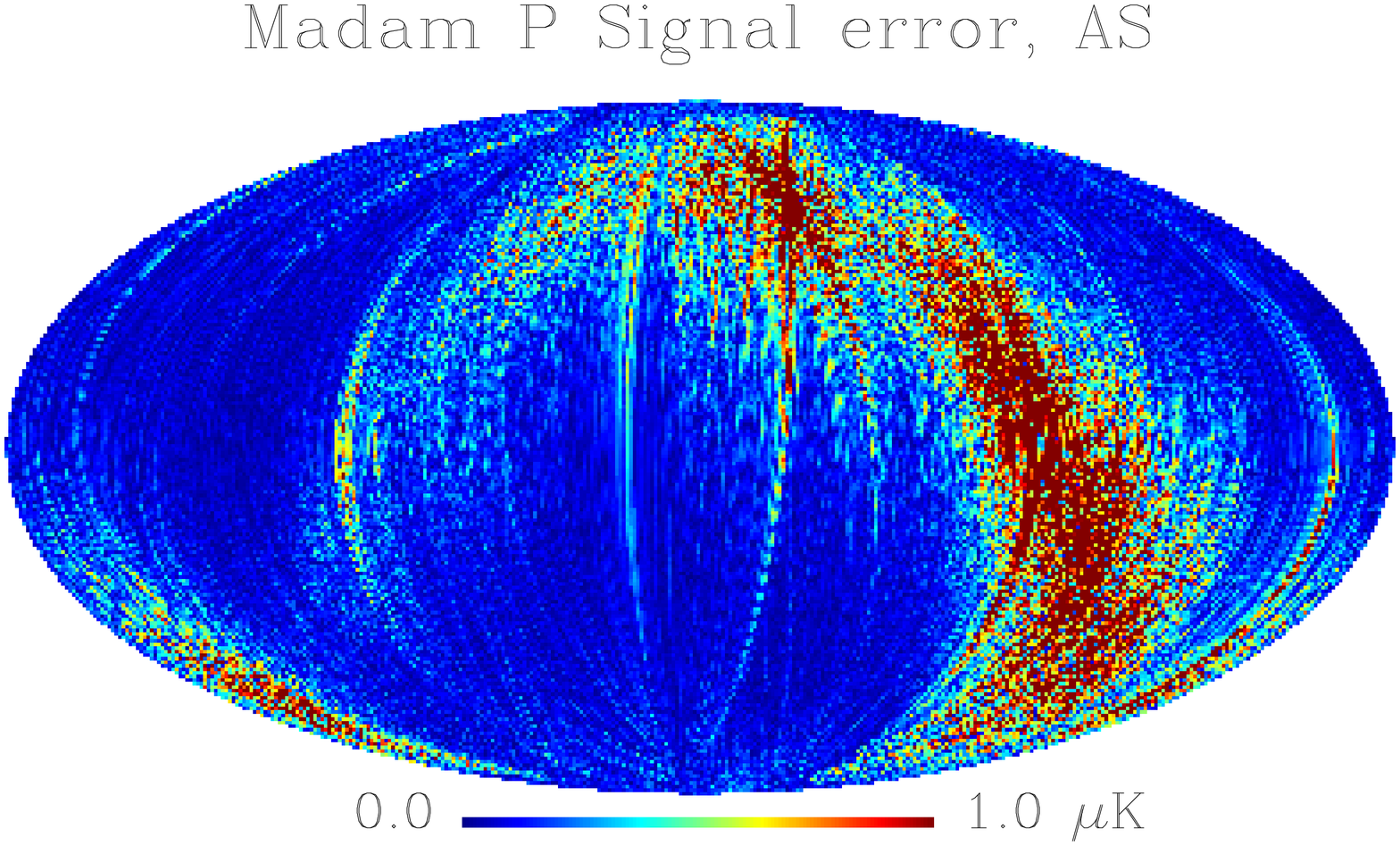}
\end{center}
\caption{Effects of beams and the sample integration in the signal
error maps. We use Madam maps (1.2\,s) here as examples. The maps
contain all sky emissions and they are in ecliptic coordinates. The
left-hand column is for the temperature and the right-hand column is
for the polarization magnitude. The rows from the top correspond to
symmetric beams \& no sampling (SN), symmetric beams \& sampling
(SS), asymmetric beams \& no sampling (AN), and asymmetric beams \&
sampling (AS). The statistics of these maps are given in Table~4.
The third row maps are the same as the second row maps of
Fig.~\ref{fig:signal_error}.} \label{fig:signal_error_madam}
\end{figure*}

\begin{table*}[!tbp]
 \global\advance\tableno by 1 \tabl {\csc Statistics of signal error maps$^{\rm a}$} \par
\setbox\tablebox=\vbox{
   \newdimen\digitwidth
   \setbox0=\hbox{\rm 0}
   \digitwidth=\wd0
   \catcode`*=\active
   \def*{\kern\digitwidth}
   \newdimen\signwidth
   \setbox0=\hbox{+}
   \signwidth=\wd0
   \catcode`!=\active
   \def!{\kern\signwidth}
\halign{\hbox to 2in{#\leaderfil}\tabskip=2em&
     \hfil#\hfil\tabskip=2em&
     \hfil#\hfil\tabskip=0.5em&
     \hfil#\hfil\tabskip=2em&
     \hfil#\hfil\tabskip=0.5em&
     \hfil#\hfil\tabskip=0.5em&
     \hfil#\hfil\tabskip=2em&
     \hfil#\hfil\tabskip=0.5em&
     \hfil#\hfil\tabskip=0.5em&
     \hfil#\hfil\tabskip=0pt\cr
\noalign{\doubleline} \omit&\multispan3\hfil\csc
$I$\hfil&\multispan3\hfil\csc $Q$\hfil& \multispan3\hfil\csc
$U$\hfil\cr \noalign{\vskip -3pt}
\omit&\multispan3\hrulefill&\multispan3\hrulefill&\multispan3\hrulefill\cr
\noalign{\vskip 3pt} \omit\hfil\csc Code
\hfil&MIN&MAX&RMS&MIN&MAX&RMS&MIN&MAX&RMS\cr \noalign{\vskip
3pt\hrule\vskip 5pt} \omit Madam (1 min)$^{\rm b}$ & *$-6.1$ &
*$3.5$ & $0.415$ & $-1.7$ & $1.5$ & $0.137$ & *$-2.4$ & $1.6$ &
$0.181$ \cr \omit Madam (1.2 s)$^{\rm b}$ & $-49.4$ & $20.5$ &
$0.890$ & $-7.8$ & $6.5$ & $0.307$ & $-11.3$ & $7.9$ & $0.421$ \cr
\omit MapCUMBA & $-56.4$ & $21.5$ & $0.910$ & $-8.5$ & $7.0$ &
$0.315$ & $-12.5$ & $8.9$ & $0.430$ \cr \omit ROMA & $-56.4$ &
$21.5$ & $0.910$ & $-8.5$ & $7.0$ & $0.315$ & $-12.5$ & $8.9$ &
$0.430$ \cr \noalign{\vskip 5pt\hrule\vskip 3pt}}}
\enddtable
\tablenote a We show here the statistics of the signal error maps
(given in $\mu$K) of different mapmaking codes. This table is for
the case of asymmetric beams and sampling off. The corresponding
maps are shown in Fig.~\ref{fig:signal_error}. They contain all sky
emissions (CMB, dipole, and foregrounds).
\par
\tablenote b We used Madam in two different configurations: with
1-minute baselines and no noise filter (Madam (1 min)), and with 1.2
s baselines and with noise filter (Madam (1.2 s)).
\par
\label{tab:signal_error_an}
\end{table*}

\begin{table*}[!tbp]
 \global\advance\tableno by 1 \tabl {\csc Statistics of Madam signal error maps$^{\rm a}$} \par
\setbox\tablebox=\vbox{
   \newdimen\digitwidth
   \setbox0=\hbox{\rm 0}
   \digitwidth=\wd0
   \catcode`*=\active
   \def*{\kern\digitwidth}
   \newdimen\signwidth
   \setbox0=\hbox{+}
   \signwidth=\wd0
   \catcode`!=\active
   \def!{\kern\signwidth}
\halign{\hbox to 2in{#\leaderfil}\tabskip=2em&
     \hfil#\hfil\tabskip=2em&
     \hfil#\hfil\tabskip=0.5em&
     \hfil#\hfil\tabskip=2em&
     \hfil#\hfil\tabskip=0.5em&
     \hfil#\hfil\tabskip=0.5em&
     \hfil#\hfil\tabskip=2em&
     \hfil#\hfil\tabskip=0.5em&
     \hfil#\hfil\tabskip=0.5em&
     \hfil#\hfil\tabskip=0pt\cr
\noalign{\doubleline} \omit&\multispan3\hfil\csc
$I$\hfil&\multispan3\hfil\csc $Q$\hfil& \multispan3\hfil\csc
$U$\hfil\cr \noalign{\vskip -3pt}
\omit&\multispan3\hrulefill&\multispan3\hrulefill&\multispan3\hrulefill\cr
\noalign{\vskip 3pt} \omit\hfil\csc TOD
\hfil&MIN&MAX&RMS&MIN&MAX&RMS&MIN&MAX&RMS\cr \noalign{\vskip
3pt\hrule\vskip 5pt} \omit Symm \& no sampling & $-32.9$ & $14.3$ &
$0.578$ & $-7.5$ & $6.1$ & $0.266$ & *$-6.1$ & $5.0$ & $0.267$ \cr
\omit Symm \& sampling & $-30.2$ & $13.1$ & $0.609$ & $-7.3$ & $6.1$
& $0.272$ & *$-5.8$ & $5.0$ & $0.272$ \cr \omit Asymm \& no sampling
& $-49.4$ & $20.5$ & $0.890$ & $-7.8$ & $6.5$ & $0.307$ & $-11.3$ &
$7.9$ & $0.421$ \cr \omit Asymm \& sampling & $-47.5$ & $21.9$ &
$0.936$ & $-7.6$ & $6.5$ & $0.315$ & $-11.7$ & $8.1$ & $0.432$ \cr
\noalign{\vskip 5pt\hrule\vskip 3pt}}}
\enddtable
\tablenote a This table shows the effects that beams and sample
integration have in the signal error maps. We show here the
statistics of Madam (1.2 s) signal error maps (given in $\mu$K). The
corresponding maps are shown in Fig.~\ref{fig:signal_error_madam}.
They contain all sky emissions (CMB, dipole, and foregrounds). The
third line of this table is the same as the second line of Table~3.
\par
\label{tab:signal_error_madam}
\end{table*}

The RMS does not provide a complete comparison of map quality. It is
weighted toward the high-$\ell$ part of the spectrum, where in any
realistic experiment we will be dominated by beam uncertainties and
detector noise. Also, since all the errors in the map are folded
into a single number, it tends to obscure the origin of the errors.
Fig.~\ref{fig:cnrdiff} shows typical angular power spectra of the
residual maps. In this plot the Madam (in this case with 1-minute
baselines and without noise filter) and MapCUMBA residual spectra
are nearly the same. The corresponding spectra of the other
mapmaking codes would fall close to them too.  To highlight the
differences we show the spectra of three difference maps between
pairs of residual maps. Green curves show that the difference
between Madam (with short baselines) and optimal residual maps is
small. Comparison of the blue and red curves shows that destriping
with 1-hour baselines (Springtide) produces RMS residuals that are
larger than those from the optimal codes, and larger than those from
destriping with 1-minute baselines (Madam with 1-minute baselines);
the differences are confined to the high-$\ell$ part of the
spectrum. At low $\ell$, the codes perform almost identically.

To examine the signal errors of our mapmaking codes we made
noiseless maps from the TOD of CMB, dipole, and foreground
emissions. We subtracted the corresponding binned noiseless maps
from noiseless output maps and obtained the signal error maps.  We
show some of our signal error maps in Figs.~\ref{fig:signal_error}
and \ref{fig:signal_error_madam} and their statistics in Tables~3
and 4. In Fig.~\ref{fig:signal_error} we show signal error maps of a
number of mapmaking methods.  These maps were made with asymmetric
beams, with sampling off. Their statistics are given in Table~3.
Optimal codes and Madam with short baselines (1.2~s) produce nearly
the same signal error, which is stronger than the signal error of
destripers with long baselines (represented here by another Madam
map with 1-minute baselines and no noise filter this time). For
optimal and Madam (1.2\,s) maps the signal error is more localized
to the vicinity of the galaxy (which has strong signal gradients)
than for long-baseline destriper maps. Fig.~\ref{fig:cnrdiff} shows
that in the high-resolution 30\,GHz maps the signal error is a small
effect compared to the residual noise or the CMB signal that we used
in this study. These results are well in line with the results of
our earlier studies (Poutanen et al.~\cite{Pou06}, Ashdown et
al.~\cite{Ash07b}). In the bottom panel of Fig.~\ref{fig:cnrdiff} we
compare the Madam signal error with the theoretical CMB $B$-mode
spectrum (10 \% tensor-to-scalar ratio). The plot shows that the
magnitude of the signal error is comparable to the magnitude of this
$B$-mode signal. Signal error can therefore limit our possibilities
to detect it. In a previous study we examined a number of techniques
to decrease the signal error (Ashdown et al.~\cite{Ash07b}). Because
the detection of the $B$-mode signal was not a goal of this paper we
did not investigate these methods for this data.

Figure~\ref{fig:signal_error_madam} and Table~4 show the effects of
beams and sample integration on the signal error maps.  We use the
Madam maps here as examples. The figure and the table show that both
switching on the sample integration and switching from symmetric to
asymmetric beams increase the signal error. This is because with
asymmetric beams also the beam orientation affects the measured
signal.

\begin{figure*}[!tbp]
\begin{center}
\includegraphics[scale=0.4]{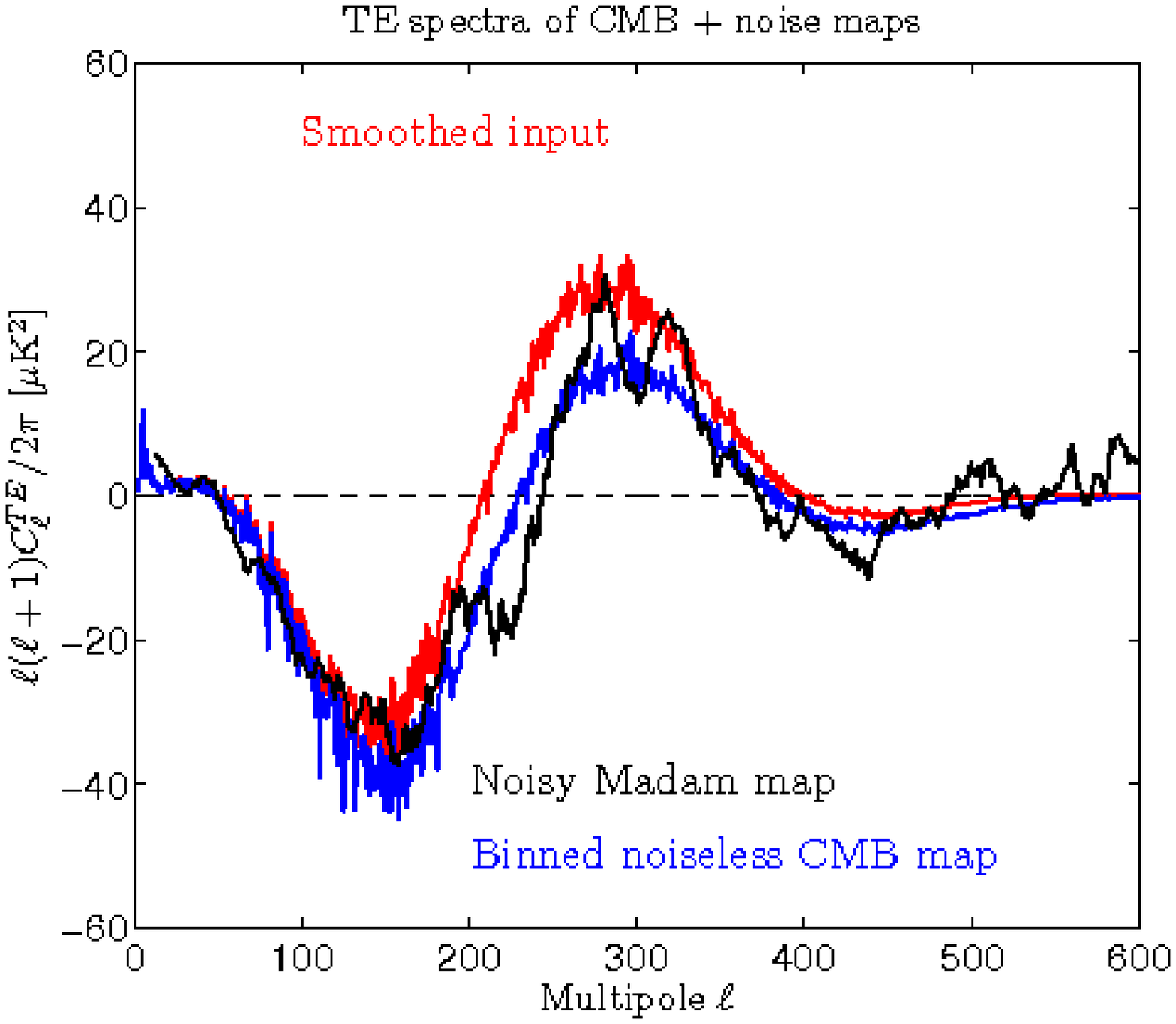}
\includegraphics[scale=0.4]{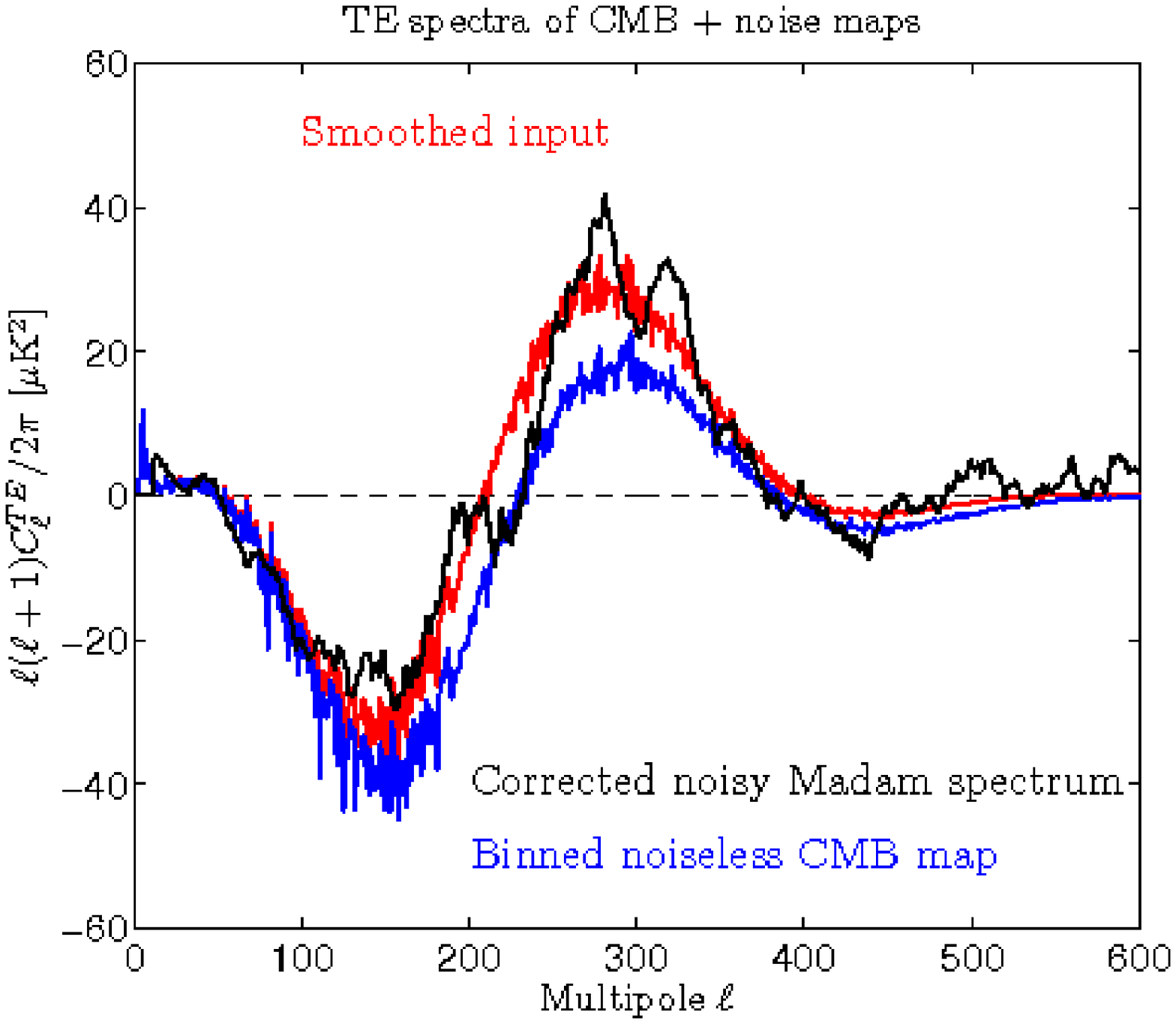}
\end{center}
\caption{Effects of beam mismatch on the noisy TE spectra.
\textit{Left hand panel:} TE spectrum of a noisy CMB map (Madam map
with 1~minute baselines and no noise filter), black curve. To reduce
its $\ell$ to $\ell$ variation, it was filtered with a sliding
average ($\Delta \ell = 20$). TE spectra of the binned noiseless CMB
map (blue curve) and the smoothed input map (red curve) are shown
too. They are the same curves as in the right-hand panel of
Fig.~\ref{fig:te_1}. The maps were made with asymmetric beams, and
sample integration was off.  \textit{Right hand panel:} Noisy Madam
TE spectrum (black curve), where the effects of beam mismatch have
been corrected using the analytical method that we developed in
Appendix.~\ref{sec:model}.} \label{fig:te_3}
\end{figure*}

Finally, we turn again to the issue of asymmetric beams, now in the
presence of CMB and detector noise.  Noise dominates the EE and BB
spectra of our observed 30\,GHz CMB maps. The TE spectrum, however,
does not have a noise bias. Therefore, the effect of the
$T\rightarrow P$ cross-coupling can be detected in the noisy TE
spectrum. We made a Madam map (with 1-minute baselines and no noise
filter) from the TOD of CMB+noise (asymmetric beams and no sample
integration). In the TE spectrum of this map, the bias due to
$T\rightarrow P$ (arising from the beam mismatch) is clearly visible
(see Fig.~\ref{fig:te_3}). We can expect that such a large bias will
lead to errors in the cosmological studies (e.g., in cosmological
parameters), if not corrected. Fig.~\ref{fig:te_3} shows that our
analytical correction method developed in Appendix~\ref{sec:model}
is able to restore the spectrum, at least on medium and large
angular scales (at $\ell \lesssim 450$).

\subsection{Cooler fluctuations} \label{subsec:cooler}

Temperature fluctuations of the {\sc Planck} sorption cooler were
described in Sect.~\ref{subsubsec:coolnoise}. These fluctuations
have an effect in the output signals (TOD) of the LFI 30\,GHz
radiometers. The typical cooler TOD waveform was shown in
Fig.~\ref{fig:cooler_tod}. The RMS of this cooler signal is
$\backsim$35~$\mu$K, which is about 1/38 of the RMS of the random
uncorrelated instrument noise (white noise).

\begin{figure} [!tbp]
\begin{center}
\resizebox{\hsize}{!}{\includegraphics[angle=90]{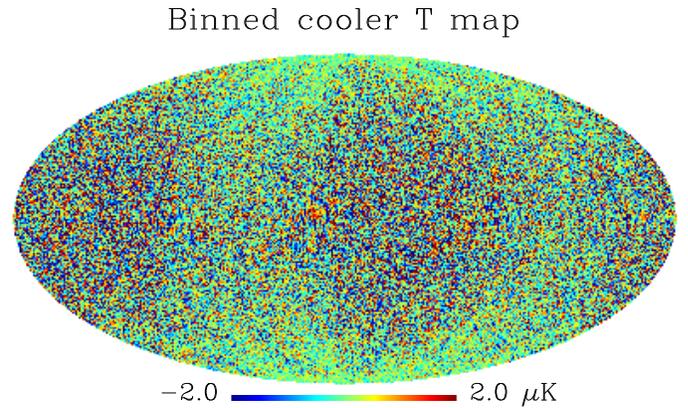}}
\end{center}
\caption{Temperature map binned from the cooler noise \hbox{TOD}.
The map is in ecliptic coordinates and it contains observations of
all four LFI 30\,GHz detectors. Its statistics are (MAX, MIN, RMS) =
(10.9, -$11.7$, 1.23)\,$\mu$K. We do not show the corresponding $Q$
and $U$ maps, but their RMSs were 0.002 and 0.007\,$\mu$K,
respectively.} \label{fig:cooler_map}
\end{figure}

It is of interest to bin the one-year cooler TODs of all four
30\,GHz detectors (identical TODs as described in
Sect.~\ref{subsubsec:coolnoise}) in a map, shown in
Fig.~\ref{fig:cooler_map}. It looks similar to a map of correlated
random noise (with faint stripes along the scan paths). Nutation of
the satellite spin axis and the fluctuation of its spin rate (see
Sect.~\ref{subsec:scanning}) randomize the regular cooler TOD signal
when we project it in the sky. Therefore all map structures that
these regularities could produce are washed out. Because a pair of
detectors sharing a horn see the same cooler signal, we might expect
no cooler effect in the polarization maps. This is not, however, the
case in reality.  A small polarization signal arises because the
polarization axes of the detectors are not exactly orthogonal within
a horn (see Table~1).

\begin{figure*}
\begin{center}
\includegraphics[scale=0.4]{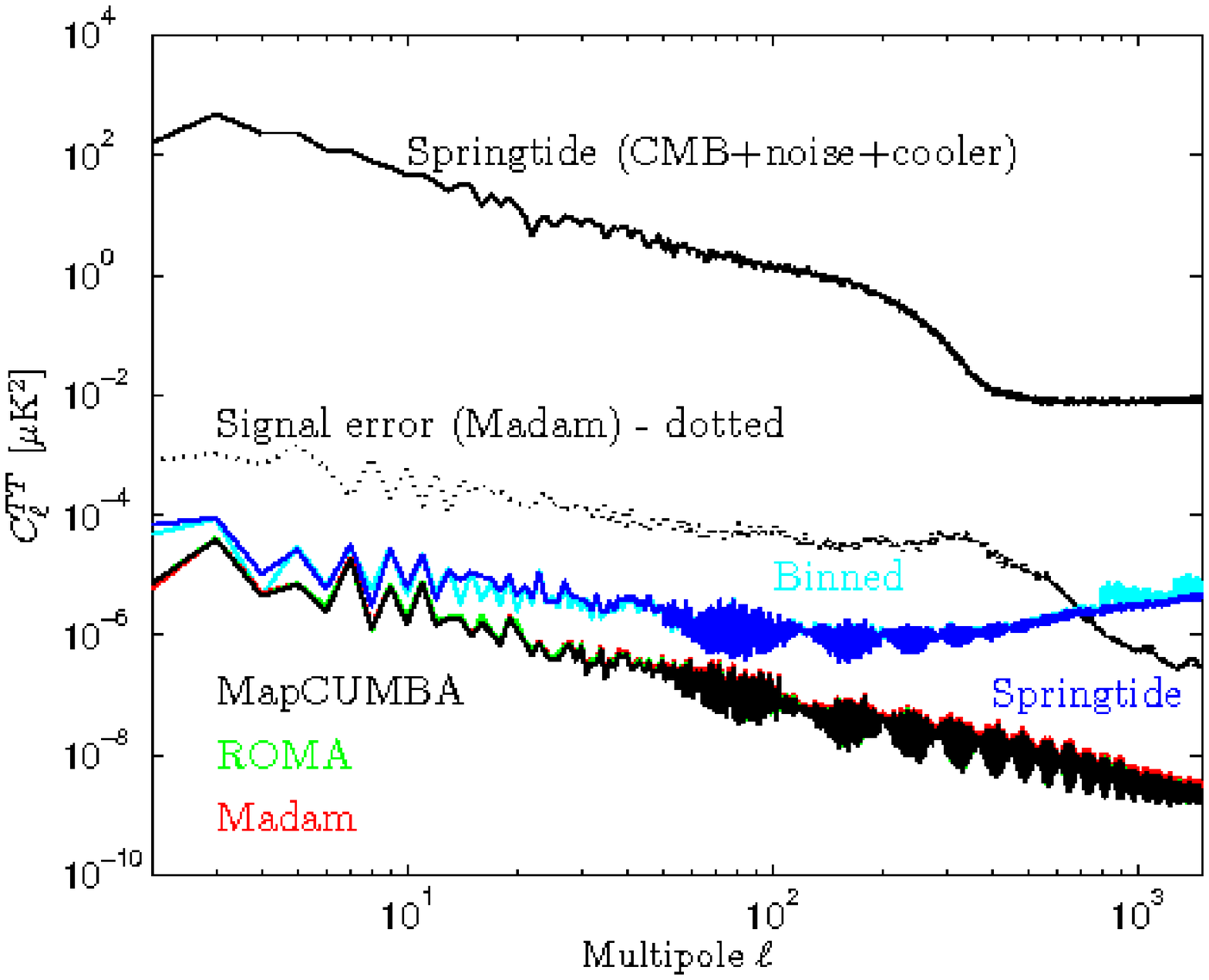}
\includegraphics[scale=0.4]{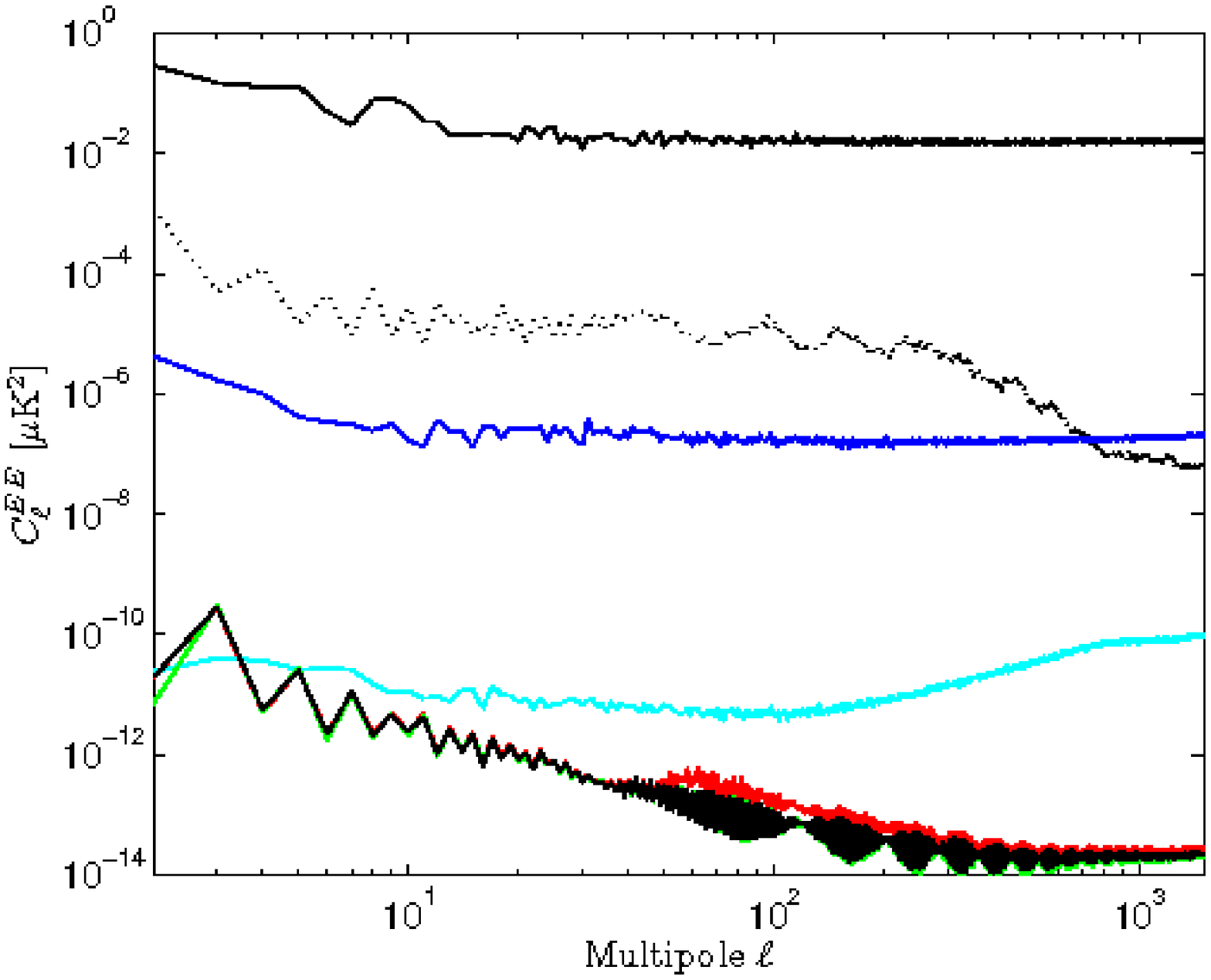}
\includegraphics[scale=0.4]{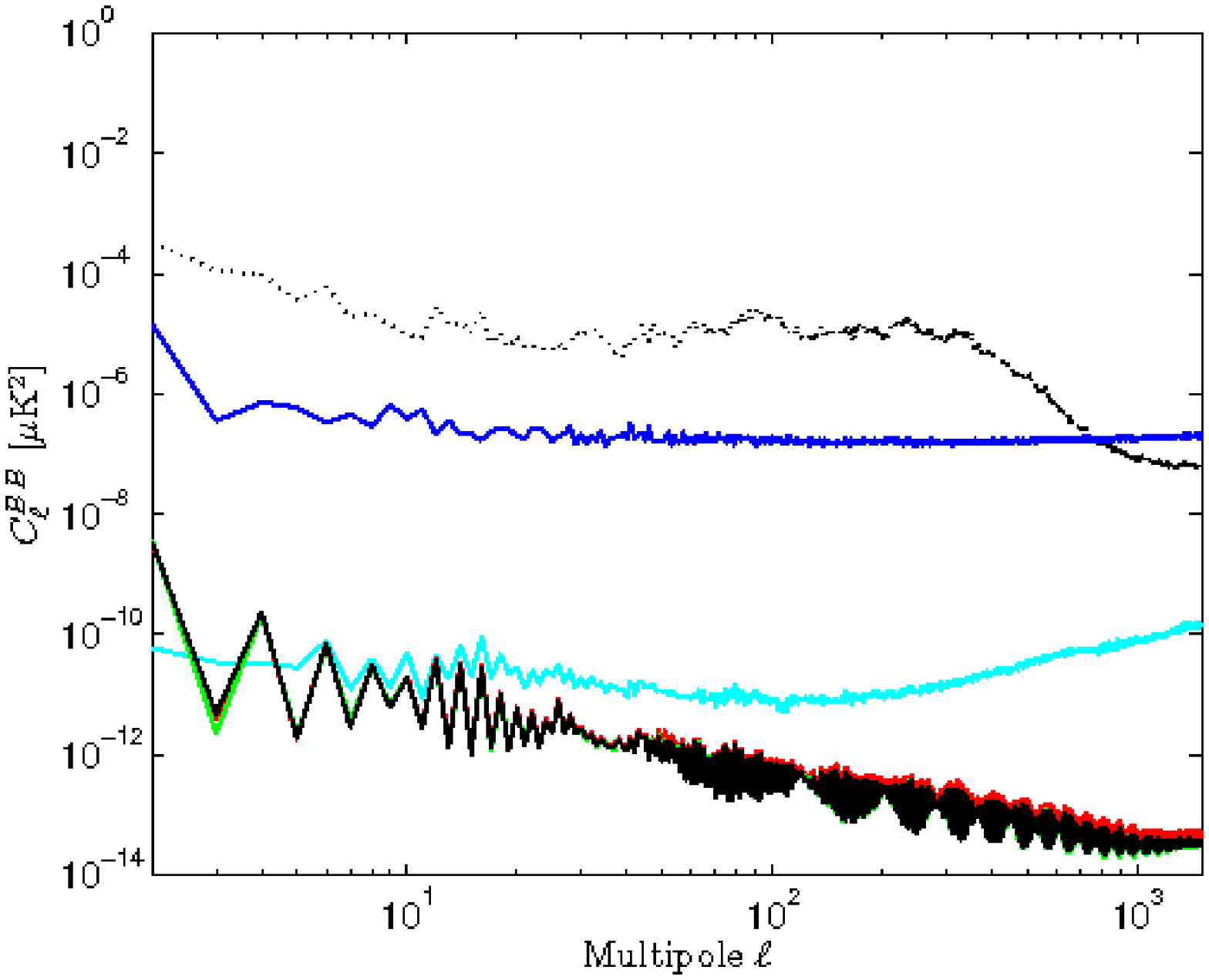}
\end{center}
\caption{Effect of sorption cooler temperature fluctuations on the
CMB maps. These plots show the TT, EE, and BB angular power spectra
of residual maps of the cooler effect. The residual maps were
computed as differences between the maps of CMB+noise+cooler and
CMB+noise. They represent one year of observations of four LFI
30\,GHz detectors. The beams were asymmetric and sample integration
was on. The light blue curves (``Binned'') show the spectrum of the
binned cooler map, whose temperature component is in
Fig.~\ref{fig:cooler_map}. Note that the MapCUMBA, ROMA, and Madam
curves are nearly on top of each other. For comparison we also show
the spectrum of the Springtide CMB+noise+cooler map and the spectrum
of the Madam signal error map (including all sky emissions,
asymmetric beams, and sample integration, dotted curve, same curve
as in Fig.~\ref{fig:cnrdiff}).} \label{fig:coolerspectra}
\end{figure*}

We made maps of CMB+noise+cooler and CMB+noise. We computed their
difference to see the residuals of the cooler fluctuations that our
mapmaking codes leave in their output maps. We computed the angular
power spectra of these residual maps and show them in
Fig.~\ref{fig:coolerspectra}. Except for Springtide at small angular
scales the residuals of the cooler fluctuations are smaller than
signal error. We can therefore conclude, that, in these simulations,
the cooler effect is a tiny signal compared to the CMB itself, or to
random instrument noise.

The period of the cooler signal is $\backsim$12~min
(Fig.~\ref{fig:cooler_tod}). Springtide with its long (1~hr)
baselines is not able to fit out any cooler power from the observed
\hbox{TOD}. Therefore the cooler signal is not suppressed in the
Springtide temperature map, but remains in the same level as in the
binned cooler map. Springtide uses noise estimates that it computes
from its ring maps. Although the TODs of this study have the same
noise spectra the white noise levels that Springtide estimates for
the rings will be different. The rings are combined using inverse
noise variance weighting, and so when rings from two detectors in
the same horn (which see the same cooler signal) are not weighted
exactly the same, the effect in polarization is enhanced over the
binned case which is generated using equal weights. We expect that
Springtide would deliver EE and BB cooler spectra that are similar
to the binned case if equal weighting would be used.

Madam and the optimal codes, which operate with short baselines
(optimal codes have one-sample-long effective baselines), are able
to set their baselines to track well the cooler signal and therefore
the cooler signal is suppressed in their output maps. Our optimal
codes and Madam assumed the same noise spectrum for all TODs. None
of them tried to include the cooler spectrum in their noise filters.

\subsection{Pointing Errors}\label{sec:pointerr}

We modified the Madam destriper in order to quantify how inaccuracy
in the knowledge of satellite pointing affects the maps. To isolate
this from other effects, we considered first the symmetric-beam,
no-sampling case. However, to see the effect on polarization
leakage, we needed a reference TOD which contained Stokes I only.
This was available only for the asymmetric-beam no-sampling case
(see Sec.~\ref{subsubsec:beam_mismatch}).

We did not have available a good model for the expected pointing
error, but to quantify the magnitude of the effect we considered two
opposite schemes: \emph{uncorrelated} pointing error with a random
offset added to the pointing of each detector sample and
\emph{strongly correlated} error for which the offset was kept
constant for an hour at a time. The first model is of course
unrealistic, but serves to indicate to which direction the effects
may change when the offset varies over shorter time scales.

The pointing errors were drawn from a Gaussian distribution so that
the RMS error for both the $\theta$ and $\phi$ directions was
$\sigma = 0\farcm5$. This translates into RMS offset of
$\sqrt{2}\sigma$.

The errors were generated independently for each horn; but the two
detectors (for the two polarization directions) in the same horn
share the same pointing and thus the same pointing error.  Thus to a
first approximation we would not expect the pointing error to
contribute to the temperature-to-polarization leakage. However,
since the two polarization directions were not aligned at exactly
$90^\circ$ to each other, there will be some effect. The different
pointing errors of the two horns should contribute to $E \rightarrow
B$ leakage.

The $\sigma = 0\farcm5$ pointing error led to misplacing 11.5\% of
the samples into wrong pixels ($N_\mathrm{side}=512$ pixels, about
$7 \arcmin$ across).

In the limit of an infinitely dense, homogeneous hit distribution, a
simple model for the pointing errors can be devised.  The errors
increase the effective size of the pixels, leading to extra
smoothing of the output map, similar to smoothing with an axially
symmetric Gaussian beam with FWHM $\sqrt{8\ln{2}}\sigma$. In the
power spectrum level this means an extra suppression factor of
$\exp{\left(-\sigma^2\ell(\ell+1)\right)}$.

We defined earlier the residual map as the difference between the
output map and the binned noiseless map. Since pointing errors
affect also the binned noiseless map, it is not enough to consider
the residual map. Therefore we consider here the {\it total error}
defined as the difference between the output map and the smoothed
input map.

Actual measurements typically include inhomogeneous distributions of
hits per pixel and within a pixel. Addition of random pointing
errors can artificially smooth this distribution, redistributing
noise more evenly among the pixels.  The expected white noise
contribution to the RMS total error is proportional to
$\sqrt{\langle 1/N_\mathrm{obs} \rangle}$.  Random pointing errors
may decrease $\langle 1/N_\mathrm{obs} \rangle$. Thus random
pointing errors may actually reduce the RMS total error in the map.
However this does not happen with correlated pointing errors.

Isolated from the redistribution of white noise are the effects on
the actual signal. We divide the signal contribution to the total
error into two parts: 1)~the pixelisation error from uneven sampling
of the pixel, defined as the difference (binned noiseless map $-$
smoothed input map); and 2)~signal error from destriping, defined as
the difference (destriped noiseless map $-$ binned noiseless map).
Pointing error affects both differences. The effects of the pointing
errors on these different error components are shown in Table~5 for
temperature maps. Interestingly, uncorrelated pointing errors also
reduced the pixelization error; although the pixel was now sampled
from a larger area, apparently the sampling was now more uniform.

The total effect of the $\sigma = 0\farcm5$ pointing error was
relatively small. In the uncorrelated case, the RMS total error in
the map decreased from 43.25\,$\mu$K to 43.21\,$\mu$\hbox{K}. In the
correlated case it increased to 43.30\,$\mu$\hbox{K}. These effects
are smaller than the difference between the optimal codes and the
destriper codes with long baselines (i.e., Madam without noise
filter, Springtide).

\begin{table*}[!tbp]
 \global\advance\tableno by 1 \tabl {\csc Effect of
satellite pointing errors on total temperature error maps$^{\rm a}$}
\par \setbox\tablebox=\vbox{
   \newdimen\digitwidth
   \setbox0=\hbox{\rm 0}
   \digitwidth=\wd0
   \catcode`*=\active
   \def*{\kern\digitwidth}
   \newdimen\signwidth
   \setbox0=\hbox{+}
   \signwidth=\wd0
   \catcode`!=\active
   \def!{\kern\signwidth}
\halign{\hbox to 2in{#\leaderfil}\tabskip=2em&
     \hfil#\hfil\tabskip=2em&
     \hfil#\hfil\tabskip=0.5em&
     \hfil#\hfil\tabskip=2em&
     \hfil#\hfil\tabskip=0.5em&
     \hfil#\hfil\tabskip=0.5em&
     \hfil#\hfil\tabskip=0pt\cr
\noalign{\doubleline} \omit&&&&\multispan3\hfil\csc Error
components\hfil\cr \noalign{\vskip -3pt}
\omit&&&&\multispan3\hrulefill\cr \noalign{\vskip 3pt}
\omit\hfil$\sigma$\hfil&Model&Total error&White noise
estimate&Residual noise$^{\rm b}$&Signal error$^{\rm
c}$&Pixelization error$^{\rm d}$\cr \noalign{\vskip 3pt\hrule\vskip
5pt}
\omit 0'   &          & *$43.2560$ & *$42.5449$ & *$42.6536$ &
**$0.5601$ & **$7.1663$ \cr
\omit 0'.5 & *uncorr. & *$-0.0432$ & *$-0.0279$ & *$-0.0318$ &
*$+0.0023$ & *$-0.0611$ \cr
\omit 0'.5 &  *corr.  & *$+0.0482$ & *$+0.0222$ & *$+0.0166$ &
*$+0.0383$ & *$+0.2047$ \cr
\noalign{\vskip 5pt\hrule\vskip 3pt}}}
\enddtable
\tablenote a We show here how the total RMS error (given in $\mu$K)
in the map changed by adding pointing errors. This table is for the
case of symmetric beams. The first row shows the case without
pointing errors. The two other rows show how the map errors {\em
changed} when adding pointing errors according to the two models
considered in Sec.~\ref{sec:pointerr}. The total map error is
divided into three parts: residual noise, signal error, and
pixelization error, and the effect of pointing error to each part is
shown.
\par
\tablenote b Residual noise is the power in the difference
(destriped map $-$ destriped noiseless map). It is dominated by
binned white noise.
\par
\tablenote c Signal error is the destriping error caused by the
signal (destriped noiseless map $-$ binned noiseless map).
\par
\tablenote d Pixelization error is caused by insufficient sampling
of the pixel temperature. We measure it from (binned noiseless map
$-$ smoothed input map).
\par
\label{tab:pointerr}
\end{table*}

\begin{table*}[!tbp]
 \global\advance\tableno by 1 \tabl {\csc Effect of
satellite pointing errors on total polarization error maps$^{\rm
a}$} \par \setbox\tablebox=\vbox{
   \newdimen\digitwidth
   \setbox0=\hbox{\rm 0}
   \digitwidth=\wd0
   \catcode`*=\active
   \def*{\kern\digitwidth}
   \newdimen\signwidth
   \setbox0=\hbox{+}
   \signwidth=\wd0
   \catcode`!=\active
   \def!{\kern\signwidth}
\halign{\hbox to 2in{#\leaderfil}\tabskip=2em&
     \hfil#\hfil\tabskip=2em&
     \hfil#\hfil\tabskip=0.5em&
     \hfil#\hfil\tabskip=2em&
     \hfil#\hfil\tabskip=0.5em&
     \hfil#\hfil\tabskip=0.5em&
     \hfil#\hfil\tabskip=0pt\cr
\noalign{\doubleline} \omit&&&&\multispan3\hfil\csc Error
components\hfil\cr \noalign{\vskip -3pt}
\omit&&&&\multispan3\hrulefill\cr \noalign{\vskip 3pt}
\omit\hfil$\sigma$\hfil&Model&Total error&White noise
estimate&Residual noise$^{\rm b}$&Signal error&Pixelization error\cr
\noalign{\vskip 3pt\hrule\vskip 5pt} \omit 0'   &          &
*$85.4361$ & *$85.0429$ & *$85.3790$ & **$0.3644$ & **$3.4872$ \cr
\omit 0'.5 & *uncorr. & *$-0.0444$ & *$-0.0543$ & *$-0.0398$ &
*$+0.0024$ & *$-0.0949$ \cr
\omit 0'.5 &  *corr.  & *$+0.1591$ &
*$+0.1428$ & *$+0.1364$ & *$+0.0439$ & *$+0.5112$ \cr
\noalign{\vskip 5pt\hrule\vskip 3pt}}}
\enddtable
\tablenote a We show here how the total RMS error in the
polarization maps changed by adding pointing errors.  Polarization
error amplitude is the square root of the square sum for the two
polarization amplitudes: $|err_{pol}| = \sqrt{err_Q^2+err_U^2}$.
\par
\tablenote b See Table~5 for the descriptions of the columns.
\par
\label{tab:pointerr2}
\end{table*}

\begin{figure*}[!tbp]
    \begin{center}
    \includegraphics*[width=1.0\textwidth]{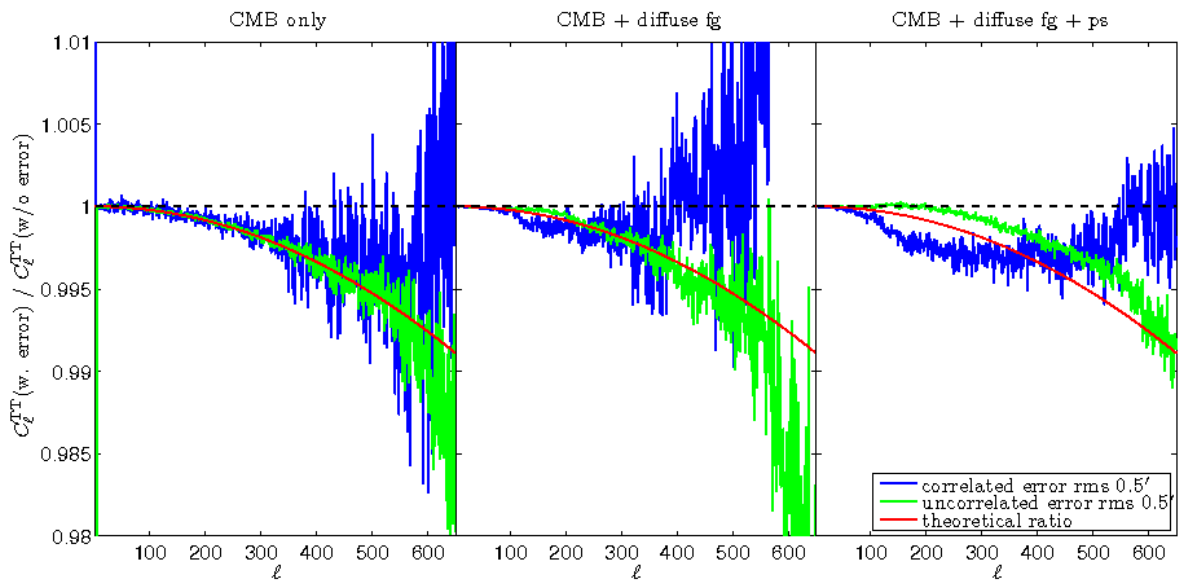}
    \end{center}
\caption{TT power spectra of the $0\farcm5$ pointing error cases
divided by the corresponding spectra without pointing errors. All
spectra are for the binned noiseless maps (fg = foreground, ps =
point source).  These plots show that for the CMB only case the
pointing errors are well approximated by an effective pixel window
(due to larger effective pixel size). However the simple model fails
in the presence of secondary signals.}
\label{fig:pointerr_and_spectra}
\end{figure*}

\begin{figure*}[!tbp]
\begin{center}
  \includegraphics*[width = 1.0\textwidth]{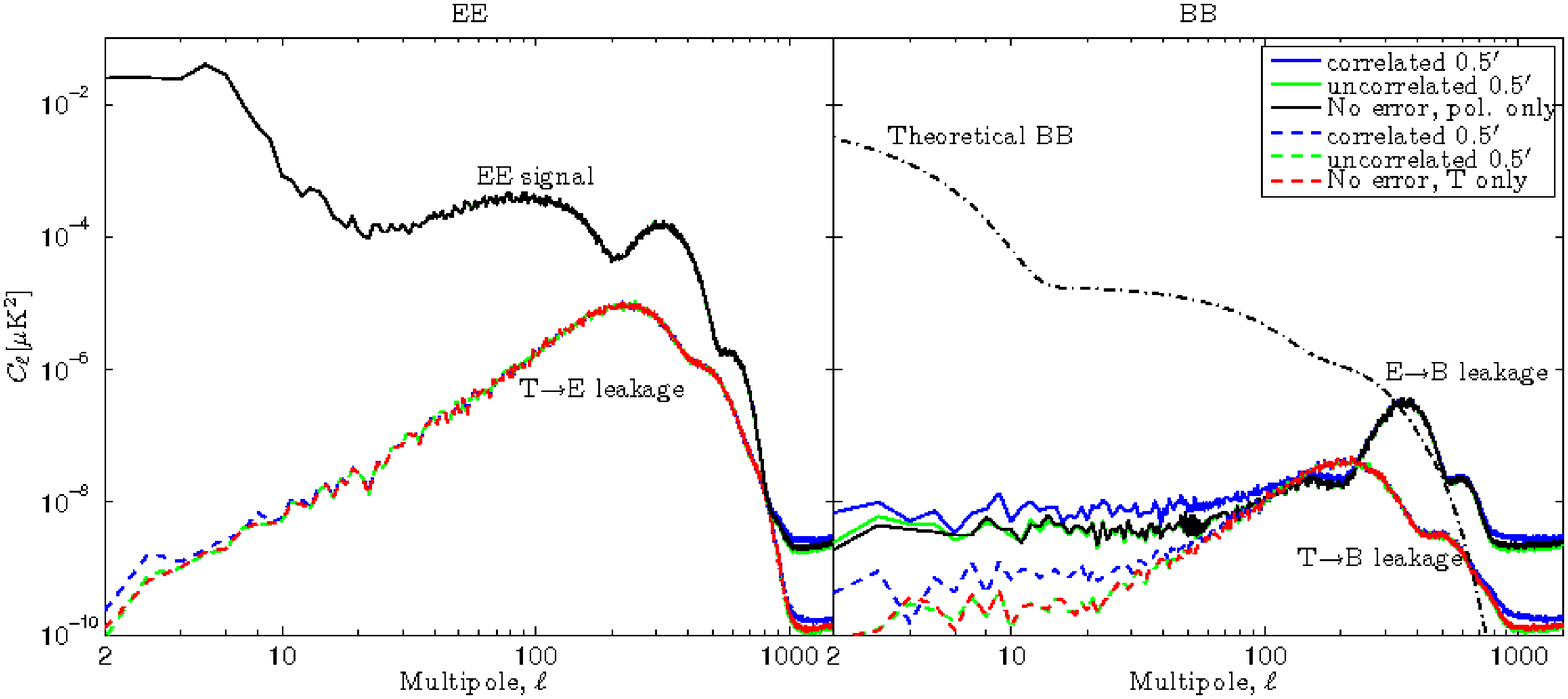}
  \caption[X]{Polarization angular power spectra of the maps made
  from the noiseless temperature-only TOD and the difference maps between
  these maps and the ones made from the noiseless TOD with polarization,
  showing the $E$-mode polarization signal, and the $T \rightarrow E$,
  $T \rightarrow B$ and the $E \rightarrow B$ leakage effects.  For
  comparison, we show a
  theoretical B mode spectrum corresponding to a 10\% tensor-to-scalar
  ratio and including lensing from $E$.
    }
  \label{fig:EE+BB_CLs}
\end{center}
\end{figure*}

We then studied the angular power spectra of noiseless maps binned
with additional pointing errors. Should the naive model derived from
an infinitely dense homogeneous hit distribution hold, the only
difference between spectra from different cases would be the added
smoothing. We found this to be the case approximately when binning
only the CMB signal. Adding diffuse foregrounds and point sources
broke the approximation to a large part, presumably due to sharper
structure and lack of statistical isotropy in these foreground
components. The decreasing trend from extra smoothing remained still
visible. In all cases the random pointing error model corresponded
to the smoothing approximation considerably better. The spectra are
displayed in Fig. \ref {fig:pointerr_and_spectra}.  Pointing errors
contribute also to the aliasing effect ($\ell$-mode coupling due to
uneven distribution of hits), which adds power to high $\ell$, but
the effect is negligible in comparison to the residual noise.

For analyzing the effect on polarization maps and power spectra we
repeated the study using the asymmetric beam TOD, since for that
case a reference TOD containing Stokes I only was available. See
Table~6. For correlated pointing errors the increase of the residual
noise error is larger by almost an order of magnitude for
polarization than for temperature (Table~5). This effect is in line
with the estimate based on white noise. The effect is probably due
to a relatively small number of pixels where the sampling of
polarization directions is not very good, and thus the noise effects
get magnified for polarization. The other effects on the map total
RMS error are of the same magnitude for polarization maps as for
temperature maps.

The $T \rightarrow P$ and $E \rightarrow B$ leakage was studied the
same way as in Sect.~\ref{subsubsec:beam_mismatch}. The effect of
pointing error on this leakage is shown in Fig.~\ref{fig:EE+BB_CLs}.
The effect on $T \rightarrow E$ leakage is negligible, and the
largest effect is on the $E \rightarrow B$ leakage, as anticipated.
This should be compared to the magnitude of the kind of $B$ mode we
might expect to detect with {\sc Planck}.  In the right-hand panel
of Fig.~\ref{fig:EE+BB_CLs} the leakage is compared to the
theoretical $B$-mode spectrum that we introduced earlier (see
Fig.~\ref{fig:bb}). It can be seen that the leakage to $B$ is small
compared to this signal for $\ell \leq 300$. On the other hand, the
pointing error effect on this leakage becomes small for $\ell \geq
200$.

Based on these results, pointing errors of $\leq 0\farcm5$ do not
appear as a major concern for the 30\,GHz channel.

\begin{figure}[!tbp]
    \begin{center}
    \resizebox{\hsize}{!}{\includegraphics{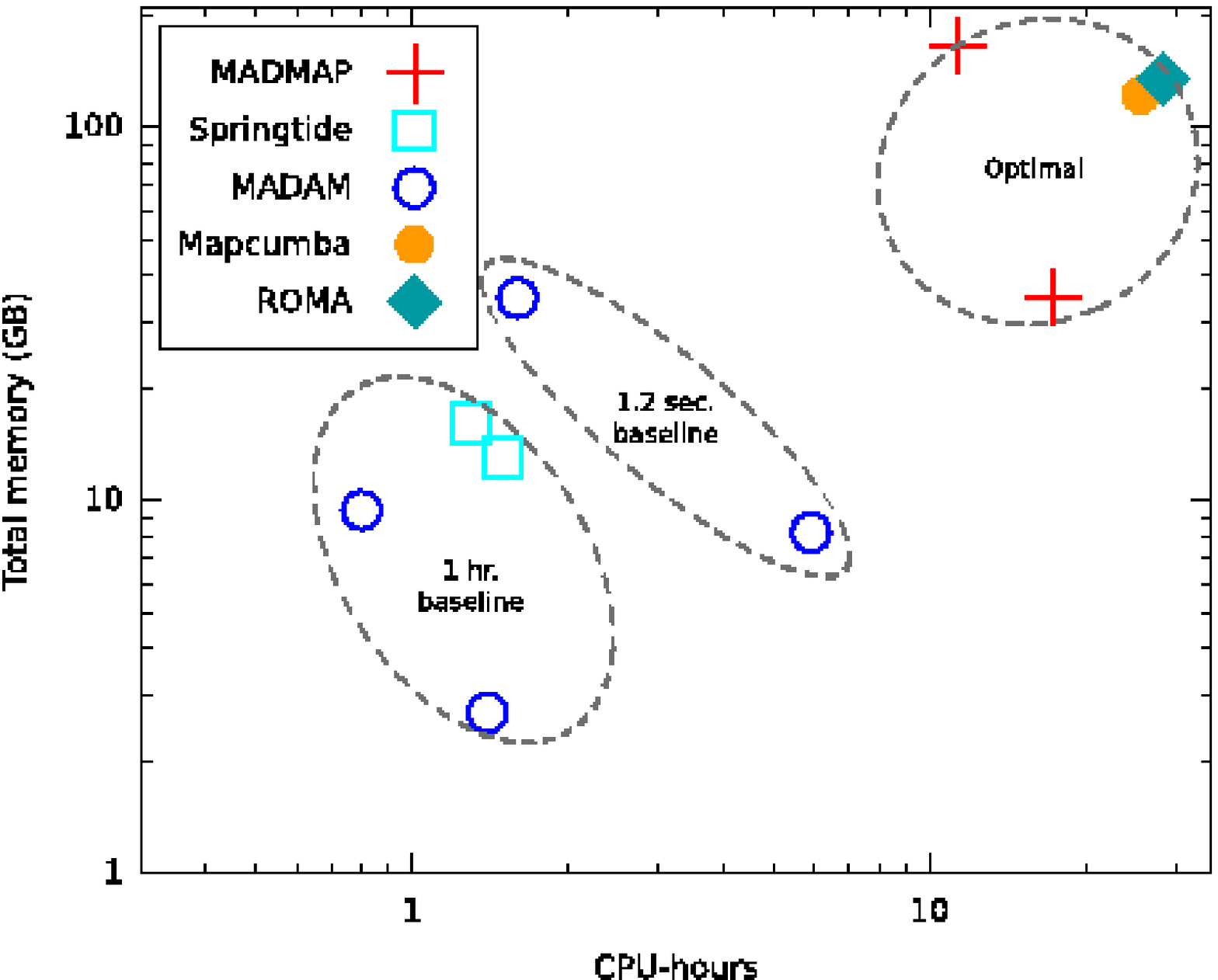}}
    \end{center}
\caption{Codes with longer baselines can run with fewer resources.
Individual codes can be tuned to be compact in memory or to run
quickly.} \label{fig:cpuh_vs_memory}
\end{figure}

\section{Computing resources}

Timing tests of codes were run on the NERSC\footnote{National Energy
Research Scientific Computing Center, http://www.nersc.gov.}
computer Jacquard, an Opteron cluster running Linux, chosen because
it is similar in architecture to the present and future machines in
the {\sc Planck} Data Processing Centers (DPCs).  Results are shown
in Table~7 and in Fig.~\ref{fig:cpuh_vs_memory}.   Different codes
have varying demands for processing power, memory, network
interconnect, and disk \hbox{I/O}.  Different computers will provide
these resources with varying performance, which can make a
significant difference in the runtime of a particular code. For this
reason, the Jacquard timing tests should be viewed as a useful
metric, but not an absolute prediction of performance. Note however
that many of the codes perform similar tasks (for example, each must
read the TOD), so a change in the performance between machines (in
this example, in disk I/O) will have less impact on the relative
performance measured between codes.  In the real-world application
of these codes, they will be tuned to the particular hardware
available.

\begin{table*}[!tbp]
\global\advance\tableno by 1 \tabl {\csc Resource Requirements$^{\rm
a}$} \par \setbox\tablebox=\vbox{
   \newdimen\digitwidth
   \setbox0=\hbox{\rm 0}
   \digitwidth=\wd0
   \catcode`*=\active
   \def*{\kern\digitwidth}
   \newdimen\signwidth
   \setbox0=\hbox{+}
   \signwidth=\wd0
   \catcode`!=\active
   \def!{\kern\signwidth}
\halign{\hbox to 2in{#\leaderfil}\tabskip=2em&
     \hfil#\hfil\tabskip=2em&
     \hfil#\hfil\tabskip=0.5em&
     \hfil#\hfil\tabskip=2em&
     \hfil#\hfil\tabskip=0.5em&
     \hfil#\hfil\tabskip=0.5em&
     \hfil#\hfil\tabskip=0pt\cr
\noalign{\doubleline} \omit&&\multispan2\hfil\csc Memory
[GB]\hfil&\multispan3\hfil\csc Time\hfil\cr \noalign{\vskip -3pt}
\omit&&\multispan2\hrulefill&\multispan3\hrulefill\cr
\noalign{\vskip 3pt} \omit\hfil\csc Code\hfil&\# Proc.&Per
Processor&Total&Wall Clock [s]&Total [CPU-hr]&I/O [CPU-hr]\cr
\noalign{\vskip 3pt\hrule\vskip 5pt} Springtide (A$^{\rm b}$)&
*16&*1.0&*16.1&*293&*1.3&**\dots$^{\rm c}$\cr Springtide (B)&
*16&*0.8&*13\phantom{.0}&*336&*1.5&**\dots\cr Madam (1-hr baseline,
A)&  *16&*0.6&**9.4&*176&*0.8&**0.27\cr Madam (1-hr baseline, B)&
*16&*0.2&**2.7&*318&*1.4&**0.64\cr Madam (1.2\,s baseline,
A)&*16&*2.2&*34.8&*362&*1.6&**0.23\cr Madam (1.2\,s baseline,
B)&*16&*0.5&**8.2&1167&*5.9&**0.29\cr MADmap (A)&
*96&*1.7&166\phantom{.0}&*427&11.4&**0.25\cr MADmap (B)&
*32&*1.1&*35.2&1953&17.4&**0.71\cr MapCUMBA&
*64&*1.9&122\phantom{.0}&1441&25.6&**9.73\cr ROMA&
*64&*2.1&134.4&1590&28.3&15.9\cr \noalign{\vskip 5pt\hrule\vskip
3pt}}}
\enddtable
\tablenote a Results are for runs on the NERSC Jacquard computer.
Typical run-to-run variations in total CPU-hours (central processing
unit) are 10\%, determined from Madam runs.  In all tests, $N_{\rm
side} = 512$ ($I, Q, U$) maps were made from 12\,months of simulated
observations from four LFI 30\,GHz detectors.
\par
\tablenote b Several codes can trade processing cost for memory
usage.  For a single code, we mark the timing run optimized for CPU
by (A) and the run optimized for memory by (B).
\par
\tablenote c In Springtide, reading the TOD from disk occurs
simultaneously with the compression of the TOD to ring-maps.  This
makes a separate evaluation of the I/O difficult.
\par
\label{tab:timing}
\end{table*}

\begin{table*}[!tbp]
\global\advance\tableno by 1 \tabl {\csc LFI 30\,GHz mapmaking
Throughput$^{\rm a}$} \par \setbox\tablebox=\vbox{
   \newdimen\digitwidth
   \setbox0=\hbox{\rm 0}
   \digitwidth=\wd0
   \catcode`*=\active
   \def*{\kern\digitwidth}
   \newdimen\signwidth
   \setbox0=\hbox{+}
   \signwidth=\wd0
   \catcode`!=\active
   \def!{\kern\signwidth}
\halign{\hbox to 2in{#\leaderfil}\tabskip=0pt&
     \hfil#\hfil\tabskip=2em&
     \hfil#\hfil\tabskip=0pt\cr
\noalign{\doubleline} \omit\hfil Code\hfil&Max. Simultaneous
Maps&Maps/day$^{\rm b}$\cr \noalign{\vskip 3pt\hrule\vskip 5pt}
Springtide (A)  &*8 &2359\cr Springtide (B)  &*8 &2057\cr Madam
(1-hr baseline, A)    &*8 &3927\cr Madam (1-hr baseline, B)    &*8
&2173\cr Madam (1.2 s baseline, A)   &*4 &*954\cr Madam (1.2 s
baseline, B)   &*8 &*592\cr MapCUMBA    &*2 &*119\cr MADmap (A) &*1
&*269\cr MADmap (B)  &*4 &*176\cr MapCUMBA &*2 &*119\cr ROMA &*1
&*108\cr \noalign{\vskip 5pt\hrule\vskip 3pt}}}
\enddtable
\tablenote a Assumptions: 1) 128 processors with total 256\,GB RAM
are allocated to mapmaking ($\sim$50\% of the LFI DPC goal
computer); 2)~code performance is taken from Table~7;  3)~computer
resources are allocated in 16-processor nodes.
\par
\tablenote b Obtained by dividing the number of seconds in a day by
the cpu usage in Table~7, taking into account the number of utilized
nodes, processors per node, and number of simultaneous maps.\par
\label{tab:throughput}
\end{table*}

In early 2009, the LFI DPC will have at least 128 processors with
1\,GB per processor (128\,GB total memory), with a  goal of
256~processors and 2\,GB per processor (512\,GB total memory), plus
10\,TB of disk space.  The HFI DPC expects to have 250~processors
and 8\,GB per processor (2000 GB\,total memory), plus 200\,TB of
disk space.

To compare the fast mapmaking capabilities of our codes, we used the
data of Table~7 to calculate the rate that 30\,GHz maps could be
produced in 24~hours, assuming that 50\% of the processors and the
memory of the LFI DPC goal computer would be available.  Results are
shown in Table~8.  The optimal codes (MADmap, MapCUMBA, ROMA) could
produce a couple of hundred maps per day, compared to between
several hundred and a couple of thousand for the destripers
(Springtide, Madam).

For the LFI and HFI DPCs, the most challenging mapmaking problem is
to make a map from 14~months of observations (the nominal {\sc
Planck} mission time) using 12\,detectors (the maximum for any
channel) at 70 and 217\,GHz.   We have estimated the total memory
and disk space required by the mapmaking codes in these extreme
cases.

For the LFI 70\,GHz channel, a naive scaling up of the size of the
TOD from the 30\,GHz timing tests (considering number of detectors,
sampling rate, and mission duration) yields total memory
requirements 8.3 times those shown in Table~7.  The Madam team made
more detailed estimates based on counting the size of allocated
arrays. For the code running in standard configuration with 1.2-s
baselines, the estimate yields a total memory requirement of $\sim
230$\,GB, compared to the naive scaling estimate of 270\,\hbox{GB}.
In split-mode, where the data are first destriped with a 1.2-s
baseline in three groups of four detectors, then combined in a
second destriping using 1-hr baselines, the memory requirement is
100\,\hbox{GB}.   For Madam with 1-hr baselines, the memory estimate
is 30\,\hbox{GB}.   In split-mode and for 1-hr baselines, the
estimate is less reliable because of uncertainties in the
performance of Madam's data compression system.   For a code using
the full detector pointing, the disk space required to store the TOD
is 0.87\,\hbox{TB}.   For the compressed satellite pointing sampled
at 1\,Hz, the requirement drops to 0.12\,\hbox{TB}.

For the HFI  217\,GHz channel, the naive scaling yields memory
requirements 21.5 times those shown in Table~7.  The Madam detailed
estimate yields 1070\,GB for 1.2-s baselines, 460\,GB for 1.2-s
baselines in split mode, and 100\,GB for 1-hr baselines.   Using
full pointing, the TOD disk requirement is 2.25\,TB, but drops to
0.32\,TB for compressed pointing.

\section{Conclusions}
In this paper we have presented results from a simulation that we
designed to determine how mapmaking codes handle four aspects of
real {\sc Planck} data: non-axially-symmetric beams, detector sample
integration, temperature fluctuations induced by the sorption
cooler, and pointing errors. Five mapmaking codes of two basic types
(destriping and optimal) were used in this study. We generated one
year long streams of observations representing four LFI 30~GHz
detectors. This simulation round (Trieste simulation) was number
four in a series of mapmaking comparison studies that the {\sc
Planck} CTP group has undertaken.

None of our mapmaking codes made an attempt to deconvolve the beam
effects from its output map. Therefore the smoothing effects of the
beam and sample integration showed up similarly in all our maps.
This is, however, a complicated smoothing, because every map pixel
has its own effective beam. We also made a thorough examination of
the temperature and polarization cross-couplings that arose from the
mismatch of the beams. These effects were also similar in all our
maps.

Our conclusions are:
\begin{enumerate}
\item Our studies showed that the temperature to polarization
cross-coupling of CMB signal caused a detectable bias in the TE
spectrum of the CMB map. In the EE spectrum the effect was small
compared to the residual noise of the map. The $E\rightarrow B$
cross-coupling of CMB produces a spurious signal whose magnitude in
intermediate and large multipoles (at $\ell \geq 300$ in this case)
seems to exceed the magnitude of the CMB $B$-mode signal that we
might expect to detect with LFI 30~GHz detectors. However, in this
range of multipoles the detection of the $B$-mode is difficult due
to map noise anyway, and at low multipoles where the signal-to-noise
ratio is higher the effect is small. Our study shows that the
spectrum biases that $T\rightarrow P$ and $E\rightarrow B$
cross-couplings cause may lead to errors in the cosmological
studies, if not corrected. We show a method in this paper that can
correct part of these effects in the spectrum domain. Map-domain
methods that address the beam convolution issues properly have been
proposed by other authors (see Sect.~\ref{subsubsec:correction}).

\item Signal error (error that mapmaking couples to the output map
from the small-scale signal structure) is the part of the output map
where beam effects appear differently in the maps of different
mapmaking codes. In optimal and short baseline Madam maps this error
is larger than in the destriper maps of long baselines (Madam and
Springtide). Signal error is, however, a small effect in a
high-resolution map if we compare it to the residual noise of the
map. Techniques to decrease the signal error were thoroughly
examined in our earlier study (Ashdown et al.~\cite{Ash07b}).

\item Based on the results of this study, cooler noise or pointing
errors of $\leq 0\farcm5$ do not appear as a major concern for the
30\,GHz channel.
\end{enumerate}

Five mapmaking codes (two destripers and three optimal codes) have
been developed and compared in four {\sc Planck} CTP simulation
rounds (Cambridge, Helsinki, Paris, and Trieste simulations). We
give the following summary of these studies:
\begin{enumerate}
\item At the end of a long process, the only essential difference
found so far between codes that affects accuracy---as assessed by
the RMS residual between an input binned map and the output map---is
baseline length, shown in Fig.~\ref{fig:kevin}.  This assumes that
all maps have been made at sufficiently high resolution that the
effects of sub-pixel structure are small (see Ashdown et
al.~\cite{Ash07b}). (Optimal codes are somewhat more affected [in an
RMS residual sense] by sub-pixel structure than are destripers, but
this is not an issue at sufficiently high map resolution.) However,
the fundamental difference between destriping and optimal codes is
in their assumptions about, and handling of, noise.   Some of the
future tests listed in Sect.~\ref{subsec:future} will probe this
difference more intensively than ones performed to date.

\item All optimal codes give essentially indistinguishable results.
Madam, a destriper with adjustable baseline length, gives the same
result as the optimal codes when the baseline is set short enough.
Reliable noise estimation is required for these codes.

\item Resource requirements for the codes vary by more than an order
of magnitude in the important categories (memory per processor,
total memory, elapsed time, and total CPU-hours), as shown in
Table~7 and Fig.~\ref{fig:cpuh_vs_memory}.   Codes set up to deal
with pointing information in multiple forms have a practical
advantage, in that the resources required in memory, I/O speed, and
processor speed can be matched to the resources available (memory,
I/O speed, and CPU time).

\item It seems inevitable that {\sc Planck} will need both a fast and
``resource light'' mapmaking code for everyday use, and an
``ultimate accuracy'' but ``resource heavy'' code, for final data
products.
\end{enumerate}

\subsection{Future Tests}\label{subsec:future}

Our final round of simulations was quite realistic compared to
earlier rounds, nevertheless there are still many instrumental and
mission effects, both subtle and blatant, that have not been
included, some of which are listed below.  Further tests of
mapmaking must be performed.

\begin{itemize}

\item Gaps in TOD.

\item High-pass filtering of TOD (e.g., to reduce low frequency noise before mapmaking, to deconvolve the bolometer frequency response in
HFI). High-pass filtering will make the noise matrix $\mathbf{C}_n$
of Eq. (\ref{eq:binned_map_in_map}) non-diagonal.

\item Leakage of galactic temperature signal to the polarization maps
due to the bandpass mismatch of the LFI radiometers.

\item Beam sidelobes.

\item Cross-polarization leakage of the detectors, and polarization angle errors.

\item Other forms of random noise (e.g., ``popcorn noise'').

\item Features in the noise power spectrum (e.g., microphonics, frequency
spikes). Note that some of these features may make the noise matrix
$\mathbf{C}_n$ of Eq. (\ref{eq:binned_map_in_map}) non-diagonal.

\item Changes in the baseline signal level (e.g., from telescope emission, background loading of
bolometers).

\item Realistic pointing errors (errors in the pointing
reconstruction)

\end{itemize}

The scope of this paper was restricted to the mapmaking from the LFI
30~GHz data. The {\sc Planck} mapmaking codes need, however, to be
able to make maps from both LFI and HFI observations. Therefore the
list includes mapmaking tests that are relevant for LFI, HFI or
both.

Future tests should be run on both a destriping code and an optimal
code.  As a practical matter, it will be efficient to run the less
resource intensive destriping code first to wring out any problems
in setting up the test.   It is no longer necessary to run tests on
multiple implementations of destriping or optimal codes, which have
been shown to give the same answer.

Generally speaking, destriping algorithms assume that the noise
power spectrum is white at high frequency.  The shape of the low
frequency power spectrum is not important, so long as the low
frequency noise may be fit by a series of offsets (or baselines).
The duration of these offsets varies from 1~second to 1~hour in the
codes we have tested here, with a trade-off between accuracy and
resource usage. For baseline durations longer than 1~minute,
destripers need no information about the noise power spectrum.   By
contrast, the optimal codes are not restricted to white noise at
high frequency, but must be informed of the shape of the noise power
spectrum.

Deconvolution of a bolometer time constant exercises a difference in
these approaches.  Bolometers have a response lag to a change in the
sky similar to the time for the beam to cross a fixed point.  This
effect smears the sky image in the direction of the scan.  A time
stream deconvolution filter, applied prior to mapmaking, can repair
this effect on the sky signal, but at the cost of correlating the
noise at small scales.  We have tested the effect of this with
simulated HFI 100\,GHz data, where the bolometer time constants are
the longest, and where the effect is the most severe (G\'orski et
al.~\cite{Gor08}).

\begin{acknowledgements}
  The work reported in this paper was done by the CTP Working Group of
  the {\sc Planck} Consortia. {\sc Planck} is a mission of the
  European Space Agency.
  The authors would like to thank Osservatorio Astronomico di Trieste (OAT) for its
  hospitality in May 2006 when the CTP Working Group met to undertake
  this work.
  This research used resources of the National Energy Research
  Scientific Computing Center, which is supported by the Office of
  Science of the U.S. Department of Energy under Contract
  No.~DE-AC02-05CH11231.
  We acknowledge the use of version 1.1 of the {\sc Planck\/} sky model,
  prepared by the members of {\sc Planck\/} Working Group 2, available at http://people.sissa.it/$\backsim$planck/reference$\_$sky (CMB
  and extra-Galactic emission), and
  http://www.cesr.fr/$\backsim$bernard/PSM/ (Galactic emission).
  We acknowledge the use of the CAMB (http://camb.info) code for generating theoretical
  CMB spectra. We thank Anna-Stiina Sirvi\"{o} for help with CAMB.
  The authors thank Aniello Mennella for providing the sorption cooler
  data.
  This work has made use of the {\sc Planck} satellite simulation
  package (Level-S), which is assembled by the Max Planck Institute
  for Astrophysics {\sc Planck} Analysis Centre (MPAC).
  This work has been partially supported by Agenzia Spaziale Italiana under ASI contract {\sc Planck} LFI Activity of Phase
  E2 and by the NASA LTSA Grant NNG04CG90G.
  This work was supported in part by the Academy of Finland grants 205800, 214598, 121703, and
  121962. RK is supported by the Jenny and Antti Wihuri Foundation.
  HKS and TP thank Waldemar von Frenckells stiftelse, HKS and TP thank
  Magnus Ehrnrooth Foundation, and EK and TP thank V\"ais\"al\"a
  Foundation for financial support.
  Some of the results in this paper have been derived using the
  HEALPix package (G\'orski et al.~\cite{Gor05a}).
  The US {\sc Planck\/} Project is supported by the NASA  Science Mission Directorate.
\end{acknowledgements}

\appendix
\section{Effects of beams in the angular power spectrum of a CMB map, an analytic
model} \label{sec:model}

In this Appendix we try to explain the mechanisms by which beam
mismatches generate cross-couplings between the Stokes $I$, $Q$, and
$U$ of a CMB map. We designed a simple analytic model to answer this
question. We start by making some simplifying assumptions:
\begin{itemize}
\item Scanning is from one ecliptic pole to another along the meridians.
\item Sky pixels are observed same number of times by all four
LFI 30\,GHz detectors.
\item All hits fall in the centers of the pixels.
\item Detector polarization angles are uniformly spaced in 180$^{\circ}$ (spacing is 45$^{\circ}$ for the four 30\,GHz detectors).
\item Detector noise is white, Gaussian, and every detector has the same noise RMS.
\item We consider only the main beams. Far sidelobes are ignored.
\item The widths of the main beams are so small that we can apply the small-scale limit and compute the convolution of the beam and sky in the plane wave Fourier space (flat sky approximation).
\item Our asymmetric main beams have elliptical Gaussian responses.
\end{itemize}

The assumptions lead to the following consequences:
\begin{itemize}
\item Ellipses are invariant to 180$^{\circ}$ rotations, therefore
the scanning direction does not matter. The results of our model
are the same in north-to-south and south-to-north scans.
\item In all parts of the sky a detector beam has the same orientation with respect to the local meridian.
\item The model does not account for the effects of the pixel sampling (pixel window function and pixelization error).
\end{itemize}

We expect that in spite of the simplifying assumptions, our model
gives a good description of the observations in most parts of the
sky. The accuracy of the model should be good in those parts of the
sky that are scanned approximately along meridians (ecliptic equator
areas). The accuracy of our model may be worse in the ecliptic pole
areas, where there is more spread of the scanning directions. In
reality, the polar regions with multiple scanning directions are a
relatively small fraction of the total sky (see
Sect.~\ref{sec:results} and the third row of Fig.~\ref{fig:maps}).

The Stokes parameters $Q$ and $U$ at a point in the sky are defined
in a reference coordinate system ($\ve{e}_{\theta}$,$\ve{e}_{\phi}$,
$\ve{n}$), where the unit vector $\ve{e}_{\theta}$ is along the
north-south meridian (increasing $\theta$), $\ve{e}_{\phi}$ is along
the increasing $\phi$, and $\ve{n}$ points to the sky (G\'orski et
al.~\cite{Gor05b}). In this reference coordinate system the
polarization angle of a detector is the angle from the local
north-south meridian to its polarization sensitive direction
(anti-clockwise rotation in the reference coordinate system). The
polarization angles of the four detectors (LFI-27a, LFI-27b,
LFI-28a, LFI-28b)\footnote{Detectors LFI-27a and LFI-27b share a
horn and the detectors LFI-28a and LFI-28b share the other horn.}
are $\psi$, $\psi$+90$^{\circ}$, $\psi$+45$^{\circ}$, and
$\psi$+135$^{\circ}$. In the north-south scanning the value of
$\psi$ is the same as the angle from the scanning direction to the
U-axis of LFI-27a (see Fig.~\ref{fig:focalplane}). This angle is
67\pdeg5 (see the caption of the same figure). For the opposite
scanning direction $\psi$ is increased by 180$^{\circ}$.

The observations from a sky pixel $p$ can be given as
 \begin{eqnarray}
\ve{y}(\ve{n}_p) = \left(\begin{array}{c}
y_1(\ve{n}_p)\\y_2(\ve{n}_p)\\y_3(\ve{n}_p)\\y_4(\ve{n}_p)\end{array}\right)
= \left(\begin{array}{c} I_1 + Q_1\cos(2\psi) + U_1\sin(2\psi)
\\I_2 - Q_2\cos(2\psi) - U_2\sin(2\psi)\\I_3 - Q_3\cos(2\psi) + U_3\sin(2\psi)\\I_4 + Q_4\cos(2\psi) -
U_4\sin(2\psi)\end{array}\right) = \nonumber\\
= \left(\begin{array}{c} I_1 + e^{-i2\psi}\cdot Z_1/2 +
e^{+i2\psi}\cdot Z^{\star}_1/2
\\I_2 - e^{-i2\psi}\cdot Z_2/2 - e^{+i2\psi}\cdot Z^{\star}_2/2\\I_3 - ie^{-i2\psi}\cdot Z_3/2
+ ie^{+i2\psi}\cdot Z^{\star}_3/2\\I_4 + ie^{-i2\psi}\cdot Z_4/2 -
ie^{+i2\psi}\dot Z^{\star}_4/2\end{array}\right).\label{eq:A1_1}
 \end{eqnarray}
Here indices (1,2,3,4) refer to the detectors (LFI-27a, LFI-27b,
LFI-28a, LFI-28b), $X^{\star}$ is the complex conjugate of $X$, $Z
\equiv Q + iU$ is the complex polarization field, and $Z^{\star}
\equiv Q - iU$. In our model the detectors measure different values
of the Stokes parameters from the same point of the sky, because
their beams are different. The observations stay the same in both
scanning directions, because the beams have 180$^{\circ}$ rotational
symmetry and a change of $\psi$ by 180$^{\circ}$ does not change
anything in Eq.~(\ref{eq:A1_1}).

Let us define a ``pointing matrix'' $\mathbf{A}$
 \beq
 \mathbf{A} \equiv \left(\begin{array}{ccc}
 1 & e^{-i2\psi} & e^{+i2\psi}\\1 & -e^{-i2\psi} & -e^{+i2\psi}\\1 & -ie^{-i2\psi} & ie^{+i2\psi}\\1 & ie^{-i2\psi} & -ie^{+i2\psi}\\
 \end{array}\right)
\label{eq:A1_2}
 \eeq
and a Stokes triplet $\ve{s}_p$ of the pixel $p$
 \beq
 \ve{s}_p \equiv
 \left(\begin{array}{c}I(\ve{n}_p)\\Z(\ve{n}_p)\\Z^{\star}(\ve{n}_p)\\\end{array}\right).
\label{eq:A1_3}
 \eeq
The unit vector $\ve{n}_p$ points to the center of the pixel.

For uniform white Gaussian noise, the Stokes map ($\hat{\ve{s}}_p$)
can be recovered from the detector observations as
(Tegmark~\cite{Teg97})
 \beq
 \hat{\ve{s}}_p =
 \left(\mathbf{A}^{\dagger}\mathbf{A}\right)^{-1}\mathbf{A}^{\dagger}\ve{y}(\ve{n}_p).
 \label{eq:A1_4}
 \eeq
Here $\mathbf{X}^{\dagger}$ is the hermitian conjugate of matrix
$\mathbf{X}$. Matrix $\mathbf{A}^{\dagger}\mathbf{A}$ is the
$\mathbf{N}_{\rm obs}$ matrix of the pixel $p$ (see the footnote of
Sect.~\ref{sec:results}). The assumptions that we made in the
beginning of this Appendix lead to diagonal $\mathbf{N}_{\rm obs}$
($\mathbf{A}^{\dagger}\mathbf{A} = diag[4,1,1]$).

We can now solve $\hat{I}(\ve{n}_p)$ and $\hat{Z}(\ve{n}_p)$ from
Eq. (\ref{eq:A1_4}).
 \bea
\hat{I}(\ve{n}_p) = \frac{\left(I_1+I_2+I_3+I_4\right)}{4} +
\frac{e^{-i2\psi}\left[(Z_1-Z_2) - i(Z_3-Z_4)\right]}{8} +\nonumber\\
+ \frac{e^{+i2\psi}\left[(Z^{\star}_1-Z^{\star}_2) +
i(Z^{\star}_3-Z^{\star}_4)\right]}{8}\nonumber\\ \label{eq:A1_5}
 \eea
 \bea
\hat{Z}(\ve{n}_p) = \frac{\left(Z_1+Z_2+Z_3+Z_4\right)}{4} +
\frac{e^{+i2\psi}\left[(I_1-I_2) + i(I_3-I_4)\right]}{2} +\nonumber\\
+ \frac{e^{+i4\psi}\left[(Z^{\star}_1+Z^{\star}_2) -
(Z^{\star}_3+Z^{\star}_4)\right]}{4}.\nonumber\\ \label{eq:A1_6}
 \eea

The first terms on the right hand sides (of Eqs.~(\ref{eq:A1_5}) and
(\ref{eq:A1_6})) provide the temperature and polarization signals
that we want. The other terms of Eq.~(\ref{eq:A1_5}) represent the
cross-coupling of polarization signal to the temperature. This
cross-coupling can usually be ignored because the coupling
coefficient is typically small and the polarization signal is weaker
than the temperature signal.

The second and the third terms of Eq.~(\ref{eq:A1_6}) correspond to
$T\rightarrow P$ and spin-flip coupling errors (Hu et
al.~\cite{Hu03}). The second term shows that the temperature signal
can pollute the polarization signal if the beams of the two
detectors sharing a horn do not match. Beams can be different from
horn to horn, but no leakage of temperature to polarization occurs
as long as the beams of a horn are identical.

The spin-flip term (the third term of Eq.~(\ref{eq:A1_6})) produces
cross-coupling between $E$- and $B$-mode polarizations. We show it
later in this Appendix. In our model spin-flip coupling occurs if
there is a horn-to-horn mismatch between the total beams. Here the
total beam means the sum of the responses of the two beams of a
horn.

Convolution of the detector beam and the sky produces the observed
temperature and polarization fields ($I_1, Z_1, I_2, Z_2, I_3, Z_3,
I_4, Z_4$). We assumed that this convolution can be computed as a
multiplication between the corresponding flat-sky Fourier-domain
quantities. This is a good approximation in a small flat patch
around the point of observation. We can therefore write $I_1(\ve{k})
= B_1(\ve{k})I(\ve{k})$, $Z_1(\ve{k}) = B_1(\ve{k})Z(\ve{k})$,
$I_2(\ve{k}) = B_2(\ve{k})I(\ve{k})$, $Z_2(\ve{k}) =
B_2(\ve{k})Z(\ve{k})$ and so on. Here $\ve{k}$ is the 2-dimensional
plane wave vector and ($B_1(\ve{k})$, $B_2(\ve{k})$,
$B_3(\ve{k})$,$B_4(\ve{k})$) are the Fourier-domain representations
of the (flat-sky) responses of the (LFI-27a, LFI-27b, LFI-28a,
LFI-28b) main beams. Because we assumed ideal elliptic Gaussian main
beams, their functional forms are well known and they are
real-valued for all widths, ellipticities and orientations (Fosalba
et al.~\cite{Fos02}). The Stokes parameters of the sky are given by
$I(\ve{k})$ and $Z(\ve{k})$.

We can now write the relation between the Stokes parameters of the
map and those of the sky (for Fourier-domain quantities):
 \beq
\left(\begin{array}{c}\hat{I}(\ve{k})\\\hat{Z}(\ve{k})\\\hat{Z}^{\star}(\ve{k})\\\end{array}\right)
= \left(\begin{array}{ccc}B_{\rm s}/4 & e^{-i2\psi}B^{\star}_{\rm
c}/8 &
e^{+i2\psi}B_{\rm c}/8\\
e^{+i2\psi}B_{\rm c}/2 & B_{\rm s}/4 &
e^{+i4\psi}B_{\rm d}/4\\
e^{-i2\psi}B^{\star}_{\rm c}/2 & e^{-i4\psi}B_{\rm d}/4 & B_{\rm
s}/4\\\end{array}\right)\cdot
\left(\begin{array}{c}I(\ve{k})\\Z(\ve{k})\\Z^{\star}(\ve{k})\\\end{array}\right).
 \label{eq:A1_7}
 \eeq
Here $B_{\rm s}$, $B_{\rm d}$, and $B_{\rm c}$ are Fourier-domain
quantities and they are defined in terms of the beam responses
 \beq
B_{\rm s}(\ve{k})\equiv
B_1(\ve{k})+B_2(\ve{k})+B_3(\ve{k})+B_4(\ve{k}) \label{eq:A1_7a}
 \eeq
  \beq
B_{\rm d}(\ve{k})\equiv
B_1(\ve{k})+B_2(\ve{k})-B_3(\ve{k})-B_4(\ve{k}) \label{eq:A1_7b}
 \eeq
 \beq
B_{\rm c}(\ve{k})\equiv \left(B_1(\ve{k})-B_2(\ve{k})\right) +
i\left(B_3(\ve{k})-B_4(\ve{k})\right). \label{eq:A1_7c}
 \eeq
 \beq
B^{\star}_{\rm c}(\ve{k})\equiv \left(B_1(\ve{k})-B_2(\ve{k})\right)
- i\left(B_3(\ve{k})-B_4(\ve{k})\right). \label{eq:A1_7d}
 \eeq

The 3$\times$3 matrix of the right-hand side of Eq. (\ref{eq:A1_7})
gives the mapping from the Stokes parameters of the sky to the
Stokes parameters of our map. We can therefore call it the Mueller
matrix. We use the symbol $\mathbf{M}(\ve{k})$ for it here. In a
real experiment each pixel of the map has its own Mueller matrix,
because detector hit counts, sampling of pixel area, and beam
orientations are different for different pixels. Our model, however,
leads to one Mueller matrix that applies in all pixels of the sky.

Next we make the connection between the polarization field
($Z(\ve{k})$) and the fields of the $E$- and $B$-mode polarization
($E(\ve{k})$ and $B(\ve{k})$) (Zaldarriaga \& Seljak~\cite{Zal97}).
We include the temperature anisotropy field ($I(\ve{k}$)) in these
equations.
 \bea
\left(\begin{array}{c}I(\ve{k})\\Z(\ve{k})\\Z^{\star}(\ve{k})\\\end{array}\right)
= \left(\begin{array}{ccc} 1 & 0 & 0\\
0 & e^{+i2\phi_{\ve{k}}} & ie^{+i2\phi_{\ve{k}}}\\
0 & e^{-i2\phi_{\ve{k}}} &
-ie^{-i2\phi_{\ve{k}}}\\\end{array}\right)\cdot
\left(\begin{array}{c}I(\ve{k})\\E(\ve{k})\\B(\ve{k})\\\end{array}\right)
= \nonumber\\
= \mathbf{R}\left(\phi_{\ve{k}}\right)\cdot
\left(\begin{array}{c}I(\ve{k})\\E(\ve{k})\\B(\ve{k})\\\end{array}\right).
 \label{eq:A1_8}
 \eea
Here the angle $\phi_{\ve{k}}$ is defined through $\ve{k} =
\left(k_x,k_y\right) =
k\left(\cos(\phi_{\ve{k}}),\sin(\phi_{\ve{k}})\right)$ and $k \equiv
|\ve{k}|$ is the magnitude of the wave vector. The definition of the
3$\times$3 matrix $\mathbf{R}\left(\phi_{\ve{k}}\right)$ is evident
from the equation.

We assume that the sky emission (like the CMB) is statistically
isotropic. Therefore the ensemble mean of its power spectrum does
not depend on the direction of $\ve{k}$ and it can be computed in a
$3\times 3$ matrix form
 \beq
 \mathbf{C}(k)\equiv \left(\begin{array}{ccc} C^{TT}(k) & C^{TE}(k) & C^{TB}(k)\\
C^{TE}(k) & C^{EE}(k) & C^{EB}(k)\\C^{TB}(k) & C^{EB}(k) &
C^{BB}(k)\\\end{array}\right) = \left\langle
\left(\begin{array}{c}I(\ve{k})\\E(\ve{k})\\B(\ve{k})\\\end{array}\right)
\left(\begin{array}{c}I(\ve{k})\\E(\ve{k})\\B(\ve{k})\\\end{array}\right)^{\dagger}\right\rangle.
 \label{eq:A1_9}
 \eeq
Here $\langle X \rangle$ is the ensemble mean of $X$. We can now
compute the ensemble mean of the power spectrum matrix of the
observed map as
 \beq
\left\langle \hat{\mathbf{C}}(k)\right\rangle =
\frac{1}{2\pi}\int_{0}^{2\pi}\left\langle
\left(\begin{array}{c}\hat{I}(\ve{k})\\\hat{E}(\ve{k})\\\hat{B}(\ve{k})\\\end{array}\right)
\left(\begin{array}{c}\hat{I}(\ve{k})\\\hat{E}(\ve{k})\\\hat{B}(\ve{k})\\\end{array}\right)^{\dagger}\right\rangle
d\phi_{\ve{k}}.
 \label{eq:A1_10}
 \eeq

Eqs. (\ref{eq:A1_7}) and (\ref{eq:A1_8}) give a relation between
($I,E,B$) of the sky and ($\hat{I},\hat{E},\hat{B}$) of our map.
With the help of this relation, we can write an equation that gives
us the power spectrum of our map if we know the power spectrum of
the sky.
 \bea
\left\langle \hat{\mathbf{C}}(k)\right\rangle =
\frac{1}{2\pi}\int_{0}^{2\pi}\mathbf{R}^{\dagger}(\phi_{\ve{k}})\cdot
\mathbf{M}(\ve{k})\cdot \mathbf{R}(\phi_{\ve{k}})\cdot \mathbf{C}(k)
\cdot \nonumber\\ \cdot\mathbf{R}(\phi_{\ve{k}})\cdot
\mathbf{M}^{\dagger}(\ve{k})\cdot
\mathbf{R}^{\dagger}(\phi_{\ve{k}})\cdot d\phi_{\ve{k}}.
 \label{eq:A1_11}
 \eea

The $3\times 3$ power spectrum matrix is symmetric. It therefore
contains 6 different component spectra $C^{XY}(k)$. We can arrange
the power spectra of the input sky in a 6-element column vector
 \beq
\ve{c}(k) \equiv
\left(\begin{array}{c}C^{TT}(k)\\C^{TE}(k)\\C^{TB}(k)\\C^{EE}(k)\\C^{EB}(k)\\C^{BB}(k)\\\end{array}\right).
 \label{eq:A1_12a}
 \eeq
Similarly we can build a 6-element column vector $\ve{\hat{c}}(k)$
of the power spectra of the observed map. Because the power spectrum
of the input sky does not depend on the direction of the wave
vector, Eq. (\ref{eq:A1_11}) allows us to construct a $6\times 6$
window matrix that, when applied to the spectrum $\ve{c}(k)$ of the
input sky, gives the spectrum $\ve{\hat{c}}(k)$ of the observed map.

We can use the above results to predict the angular power spectrum
(spherical harmonic domain) of our map, if we know the angular power
spectrum of the sky. The predicted spectrum will include the effects
of the beams, but it will not include the effects from the sampling
of the pixel areas (pixel window function and pixelization error).
The steps to compute the angular power spectrum prediction are the
following.
\begin{enumerate}
\item Assume that we know the angular power spectrum multipoles of the input sky.  Arrange them in a 6-element column vector at every multipole $\ell$ (as in Eq. (\ref{eq:A1_12a})).
\item The magnitude $k$ of the Fourier wave vector is the continuous limit of the integer $\ell$ labeling the spherical harmonics multipoles $a_{\ell m}$.
\item For a given multipole $\ell$, set $k = \ell$ in the Mueller matrix $\mathbf{M}(\ve{k})$.
\item For the same multipole, step the angle $\phi_{\ve{k}}$ from 0 to 2$\pi$ and compute the $6\times 6$ window matrix from the integral of Eq. (\ref{eq:A1_11}).
\item Apply the window matrix to the angular power spectrum vector of the input sky. The result is the angular power
spectrum vector of the observed map (of the same multipole $\ell$).
\item Repeat the previous steps to all multipoles of interest.
\end{enumerate}

Temperature to polarization cross-coupling is an important effect of
the beam mismatch. We can use the above procedure to predict this
effect too. Instead of using the full 6-element input spectrum
vector, we use a vector all of whose elements except $C^{TT}$ are
zero, and carry out the prediction steps as before.

We can invert the $6\times 6$ window matrix and compute a corrected
angular power spectrum from the spectrum of the observed map. In the
corrected spectrum the beam effects have been deconvolved out and it
is an approximation of the spectrum of the input sky.

We made a software code that implements the above
\textit{prediction} and \textit{correction} steps. Instead of
building 6-element power spectrum vectors, we assigned the full
$3\times 3$ power spectrum matrix column-wise in the vector
$\ve{c}(k)$. This leads to 9-element power spectrum vectors and
$9\times 9$ window matrices. They have unnecessary redundancy, but
this approach makes the implementation simple and straightforward.
The extra computational cost of larger vectors and matrices is
unimportant.

Using Eq. (\ref{eq:A1_11}) we can compute the $9\times 9$ window
matrix (see Appendix A of Hamimeche \& Lewis~\cite{Ham08} for
relevant equations of matrix vectorization)
 \beq
\mathbf{W}(k) = \frac{1}{2\pi}\int_{0}^{2\pi}\left(\begin{array}{ccc} S_{11}\cdot\mathbf{S} & S_{12}\cdot\mathbf{S} & S_{13}\cdot\mathbf{S}\\
S_{21}\cdot\mathbf{S} &
S_{22}\cdot\mathbf{S} & S_{23}\cdot\mathbf{S}\\
S_{31}\cdot\mathbf{S} & S_{32}\cdot\mathbf{S} &
S_{33}\cdot\mathbf{S}\\\end{array}\right)\cdot d\phi_{\ve{k}}.
 \label{eq:A1_12b}
 \eeq
Here $\mathbf{S} = \mathbf{S}(\ve{k})\equiv
\mathbf{R}^{\dagger}(\phi_{\ve{k}})\cdot \mathbf{M}(\ve{k})\cdot
\mathbf{R}(\phi_{\ve{k}})$ is a $3\times 3$ matrix (from Eq.
(\ref{eq:A1_11})) and $S_{ij} = S_{ij}(\ve{k})$ are its elements.
The predicted spectrum of the observed map (from the spectrum of the
input sky) becomes now
 \beq
 \ve{\hat{c}}(k) = \mathbf{W}(k)\cdot \ve{c}(k).
 \label{eq:A1_12c}
 \eeq
We further convolved the predicted spectra with HEALPix pixel window
function to model the smoothing due to sampling of the pixel area.

The prediction equation (Eq. (\ref{eq:A1_12c})) is simple to invert
for the correction. Before applying the inverse $\mathbf{W}(k)$ to
the spectrum of the observed map, we deconvolved the observed
spectrum with the HEALPix pixel window function.

We used this code in Sect.~\ref{sec:results} of this paper and
computed predictions from the input spectrum and corrections from
the map spectrum. These predictions we used to explain the beam
mismatch effects and beam window functions that we detected in the
spectra of our CMB maps. The correction capability of our model is
demonstrated in Fig.~\ref{fig:te_3}.

\subsection{Spin-flip coupling} \label{subsec:spin_flip}
If the (2,3) and (3,2) elements of the Mueller matrix are non-zero
(see Eq. (\ref{eq:A1_7})), $Z^{\star}$ will mix with $Z$ and vice
versa. This is the spin-flip coupling (Hu et al.~\cite{Hu03}). It
generates a $B$-mode in the map even if the input sky has no
$B$-mode power in it (like the CMB sky of this study). We can
describe this situation with the following general expression (for
the Fourier-domain quantities)
 \beq
\left(\begin{array}{c}\hat{Z}\\\hat{Z}^{\star}\\\end{array}\right) =
\left(\begin{array}{cc} a & b
\\b^{\star} & a\\\end{array}\right)\cdot
\left(\begin{array}{c}Z\\Z^{\star}\\\end{array}\right).
 \label{eq:A1_13}
 \eeq
We assume that $a$ is real as the (2,2) and (3,3) elements of the
Mueller matrix (see Eq. (\ref{eq:A1_7})). The quantity $b$ and its
complex conjugate represent the (2,3) and (3,2) elements of the
Mueller matrix. If we assume that the input sky contains no $B$
mode power (ie. $Z$ and $Z^{\star}$ arise from $E$ mode only), we
can write for the $E$ and $B$ modes of the mixed fields $\hat{Z}$
and $\hat{Z}^{\star}$ (use Eq. (\ref{eq:A1_8}) for the relation
between the $E$ and $B$ modes and the complex polarization fields)
 \bea
\left(\begin{array}{c}\hat{E}\\\hat{B}\\\end{array}\right) =
\left(\begin{array}{cc} e^{+i2\phi_{\ve{k}}} & ie^{+i2\phi_{\ve{k}}}
\\e^{-i2\phi_{\ve{k}}} & -ie^{-i2\phi_{\ve{k}}}\\\end{array}\right)^{-1}\cdot\left(\begin{array}{cc} a & b
\\b^{\star} & a\\\end{array}\right)\cdot \nonumber\\
\cdot\left(\begin{array}{cc} e^{+i2\phi_{\ve{k}}} &
ie^{+i2\phi_{\ve{k}}}
\\e^{-i2\phi_{\ve{k}}} & -ie^{-i2\phi_{\ve{k}}}\\\end{array}\right)\cdot
\left(\begin{array}{c}E\\0\\\end{array}\right). \label{eq:A1_14a}
 \eea
This equation can be put in the following form
 \beq
\left(\begin{array}{c}\hat{E}\\\hat{B}\\\end{array}\right) =
\left(\begin{array}{c}a\\0\\\end{array}\right)E +
\frac{1}{2}\left(\begin{array}{c}b^{\star}e^{+i4\phi_{\ve{k}}}+be^{-i4\phi_{\ve{k}}}\\i\left(b^{\star}e^{+i4\phi_{\ve{k}}}-be^{-i4\phi_{\ve{k}}}\right)\\\end{array}\right)E.
 \label{eq:A1_14b}
 \eeq
The equation shows that spin-flip coupling (non-zero $b$) creates
$B$ mode in the mixed field. In addition, the spin-flip coupling
influences the $E$ mode itself in the sense that $\hat{E}$ of $b =
0$ and $\hat{E}$ of $b \ne 0$ are different.

\end{document}